\documentclass[oldversion]{aa}
\usepackage{graphicx}
\usepackage{txfonts}
\usepackage{aalongtable}
\usepackage{natbib}
\bibpunct{(}{)}{;}{a}{}{,}
\begin{document}

\title{The nature of \ion{N}{v} absorbers at high redshift
\subtitle{}
  \thanks{Based on observations made with VLT/Kueyen telescope
    ESO, Paranal, Chile}
} 
\author{C. Fechner \inst{1} \and P. Richter \inst{1}}
\institute{Institut f\"{u}r Physik und Astronomie, Universit\"{a}t Potsdam,
  Haus 28, Karl-Liebknecht-Str. 24/25, 14476 Potsdam, Germany\\
  \email{cfech@astro.physik.uni-potsdam.de, prichter@astro.physik.uni-potsdam.de}}
\offprints{C.\ Fechner,\\ \email{cfech@astro.physik.uni-potsdam.de}}
\date{Received 19 June 2008 / Accepted 14 January 2009}

\abstract{
We present a study of \ion{N}{v} absorption systems at $1.5 \lesssim z \lesssim 2.5$ in the spectra of 19 QSOs, based on data obtained with the VLT/UVES instrument.
Our analysis includes both the absorbers arising from the intergalactic medium, as well as systems in the vicinity of the background quasar.

We construct detailed photoionization models to study the physical conditions and abundances in the absorbers and to constrain the spectral hardness of the ionizing radiation.

The rate of incidence for intervening \ion{N}{v} components is $d\mathcal{N}/dz = 3.38 \pm 0.43$, corresponding to $d\mathcal{N}/dX = 1.10 \pm 0.14$.
The column density distribution function is fitted by the slope $\beta = 1.89 \pm 0.22$, consistent with measurements of \ion{C}{iv} and \ion{O}{vi}.
The narrow line widths ($b_\ion{N}{v} \sim 6\,\mathrm{km\,s}^{-1}$) imply photoionization rather than collisions as the dominating ionization process.
The column densities of \ion{C}{iv} and \ion{N}{v} are correlated but show different slopes for intervening and associated absorbers, which indicates different ionizing spectra.
Associated systems are found to be more metal-rich, denser, and more compact than intervening absorbers. 
This conclusion is independent of the adopted ionizing radiation.
For the intervening \ion{N}{v} systems we find typical values of $\mathrm{[C/H]} \sim -0.6$ and $n_{\element{H}} \sim 10^{-3.6}\,\mathrm{cm}^{-3}$ and sizes of a few kpc, while for associated \ion{N}{v} absorbers we obtain $\mathrm{[C/H]} \sim +0.7$, $n_{\element{H}} \sim 10^{-2.8}\,\mathrm{cm}^{-3}$ and sizes of several 10\,pc.
The abundance of nitrogen relative to carbon [N/C] and $\alpha$-elements like oxygen and silicon [N/$\alpha$] is correlated with [N/H], indicating the enrichment by secondary nitrogen.
The larger scatter in [N/$\alpha$] in intervening systems suggests an inhomogeneous enrichment of the IGM. 
There is an anti-correlation between [N/$\alpha$] and [$\alpha$/C], which could be used to constrain the initial mass function of the carbon- and nitrogen-producing stellar population.

} 

\maketitle

\section{Introduction}

Studying elemental abundances in various environments is important for understanding the enrichment history and chemical evolution of astronomical objects.
Of particular interest are metal-poor objects, which are thought to be very old and to conserve the abundances pattern of early enrichment.
In the local Universe these are, for example, extremely metal-poor halo stars (see \citet{beerschristlieb2005} for a review).
At higher redshifts, the abundance pattern of damped Ly$\alpha$ (DLA) systems can be studied in detail \citep[reviewed by][]{wolfeetal2005}.
Metal-rich objects like quasars at very high redshift suggest that a substantial amount of metals must have been produced at very early epochs, implying star formation prior to the redshift of observation \citep[see the review by][]{hamannferland1999}.
Measured abundances can then be compared to predicted chemical yields of different stellar populations \citep[e.g.][]{woosleyweaver1995, iwamotoetal1999, hegerwoosley2002, limongichieffi2003, chieffilimongi2004, umedanomoto2005, kobayashietal2006} optionally weighted by an initial mass function (IMF) to constrain the production site of the heavy elements. 

An important tracer of the chemical evolution is the abundance of nitrogen.
Nitrogen is created by the conversion of carbon and oxygen during the CNO cycle in hydrogen-burning zones.
If the seed carbon and oxygen atoms are produced in the star itself, the nitrogen is called primary.
Secondary nitrogen is produced if carbon and oxygen have already been present in the gas that formed the star.
Since $\alpha$-elements (elements built up of $^4$He-cores, e.g.\ oxygen and silicon) are mainly produced by Type II supernova explosions of massive stars, they are released into the surrounding medium at earlier epochs than carbon and nitrogen, which are mainly created during nuclear burning within massive and intermediate-mass stars.
Stars with $M > 8\,M_{\sun}$ dominate the production of carbon at early epochs, while stars with lower masses are the main production sites for nitrogen released with a given time lag.
Because of this time delay, \citet{edmundspagel1978} first proposed to use [N/O] as an age indicator.
The sensitivity for such an age indicator depends on the duration of the time lag.
\citet{pettinietal2002b} estimate that the $\sim 250\,\mathrm{Myr}$ inferred by \citet{henryetal2000} are too short to explain all the DLA systems with low [N/$\alpha$] abundances.
However, the time delay increases if stellar rotation is taken into account shifting the production site of primary nitrogen to stars with lower masses and longer evolutionary time scales \citep{meynetmaeder2002, meynetpettini2004}.

Furthermore, [N/$\alpha$] traces the production process of nitrogen.
While the presence of only primary nitrogen leads to constant [N/$\alpha$] as found in low-metallicity DLA systems \citep[e.g.][]{prochaskaetal2002, centurionetal2003, petitjeanetal2008, pettinietal2002b, pettinietal2008}, [N/$\alpha$] increases with metallicity if secondary nitrogen contributes.
The early release of $\alpha$-elements also leads to an $\alpha$-enhancement with respect to carbon in the intergalactic medium \citep[IGM; $\mathrm{[Si,\,O/C]} > 0$, e.g.][]{aguirreetal2004, aguirreetal2008, simcoeetal2004, simcoeetal2006}, indicating enrichment by SN II explosions of massive stars \citep[e.g.][]{matteuccicalura2005, qianwasserburg2005}.
Such estimates are based on the detection of \ion{C}{iv}, \ion{Si}{iv}, and \ion{O}{vi} features in QSO absorption spectra whose measured column densities can be transformed into abundances applying appropriate photoionization corrections.
However, due to the sparse detection rate of nitrogen features the nitrogen abundance of the high-redshift IGM is rather unknown.

Though absorption features of \ion{N}{i} and/or \ion{N}{ii} are usually observed in DLA systems, nitrogen is rarely detected in optically thin (i.e.\ $N_{\ion{H}{i}} \lesssim 10^{17}\,\mathrm{cm}^{-2}$) intergalactic absorption systems.
Features of \ion{N}{v} are often identified in absorption systems proximate to the background quasar \citep[e.g.\ the recent work of][]{foxetal2008}.
Because of its rather high ionization potential ($77.5\,\mathrm{eV} = 5.7\,\mathrm{Ryd}$ to be created, $97.9\,\mathrm{eV} = 7.2\,\mathrm{Ryd}$ to be destroyed) this ion is generated by the hard radiation of the close-by QSO.
Moreover, quasars are generally metal-rich \citep[e.g.][]{dietrichetal2003} and about solar [N/H] has been measured in associated systems \citep[e.g.][]{dodoricoetal2004} making \ion{N}{v} easily detectable in proximate absorbers.
Together with other highly ionized species such as \ion{O}{vi}, \ion{N}{v} absorption is also detected in the galactic halo and high-velocity clouds \citep{savageetal1997, indebetouwshull2004b}.
\citet{foxetal2007} found 3 DLA systems exhibiting weak \ion{N}{v} features in a sample of \ion{O}{vi}-bearing DLAs \citep[see also][]{lehneretal2008}.
Recently, \citet{prochaskaetal2008} report on the detection of \ion{N}{v} absorption close to long-duration gamma-ray bursts.
The generally narrow line widths of the features imply photoionization of the gas by the gamma-ray burst afterglow.

In the IGM the metallicity and therefore the nitrogen content is lower and the ionizing UV background is substantially softer than unfiltered QSO radiation.
Thus, \ion{N}{v} features are expected to be weak and will arise mainly in the Ly$\alpha$ forest making the detection and identification difficult.
One might expect to detect \ion{N}{v} absorption preferentially when the absorbing material is exposed to a locally hardened radiation field.
This may occur in the vicinity of a foreground QSO close to the line of sight generating a so-called transverse proximity effect.
Due to the locally higher level of ionization this effect should produce a decrease in the density of neutral hydrogen, which is, however, rarely detected \citep[e.g.][]{liskewilliger2001, croft2004, schirberetal2004, galleranietal2008}.
Yet, combined analyses considering also \ion{He}{ii} \citep{jakobsenetal2003, worsecketal2007} and metal line absorption \citep{goncalvesetal2008} demonstrate the existence of the transverse proximity effect in spectral hardness.
Studying the line of sight towards HE~2347-4342 \citet{worsecketal2007} find a foreground QSO at $z=2.282$ with a transverse proper distance of $1.76\,\mathrm{Mpc}$ in the vicinity of the intergalactic \ion{N}{v} system at $z=2.2753$. 

In this paper we present a systematic search for \ion{N}{v} at $z \sim 2$ in QSO absorption spectra aiming to probe the nitrogen abundance in the IGM and to test whether \ion{N}{v} is a tracer for spectral hardness of the ionizing radiation.
Therefore we select both intervening \ion{N}{v} absorbers and systems associated to the background quasar.
Their statistical and physical properties are studied and the characteristics of the different absorber classes are extensively compared.
After presenting our selection procedure and the resulting sample in Sect.\ \ref{sample}, we investigate the statistical properties of \ion{N}{v} absorption lines like number density and column density distribution in Sect.\ \ref{statistics}.
In order to estimate elemental abundances we compute photoionization models as described in Sect.\ \ref{mod_procedure}.
The results are presented in Sect.\ \ref{results} and the implications for the enrichment of the IGM and for the presence of hard radiation sources close to \ion{N}{v} absorption systems are discussed in Sect.\ \ref{discussion}.
Our conclusions are presented in Sect.\ \ref{conclusions}.
Throughout the paper we use a cosmology with $\Omega_{\mathrm{M}} = 0.3$, $\Omega_{\mathrm{\Lambda}} = 0.7$, and $H_0 = 70\,\mathrm{km\,s}^{-1}\,\mathrm{Mpc}^{-1}$.
Abundances are given in the notation $\mathrm{[X/Y]} = \log (\mathrm{X/Y}) - \log (\mathrm{X/Y})_{\sun}$ with solar abundances taken from \citet{asplundetal2005}.

\section{Sample}\label{sample}

\begin{table}
  \caption[]{Sight lines and probed redshift ranges searched for \ion{N}{v} absorption systems and number of identified intervening and associated systems.
  }
  \label{qsos}
  $$
  \begin{array}{l c c c c c c}
    \hline\hline
    \noalign{\smallskip}
\mathrm{QSO} & z_{\mathrm{em}} & z_{\mathrm{min}} & z_{\mathrm{max}} & X & \#_{\mathrm{inter}} & \#_{\mathrm{asso}}\\     
    \noalign{\smallskip}
    \hline
    \noalign{\smallskip}
\textrm{HE~0001-2340} & 2.26 & 1.46 & 2.21 & 2.17 & 2 & 0 \\
\textrm{HE~0151-4326} & 2.79 & 1.55 & 2.73 & 3.67 & 1 & 0 \\
\textrm{HE~0940-1050} & 3.08 & 1.89 & 3.01 & 3.69 & 1 & 0 \\
\textrm{HE~1122-1648} & 2.41 & 1.46 & 2.35 & 2.65 & 0 & 0 \\
\textrm{HE~1158-1843} & 2.45 & 1.46 & 2.39 & 2.77 & 2 & 2 \\ 
\textrm{HE~1341-1020} & 2.13 & 1.46 & 2.08 & 1.77 & 0 &\phantom{^{\mathrm{a}}\,} 2\,^{\mathrm{a}} \\
\textrm{HE~1347-2457} & 2.61 & 1.46 & 2.55 & 3.29 & 1 & 0 \\
\textrm{HE~2217-2818} & 2.41 & 1.46 & 2.35 & 2.65 & 1 & 0 \\
\textrm{HE~2347-4342} & 2.89 & 1.76 & 2.83 & 3.40 & 1 & 4 \\
\textrm{PKS~0237-23 } & 2.23 & 1.46 & 2.18 & 2.08 & \phantom{^{\mathrm{b}}\,}3\,^{\mathrm{b}} & 0 \\
\textrm{PKS~0329-255} & 2.70 & 1.56 & 2.64 & 3.33 & 0 & 1 \\
\textrm{PKS~1448-232} & 2.22 & 1.49 & 2.17 & 1.96 & 3 & 0 \\
\textrm{PKS~2126-158} & 3.28 & 1.93 & 3.21 & 4.30 & 0 & 0 \\
\textrm{Q~0002-422  } & 2.77 & 1.55 & 2.71 & 3.58 & 0 & 0 \\
\textrm{Q~0109-3518 } & 2.40 & 1.46 & 2.34 & 2.61 & 1 & 0 \\
\textrm{Q~0122-380  } & 2.19 & 1.46 & 2.14 & 1.96 & 2 & 0 \\
\textrm{Q~0329-385  } & 2.44 & 1.46 & 2.38 & 2.74 & 2 & 1 \\
\textrm{Q~0420-388  } & 3.12 & 2.03 & 3.05 & 3.41 & 1 & 0 \\
\textrm{Q~0453-423  } & 2.66 & 1.46 & 2.60 & 3.46 & 0 & 1 \\
    \noalign{\smallskip}
    \hline
  \end{array}
  $$
\begin{list}{}{}
  \item[$^{\mathrm{a}}$] plus 2 associated mini-BAL systems
  \item[$^{\mathrm{b}}$] plus 2 damped Ly$\alpha$ systems with detected \ion{N}{v}
\end{list}
\end{table}
 
We search for \ion{N}{v} systems in the optical spectra of 19 QSOs taken with UVES at the VLT (ESO Large Program 166.A-0106(A)).
The spectra cover the optical range redward of the atmospheric cut-off, $\lambda \gtrsim 3500\,\mathrm{\AA}$, with high resolution ($R \approx 45\,000$) and  a typical signal-to-noise ratio of $S/N \sim 35 - 70$.
The data reduction has been performed by B.\ Aracil \citep{araciletal2004}.
We identify \ion{C}{iv} doublets and for each \ion{C}{iv} system we check for \ion{N}{v} features at the same redshift.
Therefore, our sample includes only absorption systems exhibiting \ion{N}{v} and \ion{C}{iv} simultaneously.
Moreover, we look for various additional transitions and estimate the column densities of all identified species performing Doppler profile fits.
For the complex associated system towards HE~2347-4342 we adopted the parameters estimated by \citet{fechneretal2004}.
The basic properties of the investigated lines of sight are summarized in Table \ref{qsos} including the redshift range that has been searched for intervening \ion{N}{v} systems and the corresponding absorption path length $dX = (1+z)^2/\sqrt{\Omega_{\mathrm{M}}(1+z)^3+\Omega_{\Lambda}}\,dz$.
The fitted line parameters are given in Tables \ref{coldens_inter} and \ref{coldens_asso} in the Appendix.

Systems within $|\Delta v |\le 5000\,\mathrm{km\,s}^{-1}$ are classified as associated.
Absorbers selected by this criterion may not be intrinsic to the quasar host, while systems with velocity shifts $\Delta v < -5000\,\mathrm{km\,s}^{-1}$ may be \citep[e.g.][]{richardsetal1999, misawaetal2007}.
Additional indicators to discriminate between intrinsic and intervening systems include partial coverage or time varying line profiles.
Few of our associated systems indeed show partial coverage (e.g.\ $z=2.4427$ towards HE~1158-1843) but none of the systems classified as intervening do.
\citet{misawaetal2007} investigated associated and intrinsic narrow absorption lines traced by \ion{C}{iv} in a large sample of 37 quasars, finding that roughly 15\,\% of the systems  classified as non-associated due to their velocity shift are probably intrinsic.
According to this estimate we would expect $\lesssim 3$ out of 22 non-associated \ion{N}{v} systems to be mis-classified.
This number is an upper limit since we count only systems showing \ion{C}{iv} and \ion{N}{v} absorption simultaneously whereas \citet{misawaetal2007} studied all systems traced by \ion{C}{iv} only.

The system at $z = 2.3520$ towards Q~0329-385 is probably mis-classified by the adopted velocity criterion.
Its \ion{H}{i} feature is unusually weak for exhibiting pronounced metal lines ($N_{\ion{H}{i}} \sim 10^{-13}\,\mathrm{cm}^{-2}$ in two components) and the velocity shift from the QSO emission redshift is roughly $-7670\,\mathrm{km\,s}^{-1}$.
Furthermore, photoionization modeling of this system (see Section \ref{mod_procedure}) leads to unusual high metallicity ($\mathrm{[N/H]} \sim 1.1$, when adopting an HM01 background) as well suggesting that it is probably intrinsic.
Therefore, we classify this absorber as associated \citep[see also][]{levshakovetal2008}.
Note that if we enlarge the velocity interval to select associated absorbers up to $|\Delta v | < 15\,000\,\mathrm{km\,s}^{-1}$, the one at $z=2.3520$ towards Q~0329-385 would be the only additional associated system selected.

\section{Line statistics}\label{statistics}

\subsection{Number density}

In 19 sight lines we find 21 intervening \ion{N}{v} systems in 61 individual components and 11 associated systems with 46 individual components, where we exclude two (sub-) DLAs towards PKS~0237-23 and two mini-BAL systems towards HE~1341-1020 \citep[see also][]{levshakovetal2008}.
This sample is \ion{N}{v} selected from 198 intervening and 30 associated \ion{C}{iv} systems for which \ion{N}{v} is (in principle) observable.
Thus, the fraction of intervening \ion{C}{iv} systems showing \ion{N}{v} absorption is $\sim 11\,\%$, while roughly $37\,\%$ of associated \ion{C}{iv} absorbers (within $5000\,\mathrm{km\,s}^{-1}$ from the background QSO) exhibit \ion{N}{v} as well.

The probed redshift path in total is $\Delta z = 18.06$ in the range $1.5 \le z \le 2.5$ corresponding to a total absorption length of $\Delta X = 55.50$.
Thus the rate of incidence for intervening \ion{N}{v} systems is $d\mathcal{N}_{\mathrm{sys}}/dz = 1.16 \pm 0.25$ and $d\mathcal{N}/dz = 3.38 \pm 0.43$ for individual components, corresponding to $d\mathcal{N}_{\mathrm{sys}}/dX = 0.38 \pm 0.08$ and $d\mathcal{N}/dX = 1.10 \pm 0.14$.
These numbers are calculated excluding a proximity zone of $|\Delta v | \le 5000\,\mathrm{km\,s}^{-1}$ for each line of sight.
If an enlarged proximity zone $|\Delta v | \le 15\,000\,\mathrm{km\,s}^{-1}$ is considered, the probed total redshift path and the corresponding total absorption length reduce to $\Delta z = 15.79$ and $\Delta X = 47.97$, leading to slightly higher number densities ($d\mathcal{N}_{\mathrm{sys}}/dX = 0.44 \pm 0.10$ and $d\mathcal{N}/dX = 1.27 \pm 0.61$, respectively).

Our sample is most likely incomplete due to blending with the Ly$\alpha$ forest.
Furthermore, we select only \ion{N}{v} systems exhibiting \ion{C}{iv} features and possibly neglect systems showing \ion{N}{v} but no corresponding \ion{C}{iv} absorption.
Therefore the actual rate of incidence might be higher, in particular for low column density features.
The column density distribution function for the intervening components shown in Fig.\ \ref{cddf} suggests that our sample may be complete for column densities $\log N_{\ion{N}{v}} \gtrsim 12.7$.
This subsample includes 41 individual components in 17 systems leading to $d\mathcal{N}/dz = 2.3\pm0.4$ and $d\mathcal{N}_{\mathrm{sys}}/dz = 0.9\pm0.2$, respectively, for $N_{\ion{N}{v}} \ge 4.5\cdot 10^{12}\,\mathrm{cm}^{-2}$.

The estimated number density of \ion{N}{v} components is roughly 6 times lower than that of \ion{C}{iv} as \citet{songaila2001} finds $d\mathcal{N}_{\ion{C}{iv}}/dX = 6.8 \pm 2.7$ in the same redshift range and half the number of \ion{Si}{iv} components ($d\mathcal{N}_{\ion{Si}{iv}}/dX = 3.5 \pm 1.0$), according to the values given in her Table 1.
The reduced rate of incidence for \ion{N}{v} in comparison to \ion{C}{iv} is roughly consistent with the fraction of \ion{C}{iv} systems showing \ion{N}{v} absorption ($\sim 0.11$).
This probably reflects the lower abundance of nitrogen with respect to carbon and silicon.
Due to the low nitrogen content fewer systems showing features of nitrogen are expected.

At low redshift ($z < 0.4$) \citet{danforthshull2008} detect \ion{N}{v} with a rate of incidence of $d\mathcal{N}/dX = 3\pm 1$.
Thus, \ion{N}{v} appears to be more numerous at low redshift than at $z \sim 2$.
At low redshift \ion{N}{v} is correlated with \ion{O}{vi}, which is mainly collisionally ionized due to shock heating contributing to the warm-hot intergalactic medium \citep[WHIM, see e.g.\ the review by][]{richteretal2008}.
Thus, \citet{danforthshull2008} conclude that \ion{N}{v} is also a tracer of the WHIM and therefore is collisionally ionized.
In contrast, \ion{O}{vi} as well as \ion{N}{v} are photoionized at $z \sim 2$ \citep[e.g.][discussion below]{bergeronetal2002, levshakovetal2003b, bergeronherbertfort2005, reimersetal2006}.
Therefore, \ion{N}{v} absorbers at high and low redshift may represent different populations.

\subsection{Column density distribution and line widths}

\begin{figure}
  \centering
  \resizebox{\hsize}{!}{\includegraphics[bb=55 570 370 775,clip=]{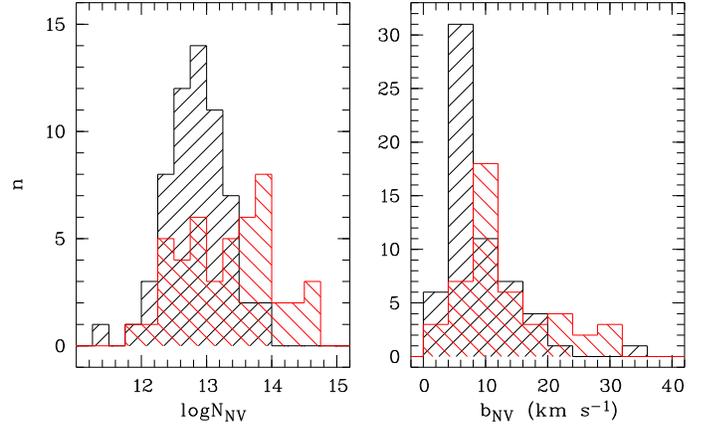}}
  \caption{Distribution of measured \ion{N}{v} column densities (left panel) and Doppler parameters (right panel) for individual components.
The black histograms present the values of the intervening systems, the red ones show the distributions for the associated \ion{N}{v} components.
  }
  \label{logN_histo}
\end{figure}

\begin{figure}
  \centering
  \resizebox{\hsize}{!}{\includegraphics[bb=40 370 285 530,clip=]{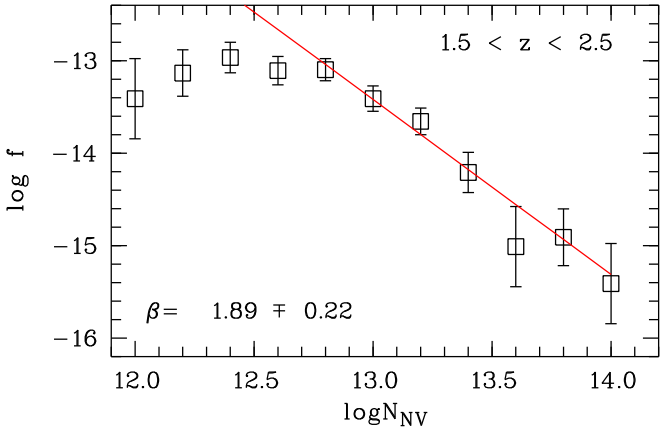}}
  \caption{Column density distribution function of intervening \ion{N}{v}.
For $\log N_{\ion{N}{v}} \ge 12.7$ die column density distribution function is fitted by a power law $f \propto N_{\ion{N}{v}}^{-\beta}$ with $\beta = 1.89\pm0.22$.
  }
  \label{cddf}
\end{figure}

The distributions of the observed \ion{N}{v} column densities of the intervening and associated components, respectively, are presented in the left panel of Fig.\ \ref{logN_histo}.
While the distribution for the intervening components is clearly peaked at $\log N_{\ion{N}{v}} \simeq 12.8$ and no components with $N_{\ion{N}{v}} > 10^{14}\,\mathrm{cm}^{-2}$ are detected, the column densities of associated systems are distributed more smoothly over a broader range with values up to $\sim 10^{14.8}\,\mathrm{cm}^{-2}$.

In case of intervening systems the distribution of the estimated Doppler parameters shown in the right panel of Fig.\ \ref{logN_histo} has a pronounced maximum at $\sim 6\,\mathrm{km\,s}^{-1}$ and a slight tail with values upto $\sim 20\,\mathrm{km\,s}^{-1}$.
If interpreted as thermal broadening, $b_{\ion{N}{v}} \sim 6\,\mathrm{km\,s}^{-1}$ corresponds to a temperature of $T \sim 30\,000\,\mathrm{K}$.
The narrow lines imply photoionization of nitrogen.
According to the computations of \citet{sutherlanddopita1993} and \citet{gnatsternberg2007}, in collisional ionization equilibrium the ionization fraction of \ion{N}{v} peaks at $T \simeq 2\cdot 10^{5}\,\mathrm{K}$, corresponding to a line width of $b \simeq 15\,\mathrm{km\,s}^{-1}$.
Only 10\,\% of the intervening components are broad enough to be collisionally ionized, clearly suggesting photoionization as dominant ionization process.
This, however, is probably a selection bias since the ionization fraction of \ion{C}{iv} in hot, collisionally ionized gas strongly decreases \citep[$f_{\ion{C}{iv}} \sim 0.03$ at $2\cdot10^5\,\mathrm{K}$;][]{gnatsternberg2007} while the dominating species is \ion{C}{v}.
Thus, our selection procedure favors photoionized intervening \ion{N}{v} absorbers.

Associated components on average show broader line widths. 
Their Doppler parameter distribution peaks at $\sim 10\,\mathrm{km\,s}^{-1}$ with a significant tail up to $\sim 30\,\mathrm{km\,s}^{-1}$.
Even though our sample is too small to draw robust conclusions there might be a bimodality in the $b$-parameter distribution showing one pronounced peak at $10\,\mathrm{km\,s}^{-1}$ corresponding to $T \sim 84\,000\,\mathrm{K}$, and a second smooth maximum at $\sim 22\,\mathrm{km\,s}^{-1}$.
The broader absorption lines may be affected by QSO outflows.
The fraction of components with $b > 15\,\mathrm{km\,s}^{-1}$, which may possibly be collisionally ionized, is $\sim 26\,\%$.

The column density distribution function is defined as the number $\mathcal{N}$ of absorbers within a given column density bin $\Delta N$ per observed absorption path length $X$, i.e.\ $f = \mathcal{N}/(\Delta N\cdot\Delta X)$ and is usually parameterized as $f = A\cdot N^{-\beta}$.
For the intervening components the column density distribution function is shown in Fig.\ \ref{cddf}.
Our sample appears to be complete down to $\log N_{\ion{N}{v}} \approx 12.7$.
Above this threshold we estimate $\beta = 1.89 \pm 0.22$ and $\log A = 11.2 \pm 2.9$.

At $z < 0.4$ the column density distribution function of \ion{N}{v} yields $\beta_{\ion{N}{v}} = 1.87\pm0.17$ \citep{danforthshull2008} in excellent agreement with the slope at $z \sim 2$.
Comparing to other species, this slope is slightly steeper than $\beta_{\ion{C}{iv}} = 1.44 \pm 0.05$  as estimated for \ion{C}{iv} absorption at $z \sim 3$ by \citet{ellisonetal2000} but fully consistent with $\beta_{\ion{C}{iv}} = 1.8 \pm 0.1$ found by \citet{songaila2001} independent of redshift in the range $1.5 < z < 5.5$.
Using an pixel optical depth-based technique \citet{songaila2005} finds $\beta_{\ion{C}{iv}} = -1.7$ considering only features with $\log N_{\ion{C}{iv}} > 13.0$ but a flatter slope of $-1.44$ if low column density absorbers are additionally taken into account and an incompleteness correction is applied.
Our sample of intervening \ion{N}{v} absorbers exhibits mainly high \ion{C}{iv} column densities (see below).
Thus, the agreement of the slopes at $N \gtrsim 10^{13}\,\mathrm{cm}^{-2}$ may suggest a common origin of both ions.
However, due to incompleteness in the low column density range we cannot conclude whether the \ion{N}{v} distribution function flattens at $\log N_{\ion{N}{v}} \lesssim 10^{13}\,\mathrm{cm}^{-2}$.
Our best-fit slope is also consistent with $\beta_{\ion{O}{vi}} = 1.7 \pm 0.5$ estimated by \citet{bergeronherbertfort2005} for \ion{O}{vi} absorbers with $13.0 < \log N_{\ion{O}{vi}} < 14.3$ at $z \sim 2.3$.
The consistency of the slope of the column density distribution function for strong absorbers of the three species \ion{C}{iv}, \ion{N}{v}, and \ion{O}{vi} indicates that they may have a common origin in the probed redshift range. 
At $z \sim 2$ this is likely photoionized gas.

\subsection{Comparison to \ion{C}{iv}}

\begin{figure*}
  \centering
  \resizebox{\hsize}{!}{\includegraphics[bb=40 630 480 775,clip=]{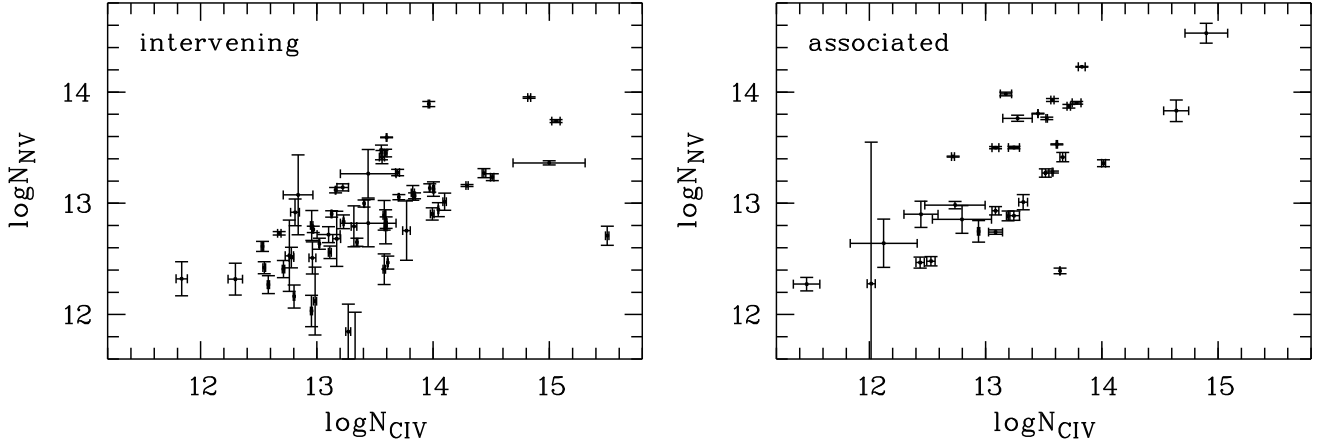}}
  \caption{Comparison of column densities of observed \ion{N}{v} and \ion{C}{iv} features for intervening (left) and associated (right panel) systems, respectively. The correlation for the intervening systems can be described by a slope of $0.40 \pm 0.07$ while the slope of the associated components is steeper yielding $0.64\pm0.10$.
  }
  \label{logN_NV_CIV}
\end{figure*}

Since the \ion{N}{v} absorbers are selected from a sample of \ion{C}{iv} systems, line parameters of \ion{N}{v} and \ion{C}{iv} can be compared directly.
Figure \ref{logN_NV_CIV} presents the measured \ion{N}{v} column densities versus \ion{C}{iv} for intervening and associated systems, respectively.
Both absorber classes show well-correlated \ion{N}{v} and \ion{C}{iv}.
While \ion{N}{v} is systematically weaker than \ion{C}{iv} in intervening absorbers, associated systems exhibit similar \ion{N}{v} and \ion{C}{iv} column densities.
Furthermore, the slope of the correlation is different for both classes of absorbers.
Fitting $N_{\ion{N}{v}} \propto N_{\ion{C}{iv}}^{\,\alpha}$ leads to $\alpha = 0.40\pm0.07$ for the intervening systems.
The slope in case of the associated systems is considerably steeper yielding $\alpha = 0.64\pm0.10$.

The different slopes are likely due to the different ionizing radiation fields the absorbing gas is exposed to. 
While associated absorbers see the hard radiation of the background QSO in addition to the general intergalactic UV background, intervening absorption systems are presumably ionized solely by the background radiation.
\citet{worsecketal2007}, however, find an intervening \ion{N}{v} system at $z = 2.2753$ towards the quasar HE~2347-4342 in the vicinity of a foreground quasar close to the line of sight, which is also part of our sample.
This system therefore is supposed to be exposed to a harder radiation field even though it is intervening.
If intervening \ion{N}{v} arise preferentially close to foreground QSOs, the \ion{N}{v}/\ion{C}{iv} ratio is expected to be rather similar for intervening and associated absorbers.
The detection of different slopes for both types of absorbers indicate that the absorber towards HE~2347-4342 may arise from a particular configuration and intervening \ion{N}{v} absorption does not directly trace hard radiation sources close to the line of sight.

However, the problem might by degenerated since associated and intervening absorbers are expected to have different abundances.
While associated systems usually show about solar metallicity and solar nitrogen abundance \citep[e.g.\ reviewed in][]{hamannferland1999}, intervening systems are expected to have typically lower abundances.
Since DLA systems believed to probe the inner parts of high-redshift galaxies are usually metal poor and underabundant nitrogen is detected, a similar abundance pattern is expected for intergalactic absorption systems even farther away from early (proto-)galactic systems.
Therefore, intervening systems are supposed to exhibit weaker \ion{N}{v} features as a consequence of their lower nitrogen content.
This, however, should affect the offset of the \ion{N}{v}/\ion{C}{iv} relation but not its slope, supporting the conclusion that intervening \ion{N}{v} systems do not strictly trace hard foreground radiation sources like quasars close to the line of sight.
In Sect.\ \ref{local_sources} we will additionally argue that no evidence of a generally harder radiation at the location of intervening \ion{N}{v} systems is found from photoionization modeling adopting several ionizing spectral energy distributions.

\begin{figure}
  \centering
  \resizebox{\hsize}{!}{\includegraphics[bb=25 530 300 770,clip=]{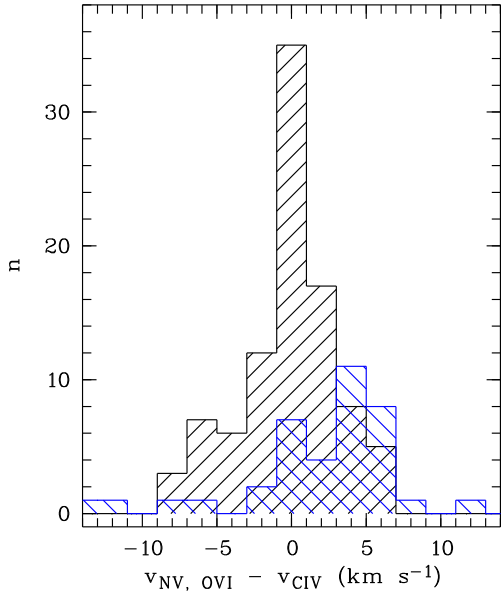}}
  \caption{Velocity offset between individual \ion{C}{iv} components and corresponding \ion{N}{v} (black) and \ion{O}{vi} features (blue) for the total sample.
While \ion{N}{v} absorption arise predominately at the same velocity as \ion{C}{iv}, \ion{O}{vi} is significantly shifted to positive velocities.
  }
  \label{velocity_shift}
\end{figure}

Inspection of the line profiles shows that the \ion{N}{v} absorption is usually well-aligned with \ion{C}{iv} while the features of \ion{O}{vi} are apparently shifted slightly to the red.
In Fig.\ \ref{velocity_shift} the distribution of the velocity difference between \ion{C}{iv} and \ion{N}{v} is shown in comparison to $v_{\ion{O}{vi}} - v_{\ion{C}{iv}}$.
While $v_{\ion{N}{v}} - v_{\ion{C}{iv}}$ clearly peaks at $0\,\mathrm{km\,s}^{-1}$, \ion{O}{vi} is shifted to positive velocities with a maximum at $\sim +4\,\mathrm{km\,s}^{-1}$.
The usual interpretation of this often observed velocity shift between \ion{O}{vi} and \ion{C}{iv} \citep[e.g.][]{reimersetal2001, carswelletal2002, simcoeetal2002} is that the lines do not arise from the same volume.
By observing multiple lines of sight of gravitationally lensed QSOs, \citet{lopezetal2007} showed that \ion{O}{vi} absorbing gas is much more extended compared to other species.
However, in a simple picture where \ion{O}{vi} and \ion{C}{iv} arise from spatially distinct volumes one would expect to find \ion{O}{vi} shifted statistically  to higher as well as lower velocities.
Checking the \ion{O}{vi} selected systems reported by \citet{carswelletal2002} and \citet{simcoeetal2002}, \ion{O}{vi} features at high redshift seem to be shifted to positive velocities with respect to \ion{C}{iv} if they are displaced at all.
At low redshift \citet{trippetal2008} have recently found a displacement between \ion{H}{i} and \ion{O}{vi} in particular for complex \ion{O}{vi} systems.
They detect a trend for \ion{O}{vi} shifted to higher velocities, too.
However, the origin of the shift remains uncertain.

In summary, we find the \ion{N}{v} features are well-aligned with \ion{C}{iv}, implying that these both species arise from the same gas phase.

\section{Photoionization modeling}\label{mod_procedure}

\begin{figure}
  \centering
  \resizebox{\hsize}{!}{\includegraphics[bb=45 505 440 775,clip=]{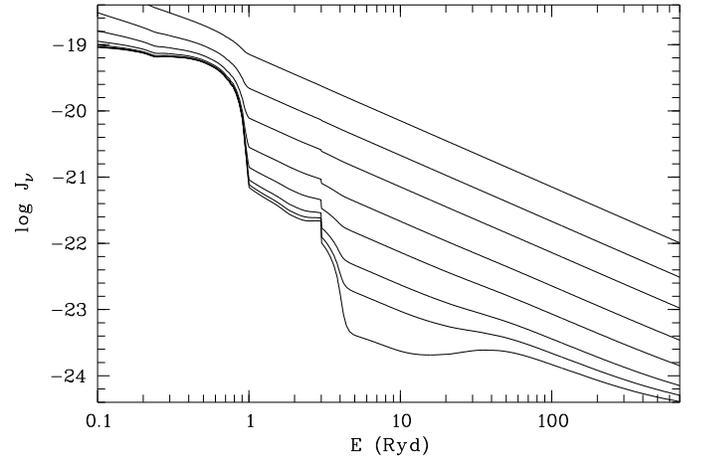}}
  \caption{Adopted ionizing spectra.
The lowest curve represents the pure \citet[][HM01]{haardtmadau2001} UV background at $z \sim 2$.
Upper curves indicate the HM01 background in addition with a QSO spectrum described as a power law $f_{\nu} \propto \nu^{\,-1.0}$ for increasing dominance of the QSO ($J_{\mathrm{QSO}}/J_{\mathrm{HM01}} = 0.1$, $0.3$, 1, 3, 10, 30, 100 at $1\,\mathrm{Ryd}$).
  }
  \label{ion_spec}
\end{figure}

For each of the systems we construct photoionization models using CLOUDY  \citep[v05.07.06;][]{ferlandetal1998} in order to estimate the nitrogen abundance.
For a given radiation field the ionization parameter $U = n_{\gamma}/n_{\element{H}}$, which is the ratio of ionizing photons and hydrogen density, can be determined by matching the observed column density ratio of two species.
The metallicity of the absorber and its relative elemental abundances are then adjusted in course of the modeling procedure.
In particular, we obtain estimates of [N/H] and [N/C] for all modeled systems.
For some absorbers additional estimates of [N/O] and/or [N/Si] can be derived.
Solar abundances are taken from \citet{asplundetal2005}.

In a first approach we adopt the UV background from \citet{haardtmadau2001} at the appropriate redshift scaled to $J_{\ion{H}{i}} = 10^{-21.15}\,\mathrm{erg\,s}^{-1}\mathrm{cm}^{-2}\mathrm{Hz}^{-1}\mathrm{sr}^{-1}$ at $1\,\mathrm{Ryd}$ \citep{scottetal2000}.
This energy distribution consists of the radiation of QSOs and galaxies whose spectra are filtered while propagating through the IGM.
Since associated absorbers are additionally exposed to the radiation of the background QSO, we add a power law spectrum $f_{\nu} \propto \nu^{\,\alpha}$ to model these systems.
The spectral index $\alpha$ is adopted from the literature if possible, otherwise we assume $\alpha = -1.0$.
The used values are listed in Table \ref{metal}.
\citet{levshakovetal2008} recently reconstructed the ionizing radiation for a few associated systems.
They find a depression of the intensity at $E > 4\,\mathrm{Ryd}$, possibly due to the \ion{He}{ii} Lyman continuum opacity of a quasar accretion disk wind.
If such a depression is present for all associated systems, a pure power law QSO continuum definitely overestimates the hardness of the radiation.
For comparison we therefore construct models adopting the pure HM01 background and can thus estimate the systematics, which may be introduced by the uncertainties of the hard QSO spectrum.

The flux of the QSO radiation reaching the absorber depends on the luminosity of the source and the distance of the absorbing material to the quasar.
Since the location of the absorber is usually unknown, several models are derived changing the intensity $4\pi J_{\nu}$ of the QSO power law at $1\,\mathrm{Ryd}$ relative to the HM01 background from 0.1, i.e.\ little contribution of the QSO to the ionizing radiation, to 100, i.e.\ the ionizing radiation is dominated by the QSO.
The tested spectral energy distributions are illustrated in Fig.\ \ref{ion_spec} showing a pure HM01 background at $z \sim 2$ and spectra combined with a $f_{\nu} \propto \nu^{-1.0}$ power law with $J_{\mathrm{QSO}}/J_{\mathrm{HM01}} = 0.1$, $0.3$, 1, 3, 10, 30, 100 at $1\,\mathrm{Ryd}$.
From the resulting model parameters a mean value of the abundances is estimated with error bars reflecting the spread of the values in the considered parameter range.

In order to test whether foreground quasars close to the line of sight provide high-energy photons generating the intergalactic \ion{N}{v} systems, models with a joint HM01+QSO power-law radiation are computed for each intervening system as well.
Generally, we assume $\alpha = -1.0$ except for 3 systems that are relatively close to the background QSO with known spectral indices from the literature.
These are the systems at $z=2.0422$ ($v \approx -17\,400\,\mathrm{km\,s}^{-1}$) towards PKS~0237-23 \citep[$\alpha = -0.48$;][]{zhengmalkan1993}, at $z = 2.1098$ ($v \approx -10\,300\,\mathrm{km\,s}^{-1}$) towards PKS~1448-232 \citep[$\alpha = -0.61$;][]{chengetal1991}, and at $z = 2.2510$ ($v \approx -16\,500\,\mathrm{km\,s}^{-1}$) towards Q~0329-385 \citep[$\alpha = -0.37$;][]{chengetal1991}.

If possible, we chose two ions of the same element to construct the model to not depend on any assumption about the relative abundances.
Models are based on \ion{C}{iii}/\ion{C}{iv} for 14\% of the intervening and 75\% of the associated systems.
\ion{Si}{iii}/\ion{Si}{iv} is used in case of 29\% of the intervening systems.
For 33\% of the intervening and one associated system we adopt \ion{C}{iv}/\ion{O}{vi} to estimate the ionization parameter assuming solar O/C abundance since no other appropriate species are observed.
Since \ion{C}{iv} and \ion{O}{vi} features are often shifted by $\sim 4\,\mathrm{km\,s}^{-1}$, both ions may not arise from the same gas phase (see Sect.\ \ref{statistics}).
Therefore, photoionization models based on the \ion{C}{iv}/\ion{O}{vi} provide only a rough estimate.
However, we have verified that our main conclusions are valid even if those models are excluded.
The identified absorption features together with the line profiles computed from the best-fit models are presented in the Appendix in Figures \ref{PKS1448_1.5855} to \ref{HE0151_2.4686} (intervening) and Figures \ref{HE1341_2.1462} to \ref{HE2347_2.9027} (associated), respectively.

Adopting different column density ratios to constrain the photoionization models might introduce systematic uncertainties to the abundance estimates since it is assumed that all species reside in the same gas phase.
In order to check whether this assumption is justified we compute models based on the ratios \ion{C}{iii}/\ion{C}{iv}, \ion{Si}{iii}/\ion{Si}{iv}, and \ion{C}{iv}/\ion{O}{vi} for the same system.
Such a comparison is possible for the intervening system at $z = 2.2510$ towards Q~0329-385 and the associated system at $z = 2.6362$ towards Q~0453-423, which exhibit features of all relevant species.
We find that the resulting abundances spread over $\lesssim 0.5\,\mathrm{dex}$.
This roughly corresponds to the size of the error bars due to the uncertain quasar contribution to the ionizing spectrum for the associated systems (see Table \ref{metal}).
Therefore, the estimated abundances could have an additional uncertainty of $\sim 0.5\,\mathrm{dex}$ due to a possible multi-phase nature of the absorbing material.

The intervening system at $z = 1.5771$ towards HE~0001-2340 has to be modeled based on the \ion{C}{ii}/\ion{C}{iv} ratio since \ion{C}{iii} and \ion{O}{vi} would both arise outside the observed spectral range and no \ion{Si}{iii} feature is detected (Fig.\ \ref{HE0001_1.5771}).
Due to the different ionization potentials ($1.79$ versus $4.74\,\mathrm{Ryd}$) \ion{C}{ii} and \ion{C}{iv} absorption may not arise from the same gas phase questioning the results.
Adopting a HM01 background, indeed, leads to rather unusual abundances ($\mathrm{[C/H]} = +0.74$, $\mathrm{[N/C]} = +0.76$, $\mathrm{[N/Si]} = +1.95$).
Since a model based on the \ion{C}{ii}/\ion{C}{iv} is not robust, we exclude this system from further discussion.
Moreover, there are 5 systems (4 intervening (19\%) and 1 associated) that cannot be modeled since they exhibit features of \ion{H}{i}, \ion{C}{iv}, and \ion{N}{v} only.
The observed species and the profiles obtained from Doppler profile fitting are presented in Figures \ref{HE0001_1.5814} to \ref{HE2347_2.8781} in the Appendix.

\section{Results}\label{results}

\begin{figure}
  \centering
  \resizebox{\hsize}{!}{\includegraphics[bb=60 419 425 770,clip=]{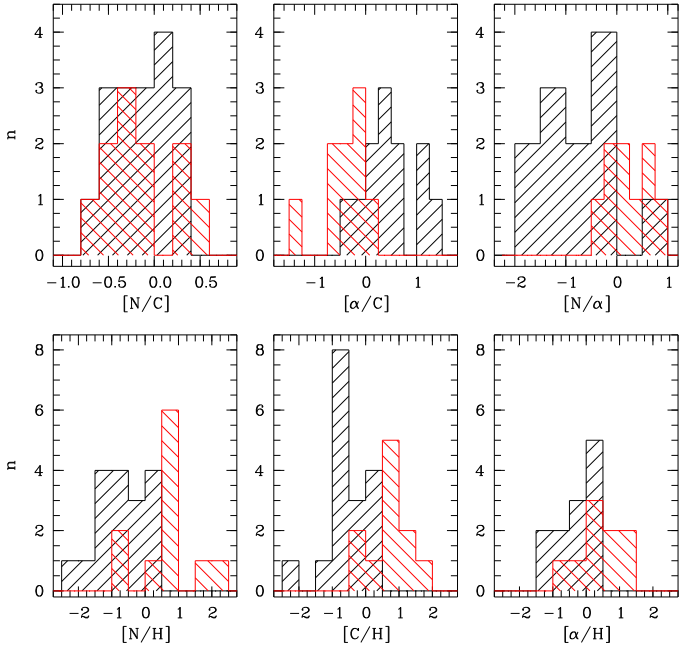}}
  \caption{Distribution of relative elemental abundances [N/C], [$\alpha$/C], and [N/$\alpha$] (upper panels from left to right) and metallicity traced by [N/H], [C/H], and [$\alpha$/H] (lower panels) for intervening (black) and associated systems (red histograms). 
  }
  \label{abund_histo}
\end{figure}

\begin{figure*}
  \centering
  \resizebox{\hsize}{!}{\includegraphics[bb=40 450 485 770,clip=]{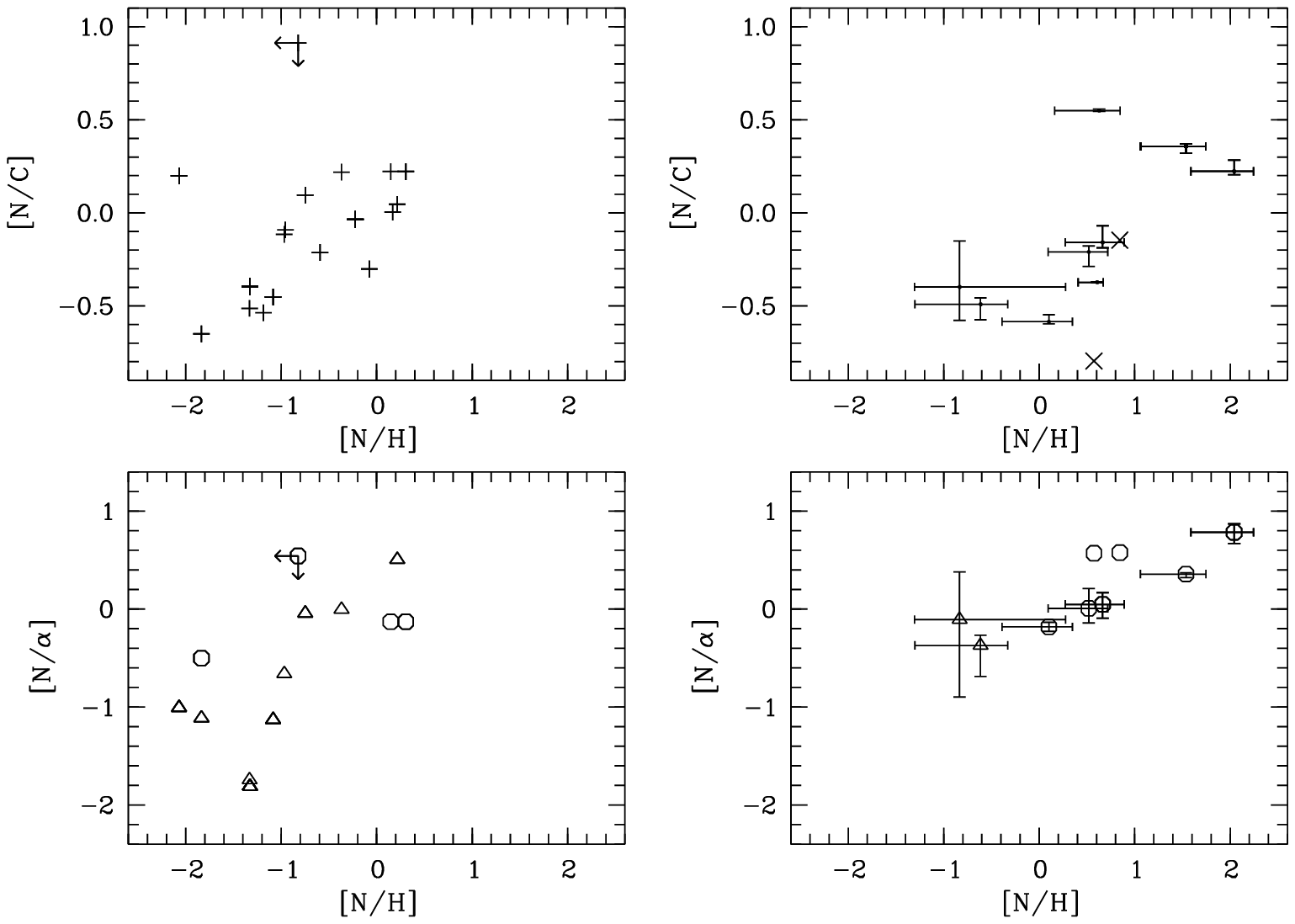}}
  \caption{Abundances derived from photoionization calculations. 
Left panels show results for the intervening absorbers adopting the HM01 background at the appropriate redshift. 
Right panels display the abundances for the associated systems using the HM01 background and an additional QSO power law spectrum (see text). 
Error bars are derived from models with varying dominance of the hard QSO radiation. 
Data points without error bars are derived assuming the HM01 background as ionizing radiation.
The lower panels present the relative abundance of nitrogen versus $\alpha$-elements like oxygen (circles) and silicon (triangles).
  }
  \label{abundances}
\end{figure*}

The abundances derived from the photoionization modeling of the intervening (associated) systems are summarized in the upper (lower) part of Table \ref{metal} and are displayed in Figures \ref{abund_histo} and \ref{abundances}.
As expected, associated systems are more metal-rich (median $\mathrm{[C/H]} \sim +0.66$) than intervening systems ($\mathrm{[C/H]} \sim -0.63$), which can be seen from the lower panels of Fig.\ \ref{abund_histo}.
Furthermore, intervening systems show enhanced abundances of $\alpha$-elements resulting in  lower [N/$\alpha$] values (median $-0.58$ in comparison to $+0.05$ for associated systems) and higher [$\alpha$/C] ($+0.41$ and $-0.29$, respectively; upper panels of Fig.\ \ref{abund_histo}).
In the following the results will be discussed in detail.

\begin{table*}
  \caption[]{Abundance estimates from photoionization modeling for the intervening (upper part) and associated systems (lower part), respectively.
  }
  \label{metal}
  $$
  \begin{array}{l c c l c c c c c c c }
    \hline\hline
    \noalign{\smallskip}
\mathrm{QSO} & z_{\mathrm{abs}} & \alpha & \mathrm{ratio} & \mathrm{[C/H]} & \mathrm{[N/H]} & \mathrm{[O/H]} & \mathrm{[Si/H]} & \mathrm{[N/C]} & [\element{N}/\alpha ] & [\alpha/\element{C}]\\     
    \noalign{\smallskip}
    \hline
    \noalign{\smallskip}
\textrm{PKS~1448-232} & ~1.5855~ & \dots & \ion{Si}{iii}/\ion{Si}{iv} &-0.82 &\phantom{<}-1.33 &\dots &+0.41&\phantom{<}-0.51&\phantom{<}-1.74&+1.23\\
\textrm{PKS~0237-23 } & 1.5966 & \dots & \ion{Si}{iii}/\ion{Si}{iv} &-0.93 &\phantom{<}-1.33 &\dots &+0.48&\phantom{<}-0.40&\phantom{<}-1.81&+1.41\\
\textrm{PKS~1448-232} & 1.7236 & \dots & \ion{Si}{iii}/\ion{Si}{iv} &-0.85 &\phantom{<}-0.97 &\dots &-0.31&\phantom{<}-0.12&\phantom{<}-0.66&+0.54\\
\textrm{HE~2217-2818} & 1.9656 & \dots & \ion{C}{iv}/\ion{O}{vi}    &-0.63 &\phantom{<}-1.09 & - &+0.05&\phantom{<}-0.45&\phantom{<}-1.13&+0.68\\
\textrm{Q~0122-380  } & 1.9744 & \dots & \ion{Si}{iii}/\ion{Si}{iv} &+0.17 &\phantom{<}+0.22 &\dots &-0.29&\phantom{<}+0.05&\phantom{<}+0.51&-0.46\\
\textrm{PKS~0237-23 } & 2.0422 & \dots & \ion{C}{iv}/\ion{O}{vi}    &+0.23 &\phantom{<}-0.08 & - &\dots&\phantom{<}-0.30&\phantom{<}\dots&\dots\\
\textrm{Q~0122-380  } & 2.0626 & \dots & \ion{C}{iv}/\ion{O}{vi}    &+0.16 &\phantom{<}+0.17 & - &\dots&\phantom{<+}0.00&\phantom{<}\dots&\dots\\
\textrm{Q~0329-385  } & 2.0764 & \dots & \ion{C}{iv}/\ion{O}{vi}    &-0.59 &\phantom{<}-0.37 & - &-0.37&\phantom{<}+0.22&\phantom{<-}0.00&+0.22\\
\textrm{PKS~1448-232} & 2.1098 & \dots & \ion{C}{iv}/\ion{O}{vi}    &-0.38 &\phantom{<}-0.59 & - &\dots&\phantom{<}-0.21&\phantom{<}\dots&\dots\\
\textrm{HE~1347-2457} & 2.1162 & \dots & \ion{C}{iv}/\ion{O}{vi}    &-0.87 &\phantom{<}-0.96 & - &\dots&\phantom{<}-0.54&\phantom{<}\dots&\dots\\
\textrm{HE~1347-2457} & 2.1162 & \dots & \ion{C}{iv}/\ion{O}{vi}    &-0.65 &\phantom{<}-1.19 & - &\dots&\phantom{<}-0.09&\phantom{<}\dots&\dots\\
\textrm{HE~0940-1050} & 2.2212 & \dots & \ion{Si}{iii}/\ion{Si}{iv} &-0.84 &\phantom{<}-0.75& \dots & -0.71&\phantom{<}+0.10&\phantom{<}-0.04&+0.14\\
\textrm{HE~1158-1843} & 2.2354 & \dots & \ion{C}{iii}/\ion{C}{iv}   &-1.74 &<-0.82 &-1.36 &\dots&<+0.92&<+0.54&+0.38\\
\textrm{Q~0420-388  } & 2.2464 & \dots & \ion{Si}{iii}/\ion{Si}{iv} &-2.27 &\phantom{<}-2.07 &\dots &-1.06&\phantom{<}+0.20&\phantom{<}-1.01&+1.21\\
\textrm{Q~0329-385  } & 2.2510 & \dots & \ion{C}{iii}/\ion{C}{iv}   &-1.19 &\phantom{<}-1.84 &-1.33 &-0.72&\phantom{<}-0.65&\phantom{<}-1.11&+0.46\\
\textrm{HE~2347-4342} & 2.2753 & \dots & \ion{C}{iv}/\ion{O}{vi}    &-0.19 &\phantom{<}-0.15 & - &\dots&\phantom{<}-0.04&\phantom{<}\dots&\dots\\
\textrm{HE~0151-4326} & 2.4686 & \dots & \ion{C}{iii}/\ion{C}{iv}   &-0.08 &\phantom{<}+0.15 &+0.28 &\dots&\phantom{<}+0.22&\phantom{<}-0.13&+0.35\\
\textrm{HE~0151-4326} & 2.4686 & \dots & \ion{C}{iii}/\ion{C}{iv}   &+0.08 &\phantom{<}+0.31 &+0.43 &\dots&\phantom{<}+0.22&\phantom{<}-0.13&+0.35\\
    \hline
    \noalign{\smallskip}
\textrm{HE~1341-1020} & 2.1462 &-1.0\phantom{0} &\ion{C}{iii}/\ion{C}{iv} &+0.82\pm^{0.14}_{0.36}&+0.66\pm^{0.23}_{0.39}&+0.62\pm^{0.25}_{0.39}&\dots&-0.21\pm^{0.03}_{0.08}&+0.01\pm^{0.20}_{0.15}&-0.22\pm^{0.07}_{0.17}\\
\textrm{HE~1341-1020} & 2.1462 &-1.0\phantom{0} &\ion{C}{iii}/\ion{C}{iv} &+0.73\pm^{0.17}_{0.35}&+0.52\pm^{0.20}_{0.43}&+0.51\pm^{0.24}_{0.52}&\dots&-0.16\pm^{0.09}_{0.03}&+0.05\pm^{0.12}_{0.14}&-0.21\pm^{0.11}_{0.03}\\
\textrm{HE~1341-1020} & 2.1475 &-1.0\phantom{0} &\ion{C}{ii}/\ion{C}{ii}^\ast &-0.44\pm^{0.86}_{0.29}&-0.83\pm^{1.11}_{0.47}& -0.72\pm^{1.15}_{0.53} &-0.73\pm^{1.47}_{0.53}&-0.40\pm^{0.25}_{0.18}&-0.11\pm^{0.49}_{0.79}&-0.29\pm^{0.61}_{0.24}\\
\textrm{Q~0329-385  } & 2.3520 &-0.37^{\mathrm{a}} &\ion{C}{iii}/\ion{C}{iv} &+1.82\pm^{0.14}_{0.44}&+2.04\pm^{0.20}_{0.45}&+1.26\pm^{0.24}_{0.46}&\dots&+0.22\pm^{0.06}_{0.02}&+0.78\pm^{0.09}_{0.11}&-0.56\pm^{0.10}_{0.03}\\
\textrm{HE~1158-1843} & 2.4278 &-1.0\phantom{0} &\ion{C}{iv}/\ion{O}{vi}  &+0.08\pm^{0.21}_{0.46}&+0.63\pm^{0.22}_{0.47}& \phantom{+}-\phantom{\pm^{0.00}_{0.00}} &\dots&+0.55\pm^{0.01}_{0.01}&\dots&\dots\\
\textrm{HE~1158-1843} & 2.4427 &-1.0\phantom{0} &\ion{C}{iii}/\ion{C}{iv} &+1.18\pm^{0.20}_{0.44}&+1.54\pm^{0.21}_{0.48}& +1.18\phantom{\pm^{0.00}_{0.00}} &\dots&+0.36\pm^{0.01}_{0.04}&+0.36\pm^{0.01}_{0.04}&\phantom{-}0.00\phantom{\pm^{0.00}_{0.00}}\\
\textrm{Q~0453-423  } & 2.6362 &-0.89^{\mathrm{a}} &\ion{C}{iii}/\ion{C}{iv} &-0.12\pm^{0.25}_{0.61}&-0.61\pm^{0.29}_{0.69}&+0.12\pm^{0.35}_{0.83}&-0.24\pm^{0.49}_{0.68}&-0.49\pm^{0.03}_{0.08}&-0.37\pm^{0.11}_{0.32}&-0.12\pm^{0.23}_{0.07}\\
\textrm{PKS~0329-255} & 2.7091 &-1.0\phantom{0} &\ion{C}{iii}/\ion{C}{iv} &+0.69\pm^{0.21}_{0.48}&+0.10\pm^{0.25}_{0.49}&+0.28\pm^{0.24}_{0.49}&\dots&-0.58\pm^{0.04}_{0.01}&-0.18\pm^{0.05}_{0.05}&-0.40\pm^{0.04}_{0.01}\\
\textrm{HE~2347-4342} & 2.8916 & \dots^{\mathrm{b}} &\ion{C}{iii}/\ion{C}{iv} &+0.99\phantom{\pm^{0.10}_{0.33}}&+0.84\phantom{\pm^{0.11}_{0.33}}&+0.27\phantom{\pm^{0.11}_{0.33}}&\dots&-0.15\phantom{\pm^{0.00}_{0.00}}&+0.58\phantom{\pm^{0.00}_{0.00}}&-0.72\phantom{\pm^{0.00}_{0.00}}\\
\textrm{HE~2347-4342} & 2.8972 & \dots^{\mathrm{b}} &\ion{C}{iii}/\ion{C}{iv} &+1.37\phantom{\pm^{0.12}_{0.38}}&+0.57\phantom{\pm^{0.13}_{0.39}}&\phantom{+}0.00\phantom{\pm^{0.13}_{0.39}}&\dots&-0.80\phantom{\pm^{0.00}_{0.00}}&+0.57\phantom{\pm^{0.00}_{0.00}}&-1.37 \phantom{\pm^{0.00}_{0.00}}\\
\textrm{HE~2347-4342} & 2.9027 & +0.56^{\mathrm{c}} &\ion{C}{iii}/\ion{C}{iv} &+0.98\pm^{0.06}_{0.20}&+0.61\pm^{0.06}_{0.20}& \phantom{+}\dots\phantom{\pm^{0.00}_{0.00}} &\dots&-0.37\pm^{0.01}_{0.01}&\dots&\dots\\
    \noalign{\smallskip}
    \hline
  \end{array}
  $$
\begin{list}{}{}
  \item[$^{\mathrm{a}}$] \citet{chengetal1991}
  \item[$^{\mathrm{b}}$] no additional power law spectrum adopted (see discussion in Section \ref{res_asso})
  \item[$^{\mathrm{c}}$] \citet{telferetal2002}
\end{list}
\end{table*}

\subsection{Intervening systems}\label{res_inter}

For the intervening systems we find metallicities in the range $-1.2 \lesssim \mathrm{[C/H]} \lesssim +0.2$ (mean $-0.46$, median $-0.63$).
The nitrogen abundances are spread over a slightly wider range $-1.8 \lesssim \mathrm{[N/H]} \lesssim +0.3$.
In particular, more than 40\,\% of the absorbers contain less nitrogen than 1/10 solar.
A clear correlation between [N/H] and the nitrogen abundance relative to carbon [N/C] is seen in the upper left panel of Fig.\ \ref{abundances}.

There are two outliers where our models lead to high [N/C] values but low overall metallicity ($\mathrm{[N/H]} \lesssim -1.0$ corresponding to $\mathrm{[C/H]} \sim -2.0$).
Both systems suffer from large uncertainties in important column density estimates.
The estimate at $\mathrm{[N/H]} \sim -0.8$ is an absorber at $z = 2.2354$ towards HE~1158-1543.
The \ion{N}{v} $\lambda\,1238$ feature of this system is completely blended. 
Thus the column density measurement relies on the second component whose profile might be disturbed by an interloper as well (see Fig.\ \ref{HE1158_2.2354}).
To be conservative the column density estimate and therefore the [N/C] and [N/H] values should be regarded as upper limits as indicated in Fig.\ \ref{abundances}.
The second outlier at $\mathrm{[N/H]} \sim -2.0$ is the system at $z = 2.2464$ towards Q~0420-388.
The model is based on the \ion{Si}{iii}/\ion{Si}{iv} ratio.
Unfortunately, both components of \ion{Si}{iv} are at least partially blended (see Fig.\ \ref{Q0420_2.2464}).
Thus, the adopted \ion{Si}{iv} column densities might be biased and the abundance estimates may suffer from large uncertainties.

Excluding the upper limit, we find a Spearman rank-order correlation coefficient of $r_s = 0.61$ for a correlation between [N/H] and [N/C].
If the second outlier is excluded as well, the correlation is tighter yielding $r_s = 0.82$.
A linear fit results in $\mathrm{[N/C]} = (0.35\pm0.07)\cdot\mathrm{[N/H]}+(0.06\pm0.06)$.

Furthermore, the abundance of nitrogen relative to $\alpha$-elements like silicon and oxygen is correlated with [N/H] (lower left panel of Fig.\ \ref{abundances}).
The rank-order coefficient is $r_s = 0.69$ and a linear fit leads to a steep slope of $0.61\pm0.20$.
The large error of the fitted slope reflects the large scatter in this correlation.
Even though individual abundance estimates may have large uncertainties, the presence of a correlation of [N/$\alpha$] with metallicity indicates that the observed absorbers are predominantly enriched with {\it secondary} nitrogen (see Sect.\ \ref{enrichment}).
However, in contrast to associated systems, the [N/$\alpha$] abundance in intervening systems is clearly sub-solar (see upper right panel of Fig.\ \ref{abund_histo}).
The median of the distribution is $[\element{N}/\alpha] = -0.58$ and values down to $\sim -1.8$ are found.
At the same time, $\alpha$-elements appear to be enhanced with respect to carbon to a median level of $[\alpha/\mathrm{C}] = +0.41$.

Two of the intervening systems with $\mathrm{[N/H]} \sim -1.3$ show rather low [N/$\alpha$] abundances ($\simeq -1.8$) below the expected range for enrichment with secondary nitrogen (see also Fig.\ \ref{Ntoalpha} discussed in detail in Sect.\ \ref{enrichment}).
These systems are at $z=1.5966$ towards PKS~0237-23 and $z=1.5855$ towards PKS~1448-232, representing the lowest redshift absorbers in our sample.
The latter is a rather unusual system obviously including multiple gas phases showing absorption features of \ion{Mg}{i}, \ion{Mg}{ii}, \ion{Al}{ii}, \ion{Al}{iii}, \ion{Fe}{ii}, \ion{Si}{ii}, \ion{Si}{iii}, \ion{Si}{iv}, \ion{C}{ii}, \ion{C}{iv}, and \ion{N}{v} and an unusual low \ion{H}{i} column density of $\log N \approx 14.4$ (see Fig.\ \ref{PKS1448_1.5855}).
A third system with $\mathrm{[N/H]} \sim -1.0$ yielding low [N/$\alpha$] is at $z=1.9656$ towards HE~2217-2818. 
Besides \ion{N}{v} and \ion{C}{iv} this system exhibits features of \ion{O}{vi} and \ion{Si}{iv}.
Thus we have to assume solar C/O to derive a model and independently use silicon as tracer for $\alpha$-elements even though similar Si/C and O/C ratios would be expected.

There are few measurements of relative abundances in the IGM, all of them suggesting enhancement of $\alpha$-elements with respect to carbon.
For the low-density IGM at $z \gtrsim 2$ \citet{aguirreetal2004} find $\mathrm{[Si/C]} = +0.77\pm0.05$.
Values in the range $+0.1 \lesssim \mathrm{[Si/C]} \lesssim +0.5$ are derived by \citet{simcoeetal2006} in metal line systems close to high-redshift galaxies at $z \sim 2.5$.
At low redshift \citet{trippetal2002} measure $\mathrm{[Si/C]} = +0.2\pm 0.1$ in an absorption system in the Virgo Supercluster.
Estimates of the oxygen abundance at $z \gtrsim 2$ lead to $+0.3 \lesssim \mathrm{[O/C]} \lesssim +1.3$ \citep{bergeronetal2002, telferetal2002b}.
\citet{simcoeetal2004} find $\mathrm{[O/C]} \sim +0.5$ for absorbers at $z \sim 2.5$ and recently \citet{aguirreetal2008} have measured $\mathrm{[O/C]} = +0.66\pm 0.21$ in the low-density IGM at similar redshifts.
Investigating 7 \ion{O}{vi} systems at $z \sim 2$ \citet{bergeronetal2002} detect no or very weak \ion{N}{v}, concluding that $\mathrm{[N/O]} < 0$ but $\mathrm{[O/C]} > 0$ in general agreement with our results.

Besides abundances the photoionization models provide informations of typical values of the absorbers' physical parameters.
The typical densities derived for intervening absorbers are $n_{\element{H}} \simeq 10^{-3.6}\,\mathrm{cm}^{-3}$ or slightly less.
Temperatures are typically $\sim 32\,000\,\mathrm{K}$ in agreement with the $b$-parameters of the \ion{N}{v} profiles discussed in Sect.\ \ref{statistics}.
Combining the model parameters it is possible to estimate the sizes of the absorbers according to
\begin{equation}\label{size}
  l = \frac{N_{\ion{N}{v}}}{n_{\element{H}}\,f_{\ion{N}{v}}\,10^{[\mathrm{N/H}]}\,(\mathrm{N/H})_{\sun}}\,\mbox{,}
\end{equation}
where $f_{\ion{N}{v}}$ denotes the ionization fraction of \ion{N}{v}.
For intervening systems typical sizes of a few to several 10\,kpc are found (median $4.8\,\mathrm{kpc}$).
About $32\,\%$ of the modeled components are smaller than $< 1.5\,\mathrm{kpc}$.
Since most of them are metal-rich (mean $[\mathrm{C/H}] = -0.26$), they may belong to the population of compact, metal-rich absorbers found by \citet{schayeetal2007}.
The authors argued these absorbers are short-lived and could therefore be responsible for the dispersion of heavy elements into the IGM.
As a consequence the enrichment of the IGM is supposed to be inhomogeneous. 
In this picture the wide spread in the relation between [N/$\alpha$] with metallicity would naturally follow from the enrichment mechanism.

\subsection{Associated systems}\label{res_asso}

The associated systems are modeled assuming a HM01 background and an additional QSO power law spectrum.
Since the distance of the absorbing material from the QSO is generally unknown, we construct models for different intensity ratios of the background and QSO spectrum in the range $0.1 \le J_{\mathrm{QSO}}/J_{\mathrm{HM01}} \le 100$ resulting in spectra like those displayed in Fig.\ \ref{ion_spec} where the slope $\alpha$ of the power law for each QSO is listed in Table \ref{metal}.
Therefore, we give error bars for the results reflecting the uncertainties of the derived parameters with respect to the influence of the QSO radiation and the unknown distance between absorber and background quasar.
Furthermore, it should be kept in mind that metal line systems proximate to the quasar may exist in time-variable conditions due to inflows, outflows, or variations of the background source.
Thus, the absorber might not be in equilibrium.
Though we assume a single-phase absorber, there may exists multiple gas phases remaining unrecognized.
In this case our models would overestimate the \ion{H}{i} column density for the considered phase and therefore underestimate the metallicity.
In the following we first will give some remarks on individual systems and then present the results from our sample of associated absorbers.

The complex associated system towards HE~2347-4342 has been analyzed to constrain the spectrum of the ionizing radiation by \citet{fechneretal2004} and more recently by \citet{levshakovetal2008}.
For our study of the \ion{N}{v} absorption we consider three groups of absorption features at $z = 2.8916$, $2.8972$, and $2.9027$ (Figs.\ \ref{HE2347_2.8916}, \ref{HE2347_2.8972}, and \ref{HE2347_2.9027}).
The latter complex has been found by \citet{fechneretal2004} to be exposed to an extremely hard radiation field, while the former two complexes are ionized by a softer spectrum, possibly by the filtered radiation of the background QSO \citep[see also][]{levshakovetal2008}.
Therefore adding a power law with $\alpha = +0.56$ \citep{telferetal2002} to the UV background will result in an ionizing spectrum too hard for the two lower redshift absorption complexes.
Such a spectrum, indeed, leads to unrealistically high metallicities $\mathrm{[C/H]} \sim +4.4$.
Testing a slightly softer spectrum with $\alpha = 0.0$ reduces the metallicity to $\mathrm{[C/H]} \sim +3.6$, which would still mean a metallicity of $\sim 4000$ solar.
Consistent with the results from \citet{fechneretal2004} and \citet{levshakovetal2008} we assume a HM01 background as ionizing radiation, which should lead to more realistic models.
The results for these models are indicated in the right panels of Fig.\ \ref{abundances} without error bars.

A special system is at $z = 2.1475$ towards HE~1341-1020, recently analyzed by \citet{levshakovetal2008}.
Features of various low and high ionized species are detected in three components (\ion{Mg}{ii}, \ion{Al}{ii}, \ion{Al}{iii}, \ion{Si}{ii}, \ion{Si}{iii}, \ion{Si}{iv}, \ion{C}{ii}, \ion{C}{iii}, \ion{C}{iv}, \ion{N}{ii}, \ion{N}{iii}, \ion{N}{v}, and \ion{O}{vi}; see Fig.\ \ref{HE1341_2.1475}).
Since all highly-ionized species are severely saturated, the \ion{N}{v} gas phase cannot be modeled reliably.
However, a feature of \ion{C}{ii}$^\ast$ $\lambda\,1335$ is detected as well.
Therefore, it is possible to estimate the electron density from the ratio of the measured \ion{C}{ii}$^\ast$/\ion{C}{ii} column densities \citep{bahcallwolf1968} using
\[ n_e = \frac{A_{21}}{\gamma_{12}}\left(\frac{N_{\ion{C}{ii}^\ast}}{N_{\ion{C}{ii}}}\right)\,\mbox{,}\]
where $A_{21}$ is the transition probability for spontaneous radiative decay and $\gamma_{12}$ is the rate coefficient for collisional excitation by electrons.
For the strongest component we measure $\log N_{\ion{C}{ii}} = 14.20 \pm 0.01$ and $\log N_{\ion{C}{ii}^\ast} = 12.49 \pm 0.03$.
Adopting $A_{21} = 2.29\cdot 10^{-6}\,\mathrm{s}^{-1}$ \citep{silvaviegas2002} and $\gamma_{12} = 1.58\cdot 10^{-7}\,\mathrm{cm}^3\,\mathrm{s}^{-1}$ \citep{trippetal1996} yields $n_e \simeq 0.278\,\mathrm{cm}^{-3}$.
This value is assumed to equal roughly the hydrogen density $n_{\element{H}}$.
Then photoionization models can be computed where the metallicity and the relative abundances are the only free parameters.
Since these models describe the low-ionization gas phase, the nitrogen abundance is constrained by the well-measured \ion{N}{ii} column density.
Remarkably, the observed features of all lowly and highly ionized species are reproduced best if the ionizing radiation is strongly dominated by the QSO power law spectrum.
The best-fit model is found for $J_{\mathrm{QSO}} = 100\cdot J_{\mathrm{HM01}}$ at $1\,\mathrm{Ryd}$ yielding $\mathrm{[C/H]} = +0.32$, $\mathrm{[N/C]} = -0.18$, $\mathrm{[Si/C]} = +0.43$, $\mathrm{[Mg/C]} = +1.01$, and $\mathrm{[Al/C]} = +0.05$, i.e.\ super-solar metallicity with slightly underabundant nitrogen and enhancement of $\alpha$-elements.

\begin{figure}
  \centering
  \resizebox{\hsize}{!}{\includegraphics[bb=50 550 515 770,clip=]{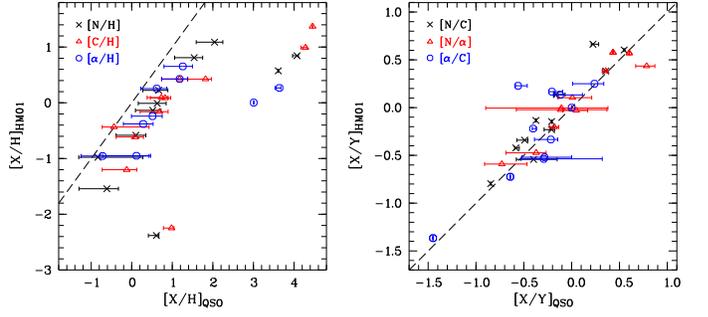}}
  \caption{Comparison of the abundances derived under the assumption of a hard ionizing spectrum (HM01 background and additional QSO power law; see text) and the HM01 background only as limit for the softest ionizing spectrum.
The left panel shows the metallicity traced by carbon, nitrogen, and the $\alpha$-elements, respectively, where the resulting metallicity is generally higher for harder spectra. The outliers are the three associated sub-systems towards HE~2347-4342.
The right panel presents the relative abundances [N/C], [N/$\alpha$], and [$\alpha$/C], which are rather independent of the hardness of the adopted ionizing radiation.
  }
  \label{compare_asso}
\end{figure}

Three of our associated systems ($z=2.4427$ towards HE~1158-1543, $z=2.6362$ towards Q~0453-423, and $z=2.7091$ towards PKS~0329-255; Figs.\ \ref{HE1158_2.4427}, \ref{Q0453_2.6362}, and \ref{PKS0329_2.7091}) are also part of a sample of associated systems studied by \citet{dodoricoetal2004}.
While the column density estimates are in excellent agreement, we generally find higher metallicities.
This discrepancy is clearly due to the different spectral energy distribution of the ionizing radiation assumed in the models.
\citet{dodoricoetal2004} adopt a composite QSO spectrum from \citet{cristianivio1990}, which is described by $f \propto \nu^{\,\alpha}$ with a spectral index of $\alpha = -1.6$ above the \ion{H}{i} Lyman limit ($E > 1\,\mathrm{Ryd}$).
This means their spectrum is softer compared to the combination of HM01 and a substantial contribution of the $\alpha = -1.0$ power law.
Generally, adopting softer ionizing spectra would shift the results presented in the right panels of Fig.\ \ref{abundances} to the upper left, i.e.\  would lead to lower metallicities (up to 1\,dex depending on the degree of softening).
This is further illustrated in Fig.\ \ref{compare_asso}, which shows the metallicities (left panel) and relative abundances (right panel) of the associated systems derived under the assumption of a hard ionizing radiation including the QSO power law spectrum versus the results for a HM01 background only.
Assuming the HM01 background as ionizing spectrum means that the associated absorbers are completely shielded from the QSO radiation.
Thus the estimated abundances are yielded in the limit of the softest ionizing radiation.
The outliers in the left panel of Fig.\ \ref{compare_asso} leading to extremely higher metallicity when adopting the hard spectra are the three systems towards HE~2347-4342 that are known to have particular characteristics and have been discussed above.
Excluding these systems, the metallicity estimated for the hard ionizing spectrum is on average $(0.7\pm0.3)\,\mathrm{dex}$ higher for each considered element.
Thus a rough estimate of the systematic error due to the uncertain spectral energy distribution of the QSO is $\sim 0.7\,\mathrm{dex}$.
However, the relative abundances are rather robust with respect to the hardness of the ionizing radiation as can be seen from the right panel of Fig.\ \ref{compare_asso}.
While for [N/C] and [$\alpha$/C] the soft radiation yields on average slightly higher values ($0.1\pm0.2$ and $0.1\pm0.3\,\mathrm{dex}$, respectively), the average offset is negligible for [N/$\alpha$] ($0.0\pm0.2\,\mathrm{dex}$).
Therefore, we are confident about our results of the relative abundances derived for the associated absorbers despite the severe uncertainties regarding the ionizing spectral energy distribution.

However, the estimated metallicity in associated systems is generally higher than in intervening absorbers (lower panels of Fig.\ \ref{abund_histo}) independent of the ionizing spectrum.
The median values derived for the associated systems are $\mathrm{[N/H]} = +0.61$ and $\mathrm{[C/H]} = +0.66$, respectively, about $1.3\,\mathrm{dex}$ higher than for the intervening systems.
If all intervening \ion{N}{v} systems were exposed to radiation harder than the standard UV background, the estimated metallicities increase by $\sim 0.9\,\mathrm{dex}$ to median values of $\mathrm{[N/H]} = +0.23$ and $\mathrm{[C/H]} = +0.26$.
This is still $0.4\,\mathrm{dex}$ lower than the values derived under the same assumptions for the associated systems.
On the other hand, if the associated systems would be exposed to a considerable softer radiation field, e.g.\ to a HM01 background as the limit for the softest spectrum (see above), the metallicity estimates are lower but the median abundances are still slightly super-solar.
We then find $\mathrm{[N/H]} = +0.11$ and $\mathrm{[C/H]} = +0.18$, roughly $0.8\,\mathrm{dex}$ larger than intervening systems exposed the same background.
Thus the finding of associated systems being statistically more metal-rich than intervening absorbers is independent of the hardness of the adopted ionizing radiation.
High metallicities of roughly solar values are usually found in associated absorbers \citep[e.g.][]{petitjeanetal1994, trippetal1996, wampleretal1996, hamannetal1997, papovichetal2000, dodoricoetal2004, gabeletal2006, gangulyetal2006} consistent with studies of QSO emission lines \citep[e.g.][]{hamannferland1999, dietrichetal2003, bradleyetal2004}.
Due to the simplifying assumption of the pure power law continuum for the QSO spectrum, our estimates tend to higher values though they are broadly consistent with  estimates from the literature in case of the softest spectra, i.e.\ when the contribution from the background QSO to the ionizing radiation is small.

The [N/$\alpha$] abundance for the associated absorbers is tightly correlated with the metallicity [N/H] (lower right panel of Fig.\ \ref{abundances}).
The Spearman rank-order coefficient yields $r_s = 0.87$ and the relation is very well fitted by $[\element{N}/\alpha] = (0.35\pm0.09)\cdot\mathrm{[N/H]}-(0.00\pm0.10)$.
A correlation between [N/H] and [N/C] is noticed in the upper right panel of Fig.\ \ref{abundances}.
The Spearman rank-order coefficient is $r_s = 0.75$ and a linear fit yields $\mathrm{[N/C]} = (0.30\pm0.14)\cdot\mathrm{[N/H]}-(0.35\pm0.13)$.
The slope, which is independent of a systematic bias in the metallicity is consistent with the best fit for the same correlation in intervening systems.

Several authors report on enhanced nitrogen abundances in absorption systems close to a background quasar and AGN outflows \citep[e.g.][]{petitjeanetal1994, petitjeansrianand1999, dodoricoetal2004, gabeletal2006, aravetal2007, fieldsetal2007}.
From the 11 associated systems investigated in this work 3 (27\,\%) show clearly enhanced nitrogen with $\mathrm{[N/C]} > 0$.
This number would increase to 4 (36\,\%) if the softer HM01 background is adopted as ionizing radiation.
\citet{misawaetal2007} defined a category of \ion{N}{v}-strong intrinsic absorbers with large \ion{N}{v} equivalent widths and rather weak \ion{C}{iv} and \ion{H}{i} features.
They classified 11 out of 28 systems (39\,\%) as \ion{N}{v}-strong.
By definition such absorbers should have a low \ion{C}{iv}/\ion{N}{v} ratio.
Our nitrogen-rich absorbers seem to belong to this category when comparing the measured column density ratios.
The mean \ion{C}{iv}/\ion{N}{v} ratio of the nitrogen-rich components is $\log\,(N_{\ion{C}{iv}}/N_{\ion{N}{v}}) = -0.25 \pm 0.35$ (median $-0.26$), which is clearly lower than the average column density ratio of the associated components with $[\mathrm{N/C}] < 0$ ($+0.55\pm 1.14$, median $+0.27$).
Moreover, our nitrogen-rich absorbers show on average weak \ion{H}{i} ($\log N_{\ion{H}{i}} \simeq 12.9$), while the other components have stronger \ion{H}{i} features ($\log N_{\ion{H}{i}} \simeq 14.5$).
Therefore, our sample of associated absorbers suggests that systems classified as \ion{N}{v}-strong according to \citet{misawaetal2007}, are indeed nitrogen-rich with $[\mathrm{N/C}] \gtrsim 0$.
Within our rather small sample these are $\sim 30\,\%$ of the associated \ion{N}{v} systems.

Regarding the physical parameters, associated absorbers are denser than the intervening yielding typically $n_{\element{H}} \simeq 10^{-2.8}\,\mathrm{cm}^{-3}$.
The distribution shows indication for a second peak at $\simeq 10^{-1.6}\,\mathrm{cm}^{-3}$, which may be an effect of the small numbers statistics.
The temperatures are $\sim 30\,000\,\mathrm{K}$ or slightly less comparable to those of the intervening systems.
Absorber sizes estimated according to Eq.\ \ref{size} are a few 10 to a few 100 pc, roughly one order of magnitude smaller than the intervening absorbers.

\section{Discussion}\label{discussion}

\subsection{Indications for hard ionizing spectra and local sources}\label{local_sources}

Comparing the column densities of related \ion{N}{v} and \ion{C}{iv}, we have argued in Sect.\ \ref{statistics} that the different slopes found for intervening and associated systems indicate a different ionizing radiation field.
This is in disagreement with the hypothesis that intervening \ion{N}{v} systems might arise preferentially close to radiation sources locally hardening the UV background.
In order to test this conclusion independently we compare the results obtained with the soft HM01 models to those assuming an additional contribution of a power law spectrum $\propto \nu^{\,\alpha}$ with spectral index $\alpha = -1.0$.
The harder models are computed for different intensity ratios $0.1 \le J_{\mathrm{QSO}}/J_{\mathrm{HM01}} \le 100$ at 1\,Ryd and are displayed in Fig.\ \ref{ion_spec}.
We find that the intervening systems are generally consistent with ionization by the soft HM01 background.

Furthermore, for 7 out of 16 intervening systems ($\sim 44\,\%$) a very hard spectral energy distribution can be excluded.
Any power law contribution to the radiation ionizing the system at $z=1.5855$ towards PKS~1448-232 leads to very high metallicities and inconsistent models.
Similar problems arise for spectra with $J_{\mathrm{QSO}}/J_{\mathrm{HM01}} > 0.1$ for the systems at $z=2.0626$ towards Q~0122-380 and $z=2.0764$ towards Q~0329-385.
The system at $z=1.7236$ towards PKS~1448-232 cannot be modeled with very hard spectra $J_{\mathrm{QSO}}/J_{\mathrm{HM01}} > 3.0$.
Absorption features of \ion{C}{ii} and \ion{Si}{ii}, respectively, are overestimated for systems $z=2.2510$ towards Q~0329-385, $z=1.9744$ towards Q~0122-380, and $z=2.2212$ towards HE~0940-1050 when applying some of the tested energy distributions.
While the former are inconsistent with $J_{\mathrm{QSO}}/J_{\mathrm{HM01}} > 0.1$, the latter can be modeled with harder spectra up to $J_{\mathrm{QSO}}/J_{\mathrm{HM01}} \sim 0.3$.

To summarize, none of the intervening \ion{N}{v} systems requires a harder ionizing radiation field than the HM01 background to be modeled consistently.
In contrary, for about half of the systems the contribution from a hard power law spectrum has to be low ($J_{\mathrm{QSO}}/J_{\mathrm{HM01}} \lesssim 0.3$) to produce consistent models.
Therefore, intervening \ion{N}{v} absorption does not arise predominately in the vicinity of foreground QSOs where the radiation field is supposed to be harder, but rather in "normal" intergalactic absorption systems.

\subsection{Implications for metal enrichment}\label{enrichment}

\begin{figure*}
  \centering
  \resizebox{\hsize}{!}{\includegraphics[bb=45 425 472 585,clip=]{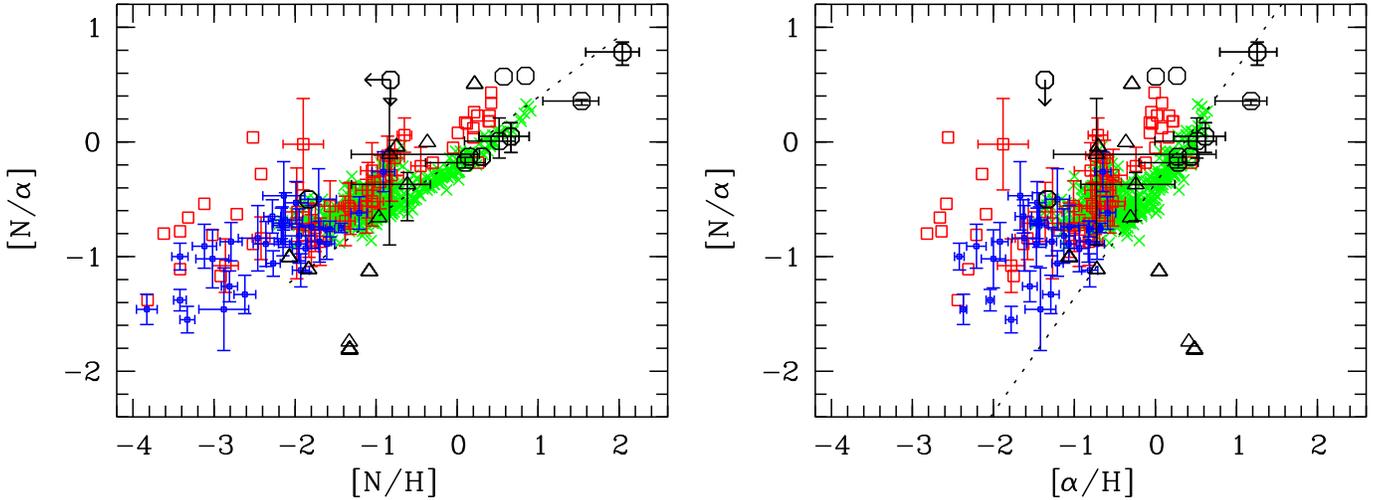}}
  \caption{Abundance of [N/$\alpha$] versus metallicity [N/H] (left panel) and [$\alpha$/H] (right panel), respectively.
The results from this work (black circles/triangles indicate oxygen/silicon estimates) for the total sample of intergalactic \ion{N}{v} systems (without error bars) and systems associated to a background QSO (with error bars, but see text) are compared to values measured in damped Ly$\alpha$ systems \citep[blue dots; following the compilation given by \citet{pettinietal2008} and adding the measurement of][]{richteretal2005}, extragalactic \ion{H}{ii} regions \citep[green crosses;][]{kobulnickyskillman1996, fergusonetal1998, vanzeeetal1998, izotovthuan1999, vanzee2000, melbourneetal2004, vanzeehaynes2006, malleryetal2007}, and metal-poor stars\citep[red squares;][]{israelianetal2004, spiteetal2005}.
The dotted line in the left panel represents a linear fit $[\element{N}/\alpha] = (0.53\pm0.09)\cdot\mathrm{[N/H]} -(0.14\pm0.10)$ to the measurements in intervening and associated systems.
In the right panel the dotted line indicates the expected distribution for secondary nitrogen.
  }
  \label{Ntoalpha}
\end{figure*}

Nitrogen is mainly produced in intermediate mass stars and is therefore released later into the interstellar and intergalactic space compared to $\alpha$-elements whose main production sites are Type II supernova explosions.
Thus, the relative abundances of these elements in astronomical objects depend on the overall level of enrichment.
In low-metallicity environments $\alpha$-elements are usually enhanced with respect to carbon while nitrogen is underabundant.
We therefore expect to find different distributions of [$\alpha$/C] and [N/$\alpha$] for the intervening absorbers, which are generally metal-poor ($\mathrm{[C/H]} < 0$) and the metal-rich ($\mathrm{[C/H]} \gtrsim 0$) associated systems.
Comparing the abundance estimates in the upper panels of Fig.\ \ref{abund_histo}, intervening absorbers have generally lower [N/$\alpha$] and higher [$\alpha$/C] values (median $-0.58$ and $+0.41$, respectively), whereas for associated systems we find about solar [N/$\alpha$] (median $+0.05$) and slightly subsolar $[\alpha/\element{C}] \simeq -0.29$.
Thus, we confirm that [$\alpha$/C] and [N/$\alpha$] trace the overall metallicity in the sense that metal-poor systems show enhanced $\alpha$-abundances compared to carbon and nitrogen.

In Fig.\ \ref{Ntoalpha} we compare our estimates of [N/$\alpha$] versus metallicity (traced by [N/H] (left panel) and [$\alpha$/H] (right panel), respectively) to results collected from the literature.
The abundance of nitrogen has been measured in various objects such as DLAs \citep[e.g.][]{prochaskaetal2002, centurionetal2003, petitjeanetal2008, pettinietal2002b, pettinietal2008}, extragalactic \ion{H}{ii} regions \citep[e.g.][]{kobulnickyskillman1996, fergusonetal1998, vanzeeetal1998, izotovthuan1999, vanzee2000, melbourneetal2004, vanzeehaynes2006, malleryetal2007}, and stars in the halo of the Milky Way \citep{israelianetal2004, spiteetal2005}.
At metallicities $\mathrm{[N/H]} \gtrsim -1$ (or $[\alpha/\element{H}] \gtrsim -0.4$) the nitrogen abundance appears to be dominated by secondary nitrogen and [N/$\alpha$] increases with metallicity.
At lower metallicity there is a plateau of nearly constant [N/$\alpha$], which is attributed to primary nitrogen.
Measurements in extragalactic \ion{H}{ii} regions in the local Universe and in metal-poor stars in the Milky Way find $[\element{N}/\alpha] \sim -0.7$ as plateau value.
Abundance estimates in metal-poor DLA systems lead to even lower values and a second plateau near $[\element{N}/\alpha] \sim -1.5$ is discussed \citep{prochaskaetal2002, centurionetal2003, richteretal2005, petitjeanetal2008, pettinietal2008}.
Our abundance estimates generally follow the distribution from the literature approximately down to $[\element{N}/\alpha] \sim -1.0$, except two severe outliers that have been discussed in Sect.\ \ref{res_inter}.

The estimated [N/$\alpha$] abundances at $\mathrm{[N/H]} \gtrsim -2.0$ can be fitted with $[\element{N}/\alpha] = (0.53\pm0.09)\cdot\mathrm{[N/H]} -(0.14\pm0.10)$ considering both intervening and associated systems.
The relation is indicated as dotted line in the left panel of Fig.\ \ref{Ntoalpha}.
It is in very good agreement with the measurements in extragalactic \ion{H}{ii} regions and halo stars in the same metallicity range.
Therefore, the majority of the analyzed absorption systems appears to be enriched with secondary nitrogen.
The expected distribution in case of enrichment with secondary nitrogen is represented by the dotted line in the right panel of Fig.\ \ref{Ntoalpha}.
In particular, the associated systems tracing metal-rich environments ($> 0.1$ solar) follow a tight relation. 
The estimated values for the intervening systems show a larger scatter, indicating inhomogeneous enrichment with nitrogen.
This can be interpreted in the sense of \citet{schayeetal2007} who argue that the IGM is enriched by short-lived metal-rich clouds leading to inhomogeneous enrichment with heavy elements.

The second plateau of primary nitrogen at $\mathrm{[N/H]} < -2.8$ in DLA systems is not probed with the present sample since the lower, non-DLA \ion{H}{i} column densities of the intervening absorbers and our \ion{C}{iv}-based selection procedure favor metal-rich absorption systems.
For a typical absorber with $\log n_{\element{H}} \sim -3.5$, $\log N_{\ion{H}{i}}\lesssim 16$, and $[\alpha/\element{H}] \sim -2.8$ at $z \sim 2$ we would expect $\log N_{\ion{N}{v}} \lesssim 11.2$ if $[\element{N}/\alpha] \sim -1.5$, which is below our detection limit.
In order to investigate the nitrogen abundances of intervening absorbers at this metallicity level the dominant ionization stage \ion{N}{iv} has to be studied, which arises in the UV \citep[rest wavelength $765.148\,\mathrm{\AA}$;][]{verneretal1994}. 
For this species a roughly $0.5\,\mathrm{dex}$ higher column density is expected.

Out of the 11 modeled intervening systems 5 yield $[\element{N}/\alpha] < -1.0$.
Except the unusual multi-phase absorber at $z = 1.5855$ towards PKS~1448-232 all these systems show rather strong \ion{H}{i} absorption. 
From the absorbers with $[\element{N}/\alpha] > -1.0$ only 2 have comparable strong \ion{H}{i} features.
However, 3 of the 5 low-[N/$\alpha$] systems are at low redshifts.
The combination of low-redshift and strong \ion{H}{i} absorption means that the \ion{H}{i} column density relies on the measurement for a saturated Ly$\alpha$ feature since higher orders of the Lyman series are beyond the observed spectral range.
We therefore might underestimate the hydrogen content of these absorbers and overestimate the metallicity.
Moreover, a possible multi-phase structure of the absorbers may bias the estimated metallicities.
In Sect.\ \ref{mod_procedure} we have estimated the corresponding uncertainty in the derived abundances to be $\lesssim 0.5\,\mathrm{dex}$.
Furthermore, the ionizing radiation might be harder than assumed.
Studies of metal lines systems suggest that the UV background at $z \lesssim 2.0$ is significantly harder than at higher redshifts \citep{fechneretal2006a, agafonovaetal2007}.
If a power law spectrum with spectral index $-1.0$ is added to the appropriate HM01 background, the outliers in Fig.\ \ref{Ntoalpha} shift to larger [N/H] and [N/$\alpha$] but are still below the expected parameter range.

\begin{figure}
  \centering
  \resizebox{\hsize}{!}{\includegraphics[bb=275 130 485 290,clip=]{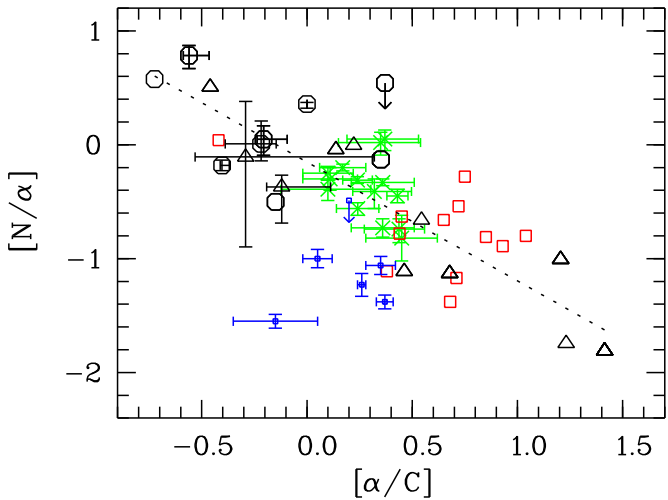}}
  \caption{Abundance of [N/$\alpha$] versus [$\alpha$/C].
The results from this work (black circles/triangles indicate oxygen/silicon) for the total sample of intergalactic (without error bars) and associated (with error bars) absorbers are compared to values measured in damped Ly$\alpha$ systems \citep[blue dots;][]{dodoricomolaro2004, richteretal2005, ernietal2006, pettinietal2008}, extragalactic \ion{H}{ii} regions in metal-poor galaxies \citep[green crosses; ][and references therein]{garnettetal1999, kobulnickyskillman1998}  and metal-poor halo stars \citep[red squares;][]{spiteetal2005}.
The dotted line represents a linear fit to our data points yielding $[\element{N}/\alpha] = -(1.04 \pm 0.13)\cdot [\alpha/\element{C}] - (0.15 \pm 0.08)$.
  }
  \label{alphatoC}
\end{figure}

The nitrogen-poor absorbers exhibit large [$\alpha$/C] abundances (and low [N/C]).
Since both, low [N/$\alpha$] and large [$\alpha$/C] are expected for low metallicity systems, an anti-correlation between these quantities is expected.
Fig.\ \ref{alphatoC} shows [N/$\alpha$] versus [$\alpha$/C] and a clear anti-correlation in seen.
Considering the total sample of systems with $\alpha$-element abundance measurements in the range $-1.0 < [\alpha/\element{C}] < +1.5$, the rank-order coefficient is $r_s = -0.82$.
The anti-correlation can be fitted with a straight line yielding $[\element{N}/\alpha] = -(1.04 \pm 0.13)\cdot [\alpha/\element{C}] - (0.15 \pm 0.08)$ indicated as dotted line.
If a harder ionizing radiation is assumed for the low-redshift systems, the abundances are adjusted but do not modify the fitted relation.

The abundances of metal-poor halo stars adopted from \citet{spiteetal2005} appear to be consistent with this relation even though the values cluster at $[\alpha/\element{C}] \sim +0.7$.
\citet{akermanetal2004} find $[\mathrm{O/C}] \sim +0.5$ as a typical value in metal-poor halo stars depending on [O/H].
The authors relate the increasing [O/C] value with decreasing oxygen abundance to the metallicity dependence of the carbon yields of massive stars with mass loss. 
However, the origin of carbon in the Milky Way is complex.
While some authors argue the main production site of carbon should be massive stars \citep[e.g.][where the latter also include extragalactic \ion{H}{ii} regions]{gustafssonetal1999, henryetal2000}, others find carbon has to be produced mainly in low- and intermediate-mass stars in the Milky Way and other galaxies \citep{chiappinietal2003b, chiappinietal2003}.
More recently, \citet{bensbyfeltzing2006} suggest that massive stars dominate the carbon production at low metallicities while intermediate- and low-mass stars become important at higher metallicities \citep[see also][]{carigietal2005}.

\citet{kobulnickyskillman1998} find a relation between [N/O] and [O/C] in metal-poor galaxies.
Their data points together with measurements obtained by \citet{garnettetal1999} are also presented in Fig.\ \ref{alphatoC}.
They cover the range $0.0 \lesssim [\alpha/\element{C}] \lesssim +0.5$ and follow well the fitted relation.
The authors conclude that the correlation reflects the global enrichment of the interstellar medium resulting from the star formation history in each investigated galaxy.
The slope of the relation is supposed to depend on the initial mass function fixing the ratio of carbon and nitrogen producing stars.
\citet{garnettetal1999} find that [N/C] is independent of [O/H] while [O/C] depends on the metallicity in extragalactic \ion{H}{ii} regions.
Therefore, carbon and nitrogen are supposed to be produced in similar stellar mass ranges.

If in low-metallicity environments carbon is mainly generated in massive stars,  those stars are expected to contribute most of the carbon in the IGM.
The estimate of $[\mathrm{Si/C}] \simeq 0.77$ with small scatter in the low-density IGM \citep{aguirreetal2004} indicates no inhomogeneity in [Si/C], suggesting a common origin of silicon and carbon, presumably massive stars.
Considering very massive stars ($500 - 1000\,M_{\sun}$) exploding as core-collapse supernovae, \citet{ohkuboetal2006} reproduced the observed [Si/C] value.
On the other hand, \citet{matteuccicalura2005} studied the chemical enrichment due to very massive Population III stars, and according to their calculations pair-instability supernovae cannot account for the observed [Si/C] at $z\sim 5$ derived by \citet{songaila2001}.
Furthermore, enrichment by Pop III stars predict far too little nitrogen to be consistent with the abundance ratios in DLA systems.
Therefore, a significant contribution to the carbon and nitrogen abundance by intermediate- and low-mass Pop II stars is required.

DLA systems with reliable estimates of the carbon abundance are rare \citep{dodoricomolaro2004, richteretal2005, ernietal2006, pettinietal2008} since the \ion{C}{ii} features are usually saturated.
The existing measurements are indicated in Fig.\ \ref{alphatoC} as well.
Though there are only 5 data points and one upper limit, [N/$\alpha$] in DLAs appears to be systematically lower than in intergalactic systems with the same [$\alpha$/C].
Since the DLAs with existing carbon measurements are the most metal-poor \citep[e.g.][]{ernietal2006}, their [N/$\alpha$] is believed to represent the value for enrichment with primary nitrogen and thus is supposed to be constant.
The discrepancy to the [N/$\alpha$]-[$\alpha$/C]-relation for intervening and associated systems therefore confirms that the latter are enriched by secondary nitrogen.

Summarizing, we speculate about the origin of carbon and nitrogen in the IGM.
Provided that the nitrogen abundance is dominated by secondary nitrogen as concluded from Fig.\ \ref{Ntoalpha}, the absorbers are supposed to be enriched by intermediate mass stars.
That means the enrichment should have occurred only recently due to the time delay between the release of $\alpha$-elements created in massive stars and nitrogen produced in intermediate-mass stars.
Carbon can be produced in stars of very different masses.
Therefore, the carbon in the observed systems might be dominated by material produced in intermediate-mass stars as well.
Thus, the correlation might indicate a common origin of carbon and nitrogen as suggested by \citet{garnettetal1999}.
However, the yields of C and O from massive stars are affected by the mass loss during the stellar evolution which in turn depends on the metallicity.
On the other hand, if [N/C] correlates with [N/H] as indicated by our measurements, a different origin of nitrogen and carbon is suggested with carbon released earlier.
From the observed correlations only we speculate that $\alpha$-elements presumably are released into the IGM first, then carbon, and most recently nitrogen, thus in an order that reflect the decreasing masses of their stellar production sites.
However, resulting abundances then depend on the IMF making the interpretation of observed abundance ratios a complex problem \citep[e.g.][]{matteuccichiappini2005}.
For a detailed interpretation models applying nucleosynthesis yields for different types of the IMF would be required, which is beyond the scope of this work.

\section{Summary and conclusions}\label{conclusions}

We have performed a survey for \ion{N}{v} absorption systems in the spectra of 19 QSOs observed with VLT/UVES.
The \ion{N}{v} systems are selected from 198 intervening and 30 associated metal line systems exhibiting \ion{C}{iv} features, where absorbers within $|\Delta v| < 5000\,\mathrm{km\,s}^{-1}$ from the QSO emission redshift are classified as associated.\\

(1) In total, 21 intervening systems with 61 individual components and 11 associated systems with 46 individual components are identified in the redshift range $1.5 \lesssim z \lesssim 2.5$.
Thus, the fraction of \ion{C}{iv} systems exhibiting \ion{N}{v} features is roughly 11\,\% and 37\,\% for intervening and associated absorbers, respectively. 
The rate of incident for intervening \ion{N}{v} systems is $d\mathcal{N}/dz = 3.38\pm0.43$, corresponding to $d\mathcal{N}/dX = 1.10\pm0.14$.\\

(2) The column density distribution function can be described by the slope $\beta = 1.89\pm0.22$. 
This slope is consistent with measurements for \ion{C}{iv} and \ion{O}{vi}, suggesting a common origin of these species, which is photoionized gas. 
Further evidence for photoionization is provided by the narrow line width of the \ion{N}{v} components ($b_{\ion{N}{v}} \sim 6\,\mathrm{km\,s}^{-1}$ for the intervening systems), implying temperatures of $T \sim 3\cdot10^{4}\,\mathrm{K}$ too low for collisional ionization that produces \ion{N}{v} at much higher temperatures of $T \sim 2\cdot10^{5}\,\mathrm{K}$.
The \ion{N}{v} features in the individual systems are detected at the same position as \ion{C}{iv} while \ion{O}{vi} is shifted in velocity space to slightly higher values, typically by $v_{\ion{O}{vi}} - v_{\ion{C}{iv}} \simeq +4\,\mathrm{km\,s}^{-1}$. 
Therefore, \ion{N}{v} is apparently in the same gas phase as \ion{C}{iv} despite its higher ionization potential.\\

(3) The column densities of the \ion{N}{v} and \ion{C}{iv} components are closely correlated with different slopes for intervening and associated absorbers where intervening systems exhibit generally weaker \ion{N}{v} features than associated systems. 
The different slopes indicate a different spectral energy distribution of the ionizing radiation. 
While associated absorbers are exposed to the hard spectrum of the background QSO, the intervening systems are ionized by the substantially softer radiation of the intergalactic UV background \citep[][HM01]{haardtmadau2001}. 
In conclusion, intervening \ion{N}{v} features arise in rather "normal" intergalactic metal absorption systems and do not trace locally hardened radiation. 
This interpretation is supported by extensive photoionization modeling testing several spectral energy distributions. 
No indications for ionizing spectra substantially harder than the HM01 background are found in case of intervening systems.\\

(4) Detailed photoionization models are derived for each system in order to estimate the elemental abundances of the absorbers. 
We estimate a systematic error of roughly $0.7\,\mathrm{dex}$ for the individual metallicities of the associated absorbers due to the uncertain spectral energy distribution of the background QSO.
The resulting metallicities for intervening systems are typically low with a median of $\mathrm{[C/H]} = -0.63$. 
$\alpha$-elements like oxygen and silicon are enhanced yielding median values $[\alpha/\element{C}] = +0.41$ and $[\element{N}/\alpha] = -0.58$, respectively. 
The typical density is $n_{\element{H}} \simeq 10^{-3.6}\,\mathrm{cm}^{-3}$ and sizes of a few to several 10\,kpc with a median $\sim 5\,\mathrm{kpc}$ are found.
In contrast, associated systems are more metal-rich independently of the adopted radiation field, having solar metallicity or even higher (median $\mathrm{[C/H]} = +0.66$) and nearly solar relative abundances ($[\alpha/\element{C}] = -0.29$ and $[\element{N}/\alpha] = -0.05$, respectively).
These absorbers are denser, yielding $n_{\element{H}} \gtrsim 10^{-2.8}\,\mathrm{cm}^{-3}$, and about one order of magnitude smaller with typical sizes of 10 to a few 100\,pc.\\

(5) For both intervening and associated systems [N/C] and [N/$\alpha$] are correlated with metallicity where nitrogen is more abundant in metal-rich absorbers.
The increase of [N/$\alpha$] with [N/H] suggests that the nitrogen content is dominated by secondary nitrogen. 
The best-fit slope is $\sim 0.5$ consistent with abundance measurements in extragalactic \ion{H}{ii} regions and metal-poor halo stars. 
The values inferred for intervening systems show a larger scatter around this relation, indicating more inhomogeneous enrichment of the IGM. 
With our sample we only probe systems with $\mathrm{[N/H]} \gtrsim -2.0$. 
In order to detect nitrogen in lower-metallicity, intervening systems \ion{N}{iv} has to be surveyed whose main transition would arise in the UV at $z=2$.\\

(6) The relative abundances [$\alpha$/C] and [N/$\alpha$] trace the overall metallicity in the sense that metal-poor systems show enhanced $\alpha$-abundances compared to carbon and nitrogen. 
[N/$\alpha$] is anti-correlated with [$\alpha$/C].
A linear fit yields $[\element{N}/\alpha] = -(1.04 \pm 0.13)\cdot [\alpha/\element{C}] - (0.15 \pm 0.08)$. 
This correlation is consistent with abundance measurements in metal-poor halo stars and extragalactic \ion{H}{ii} regions, whereas the most metal-poor DLA systems do not fit the relation, probably reflecting the difference between primary and secondary nitrogen enrichment. 
Since the investigated sample is selected by \ion{N}{v}, the systems are expected to be rather nitrogen-rich and therefore dominated by secondary nitrogen. 
The relation between [N/$\alpha$] and [$\alpha$/C] could be used in future studies to constrain the initial mass function of the carbon-and nitrogen-producing stellar population.\\

\begin{acknowledgements}
\end{acknowledgements}

\bibliographystyle{aa}
\bibliography{0421}

\begin{thebibliography}{111}
\expandafter\ifx\csname natexlab\endcsname\relax\def\natexlab#1{#1}\fi

\bibitem[{{Agafonova} {et~al.}(2007){Agafonova}, {Levshakov}, {Reimers},
  {Fechner}, {Tytler}, {Simcoe}, \& {Songaila}}]{agafonovaetal2007}
{Agafonova}, I.~I., {Levshakov}, S.~A., {Reimers}, D., {et~al.} 2007, \aap,
  461, 893

\bibitem[{{Aguirre} {et~al.}(2008){Aguirre}, {Dow-Hygelund}, {Schaye}, \&
  {Theuns}}]{aguirreetal2008}
{Aguirre}, A., {Dow-Hygelund}, C., {Schaye}, J., \& {Theuns}, T. 2008, \apj,
  689, 851

\bibitem[{{Aguirre} {et~al.}(2004){Aguirre}, {Schaye}, {Kim}, {Theuns},
  {Rauch}, \& {Sargent}}]{aguirreetal2004}
{Aguirre}, A., {Schaye}, J., {Kim}, T., {et~al.} 2004, \apj, 602, 38

\bibitem[{{Akerman} {et~al.}(2004){Akerman}, {Carigi}, {Nissen}, {Pettini}, \&
  {Asplund}}]{akermanetal2004}
{Akerman}, C.~J., {Carigi}, L., {Nissen}, P.~E., {Pettini}, M., \& {Asplund},
  M. 2004, \aap, 414, 931

\bibitem[{{Aracil} {et~al.}(2004){Aracil}, {Petitjean}, {Pichon}, \&
  {Bergeron}}]{araciletal2004}
{Aracil}, B., {Petitjean}, P., {Pichon}, C., \& {Bergeron}, J. 2004, \aap, 419,
  811

\bibitem[{{Arav} {et~al.}(2007){Arav}, {Gabel}, {Korista}, {Kaastra}, {Kriss},
  {Behar}, {Costantini}, {Gaskell}, {Laor}, {Kodituwakku}, {Proga}, {Sako},
  {Scott}, \& {Steenbrugge}}]{aravetal2007}
{Arav}, N., {Gabel}, J.~R., {Korista}, K.~T., {et~al.} 2007, \apj, 658, 829

\bibitem[{{Asplund} {et~al.}(2005){Asplund}, {Grevesse}, \&
  {Sauval}}]{asplundetal2005}
{Asplund}, M., {Grevesse}, N., \& {Sauval}, A.~J. 2005, in ASP Conf. Ser. 336:
  Cosmic Abundances as Records of Stellar Evolution and Nucleosynthesis, ed.
  T.~G. {Barnes}, III \& F.~N. {Bash}, 25--+, astro-ph/0410214

\bibitem[{{Bahcall} \& {Wolf}(1968)}]{bahcallwolf1968}
{Bahcall}, J.~N. \& {Wolf}, R.~A. 1968, \apj, 152, 701

\bibitem[{{Beers} \& {Christlieb}(2005)}]{beerschristlieb2005}
{Beers}, T.~C. \& {Christlieb}, N. 2005, \araa, 43, 531

\bibitem[{{Bensby} \& {Feltzing}(2006)}]{bensbyfeltzing2006}
{Bensby}, T. \& {Feltzing}, S. 2006, \mnras, 367, 1181

\bibitem[{{Bergeron} {et~al.}(2002){Bergeron}, {Aracil}, {Petitjean}, \&
  {Pichon}}]{bergeronetal2002}
{Bergeron}, J., {Aracil}, B., {Petitjean}, P., \& {Pichon}, C. 2002, \aap, 396,
  L11

\bibitem[{{Bergeron} \& {Herbert-Fort}(2005)}]{bergeronherbertfort2005}
{Bergeron}, J. \& {Herbert-Fort}, S. 2005, in IAU Colloq. 199: Probing Galaxies
  through Quasar Absorption Lines, ed. P.~{Williams}, C.-G. {Shu}, \&
  B.~{Menard}, 265--280

\bibitem[{{Bradley} {et~al.}(2004){Bradley}, {Kaiser}, \&
  {Baan}}]{bradleyetal2004}
{Bradley}, L.~D., {Kaiser}, M.~E., \& {Baan}, W.~A. 2004, \apj, 603, 463

\bibitem[{{Carigi} {et~al.}(2005){Carigi}, {Peimbert}, {Esteban}, \&
  {Garc{\'{\i}}a-Rojas}}]{carigietal2005}
{Carigi}, L., {Peimbert}, M., {Esteban}, C., \& {Garc{\'{\i}}a-Rojas}, J. 2005,
  \apj, 623, 213

\bibitem[{{Carswell} {et~al.}(2002){Carswell}, {Schaye}, \&
  {Kim}}]{carswelletal2002}
{Carswell}, B., {Schaye}, J., \& {Kim}, T.-S. 2002, \apj, 578, 43

\bibitem[{{Centuri{\'o}n} {et~al.}(2003){Centuri{\'o}n}, {Molaro}, {Vladilo},
  {P{\'e}roux}, {Levshakov}, \& {D'Odorico}}]{centurionetal2003}
{Centuri{\'o}n}, M., {Molaro}, P., {Vladilo}, G., {et~al.} 2003, \aap, 403, 55

\bibitem[{{Cheng} {et~al.}(1991){Cheng}, {Gaskell}, \&
  {Koratkar}}]{chengetal1991}
{Cheng}, F.~H., {Gaskell}, C.~M., \& {Koratkar}, A.~P. 1991, \apj, 370, 487

\bibitem[{{Chiappini} {et~al.}(2003{\natexlab{a}}){Chiappini}, {Matteucci}, \&
  {Meynet}}]{chiappinietal2003b}
{Chiappini}, C., {Matteucci}, F., \& {Meynet}, G. 2003{\natexlab{a}}, \aap,
  410, 257

\bibitem[{{Chiappini} {et~al.}(2003{\natexlab{b}}){Chiappini}, {Romano}, \&
  {Matteucci}}]{chiappinietal2003}
{Chiappini}, C., {Romano}, D., \& {Matteucci}, F. 2003{\natexlab{b}}, \mnras,
  339, 63

\bibitem[{{Chieffi} \& {Limongi}(2004)}]{chieffilimongi2004}
{Chieffi}, A. \& {Limongi}, M. 2004, \apj, 608, 405

\bibitem[{{Cristiani} \& {Vio}(1990)}]{cristianivio1990}
{Cristiani}, S. \& {Vio}, R. 1990, \aap, 227, 385

\bibitem[{{Croft}(2004)}]{croft2004}
{Croft}, R.~A.~C. 2004, \apj, 610, 642

\bibitem[{{Danforth} \& {Shull}(2008)}]{danforthshull2008}
{Danforth}, C.~W. \& {Shull}, J.~M. 2008, \apj, 679, 194

\bibitem[{{Dietrich} {et~al.}(2003){Dietrich}, {Hamann}, {Shields},
  {Constantin}, {Heidt}, {J{\"a}ger}, {Vestergaard}, \&
  {Wagner}}]{dietrichetal2003}
{Dietrich}, M., {Hamann}, F., {Shields}, J.~C., {et~al.} 2003, \apj, 589, 722

\bibitem[{{D'Odorico} {et~al.}(2004){D'Odorico}, {Cristiani}, {Romano},
  {Granato}, \& {Danese}}]{dodoricoetal2004}
{D'Odorico}, V., {Cristiani}, S., {Romano}, D., {Granato}, G.~L., \& {Danese},
  L. 2004, \mnras, 351, 976

\bibitem[{{D'Odorico} \& {Molaro}(2004)}]{dodoricomolaro2004}
{D'Odorico}, V. \& {Molaro}, P. 2004, \aap, 415, 879

\bibitem[{{Edmunds} \& {Pagel}(1978)}]{edmundspagel1978}
{Edmunds}, M.~G. \& {Pagel}, B.~E.~J. 1978, \mnras, 185, 77P

\bibitem[{{Ellison} {et~al.}(2000){Ellison}, {Songaila}, {Schaye}, \&
  {Pettini}}]{ellisonetal2000}
{Ellison}, S.~L., {Songaila}, A., {Schaye}, J., \& {Pettini}, M. 2000, \aj,
  120, 1175

\bibitem[{{Erni} {et~al.}(2006){Erni}, {Richter}, {Ledoux}, \&
  {Petitjean}}]{ernietal2006}
{Erni}, P., {Richter}, P., {Ledoux}, C., \& {Petitjean}, P. 2006, \aap, 451, 19

\bibitem[{{Fechner} {et~al.}(2004){Fechner}, {Baade}, \&
  {Reimers}}]{fechneretal2004}
{Fechner}, C., {Baade}, R., \& {Reimers}, D. 2004, \aap, 418, 857

\bibitem[{{Fechner} {et~al.}(2006){Fechner}, {Reimers}, {Songaila}, {Simcoe},
  {Rauch}, \& {Sargent}}]{fechneretal2006a}
{Fechner}, C., {Reimers}, D., {Songaila}, A., {et~al.} 2006, \aap, 455, 73

\bibitem[{{Ferguson} {et~al.}(1998){Ferguson}, {Gallagher}, \&
  {Wyse}}]{fergusonetal1998}
{Ferguson}, A.~M.~N., {Gallagher}, J.~S., \& {Wyse}, R.~F.~G. 1998, \aj, 116,
  673

\bibitem[{{Ferland} {et~al.}(1998){Ferland}, {Korista}, {Verner}, {Ferguson},
  {Kingdon}, \& {Verner}}]{ferlandetal1998}
{Ferland}, G.~J., {Korista}, K.~T., {Verner}, D.~A., {et~al.} 1998, \pasp, 110,
  761

\bibitem[{{Fields} {et~al.}(2007){Fields}, {Mathur}, {Krongold}, {Williams}, \&
  {Nicastro}}]{fieldsetal2007}
{Fields}, D.~L., {Mathur}, S., {Krongold}, Y., {Williams}, R., \& {Nicastro},
  F. 2007, \apj, 666, 828

\bibitem[{{Fox} {et~al.}(2008){Fox}, {Bergeron}, \& {Petitjean}}]{foxetal2008}
{Fox}, A.~J., {Bergeron}, J., \& {Petitjean}, P. 2008, \mnras, 388, 1557

\bibitem[{{Fox} {et~al.}(2007){Fox}, {Petitjean}, {Ledoux}, \&
  {Srianand}}]{foxetal2007}
{Fox}, A.~J., {Petitjean}, P., {Ledoux}, C., \& {Srianand}, R. 2007, \aap, 465,
  171

\bibitem[{{Gabel} {et~al.}(2006){Gabel}, {Arav}, \& {Kim}}]{gabeletal2006}
{Gabel}, J.~R., {Arav}, N., \& {Kim}, T.-S. 2006, \apj, 646, 742

\bibitem[{{Gallerani} {et~al.}(2008){Gallerani}, {Ferrara}, {Fan}, \&
  {Choudhury}}]{galleranietal2008}
{Gallerani}, S., {Ferrara}, A., {Fan}, X., \& {Choudhury}, T.~R. 2008, \mnras,
  386, 359

\bibitem[{{Ganguly} {et~al.}(2006){Ganguly}, {Sembach}, {Tripp}, {Savage}, \&
  {Wakker}}]{gangulyetal2006}
{Ganguly}, R., {Sembach}, K.~R., {Tripp}, T.~M., {Savage}, B.~D., \& {Wakker},
  B.~P. 2006, \apj, 645, 868

\bibitem[{{Garnett} {et~al.}(1999){Garnett}, {Shields}, {Peimbert},
  {Torres-Peimbert}, {Skillman}, {Dufour}, {Terlevich}, \&
  {Terlevich}}]{garnettetal1999}
{Garnett}, D.~R., {Shields}, G.~A., {Peimbert}, M., {et~al.} 1999, \apj, 513,
  168

\bibitem[{{Gnat} \& {Sternberg}(2007)}]{gnatsternberg2007}
{Gnat}, O. \& {Sternberg}, A. 2007, \apjs, 168, 213

\bibitem[{{Gon{\c c}alves} {et~al.}(2008){Gon{\c c}alves}, {Steidel}, \&
  {Pettini}}]{goncalvesetal2008}
{Gon{\c c}alves}, T.~S., {Steidel}, C.~C., \& {Pettini}, M. 2008, \apj, 676,
  816

\bibitem[{{Gustafsson} {et~al.}(1999){Gustafsson}, {Karlsson}, {Olsson},
  {Edvardsson}, \& {Ryde}}]{gustafssonetal1999}
{Gustafsson}, B., {Karlsson}, T., {Olsson}, E., {Edvardsson}, B., \& {Ryde}, N.
  1999, \aap, 342, 426

\bibitem[{{Haardt} \& {Madau}(2001)}]{haardtmadau2001}
{Haardt}, F. \& {Madau}, P. 2001, in Clusters of Galaxies and the High Redshift
  Universe Observed in X-rays, ed. D.~M. {Neumann} \& J.~T.~T. {Van}, 64

\bibitem[{{Hamann} {et~al.}(1997){Hamann}, {Barlow}, {Junkkarinen}, \&
  {Burbidge}}]{hamannetal1997}
{Hamann}, F., {Barlow}, T.~A., {Junkkarinen}, V., \& {Burbidge}, E.~M. 1997,
  \apj, 478, 80

\bibitem[{{Hamann} \& {Ferland}(1999)}]{hamannferland1999}
{Hamann}, F. \& {Ferland}, G. 1999, \araa, 37, 487

\bibitem[{{Heger} \& {Woosley}(2002)}]{hegerwoosley2002}
{Heger}, A. \& {Woosley}, S.~E. 2002, \apj, 567, 532

\bibitem[{{Henry} {et~al.}(2000){Henry}, {Edmunds}, \& {K{\"
  o}ppen}}]{henryetal2000}
{Henry}, R.~B.~C., {Edmunds}, M.~G., \& {K{\" o}ppen}, J. 2000, \apj, 541, 660

\bibitem[{{Indebetouw} \& {Shull}(2004)}]{indebetouwshull2004b}
{Indebetouw}, R. \& {Shull}, J.~M. 2004, \apj, 607, 309

\bibitem[{{Israelian} {et~al.}(2004){Israelian}, {Ecuvillon}, {Rebolo},
  {Garc{\'{\i}}a-L{\'o}pez}, {Bonifacio}, \& {Molaro}}]{israelianetal2004}
{Israelian}, G., {Ecuvillon}, A., {Rebolo}, R., {et~al.} 2004, \aap, 421, 649

\bibitem[{{Iwamoto} {et~al.}(1999){Iwamoto}, {Brachwitz}, {Nomoto},
  {Kishimoto}, {Umeda}, {Hix}, \& {Thielemann}}]{iwamotoetal1999}
{Iwamoto}, K., {Brachwitz}, F., {Nomoto}, K., {et~al.} 1999, \apjs, 125, 439

\bibitem[{{Izotov} \& {Thuan}(1999)}]{izotovthuan1999}
{Izotov}, Y.~I. \& {Thuan}, T.~X. 1999, \apj, 511, 639

\bibitem[{{Jakobsen} {et~al.}(2003){Jakobsen}, {Jansen}, {Wagner}, \&
  {Reimers}}]{jakobsenetal2003}
{Jakobsen}, P., {Jansen}, R.~A., {Wagner}, S., \& {Reimers}, D. 2003, \aap,
  397, 891

\bibitem[{{Kobayashi} {et~al.}(2006){Kobayashi}, {Umeda}, {Nomoto}, {Tominaga},
  \& {Ohkubo}}]{kobayashietal2006}
{Kobayashi}, C., {Umeda}, H., {Nomoto}, K., {Tominaga}, N., \& {Ohkubo}, T.
  2006, \apj, 653, 1145

\bibitem[{{Kobulnicky} \& {Skillman}(1996)}]{kobulnickyskillman1996}
{Kobulnicky}, H.~A. \& {Skillman}, E.~D. 1996, \apj, 471, 211

\bibitem[{{Kobulnicky} \& {Skillman}(1998)}]{kobulnickyskillman1998}
{Kobulnicky}, H.~A. \& {Skillman}, E.~D. 1998, \apj, 497, 601

\bibitem[{{Lehner} {et~al.}(2008){Lehner}, {Howk}, {Prochaska}, \&
  {Wolfe}}]{lehneretal2008}
{Lehner}, N., {Howk}, J.~C., {Prochaska}, J.~X., \& {Wolfe}, A.~M. 2008,
  \mnras, 390, 2

\bibitem[{{Levshakov} {et~al.}(2003){Levshakov}, {Agafonova}, {Reimers}, \&
  {Baade}}]{levshakovetal2003b}
{Levshakov}, S.~A., {Agafonova}, I.~I., {Reimers}, D., \& {Baade}, R. 2003,
  \aap, 404, 449

\bibitem[{{Levshakov} {et~al.}(2008){Levshakov}, {Agafonova}, {Reimers}, {Hou},
  \& {Molaro}}]{levshakovetal2008}
{Levshakov}, S.~A., {Agafonova}, I.~I., {Reimers}, D., {Hou}, J.~L., \&
  {Molaro}, P. 2008, \aap, 483, 19

\bibitem[{{Limongi} \& {Chieffi}(2003)}]{limongichieffi2003}
{Limongi}, M. \& {Chieffi}, A. 2003, \apj, 592, 404

\bibitem[{{Liske} \& {Williger}(2001)}]{liskewilliger2001}
{Liske}, J. \& {Williger}, G.~M. 2001, \mnras, 328, 653

\bibitem[{{Lopez} {et~al.}(2007){Lopez}, {Ellison}, {D'Odorico}, \&
  {Kim}}]{lopezetal2007}
{Lopez}, S., {Ellison}, S., {D'Odorico}, S., \& {Kim}, T.-S. 2007, \aap, 469,
  61

\bibitem[{{Mallery} {et~al.}(2007){Mallery}, {Kewley}, {Rich}, {Salim},
  {Charlot}, {Tremonti}, {Seibert}, {Small}, {Wyder}, {Barlow}, {Forster},
  {Friedman}, {Martin}, {Morrissey}, {Neff}, {Schiminovich}, {Bianchi},
  {Donas}, {Heckman}, {Lee}, {Madore}, {Milliard}, {Szalay}, {Welsh}, \&
  {Yi}}]{malleryetal2007}
{Mallery}, R.~P., {Kewley}, L., {Rich}, R.~M., {et~al.} 2007, \apjs, 173, 482

\bibitem[{{Matteucci} \& {Calura}(2005)}]{matteuccicalura2005}
{Matteucci}, F. \& {Calura}, F. 2005, \mnras, 360, 447

\bibitem[{{Matteucci} \& {Chiappini}(2005)}]{matteuccichiappini2005}
{Matteucci}, F. \& {Chiappini}, C. 2005, Publications of the Astronomical
  Society of Australia, 22, 49

\bibitem[{{Melbourne} {et~al.}(2004){Melbourne}, {Phillips}, {Salzer},
  {Gronwall}, \& {Sarajedini}}]{melbourneetal2004}
{Melbourne}, J., {Phillips}, A., {Salzer}, J.~J., {Gronwall}, C., \&
  {Sarajedini}, V.~L. 2004, \aj, 127, 686

\bibitem[{{Meynet} \& {Maeder}(2002)}]{meynetmaeder2002}
{Meynet}, G. \& {Maeder}, A. 2002, \aap, 381, L25

\bibitem[{{Meynet} \& {Pettini}(2004)}]{meynetpettini2004}
{Meynet}, G. \& {Pettini}, M. 2004, in IAU Symposium, Vol. 215, Stellar
  Rotation, ed. A.~{Maeder} \& P.~{Eenens}, 579--+

\bibitem[{{Misawa} {et~al.}(2007){Misawa}, {Charlton}, {Eracleous}, {Ganguly},
  {Tytler}, {Kirkman}, {Suzuki}, \& {Lubin}}]{misawaetal2007}
{Misawa}, T., {Charlton}, J.~C., {Eracleous}, M., {et~al.} 2007, \apjs, 171, 1

\bibitem[{{Ohkubo} {et~al.}(2006){Ohkubo}, {Umeda}, {Maeda}, {Nomoto},
  {Suzuki}, {Tsuruta}, \& {Rees}}]{ohkuboetal2006}
{Ohkubo}, T., {Umeda}, H., {Maeda}, K., {et~al.} 2006, \apj, 645, 1352

\bibitem[{{Papovich} {et~al.}(2000){Papovich}, {Norman}, {Bowen}, {Heckman},
  {Savaglio}, {Koekemoer}, \& {Blades}}]{papovichetal2000}
{Papovich}, C., {Norman}, C.~A., {Bowen}, D.~V., {et~al.} 2000, \apj, 531, 654

\bibitem[{{Petitjean} {et~al.}(2008){Petitjean}, {Ledoux}, \&
  {Srianand}}]{petitjeanetal2008}
{Petitjean}, P., {Ledoux}, C., \& {Srianand}, R. 2008, \aap, 480, 349

\bibitem[{{Petitjean} {et~al.}(1994){Petitjean}, {Rauch}, \&
  {Carswell}}]{petitjeanetal1994}
{Petitjean}, P., {Rauch}, M., \& {Carswell}, R.~F. 1994, \aap, 291, 29

\bibitem[{{Petitjean} \& {Srianand}(1999)}]{petitjeansrianand1999}
{Petitjean}, P. \& {Srianand}, R. 1999, \aap, 345, 73

\bibitem[{{Pettini} {et~al.}(2002){Pettini}, {Ellison}, {Bergeron}, \&
  {Petitjean}}]{pettinietal2002b}
{Pettini}, M., {Ellison}, S.~L., {Bergeron}, J., \& {Petitjean}, P. 2002, \aap,
  391, 21

\bibitem[{{Pettini} {et~al.}(2008){Pettini}, {Zych}, {Steidel}, \&
  {Chaffee}}]{pettinietal2008}
{Pettini}, M., {Zych}, B.~J., {Steidel}, C.~C., \& {Chaffee}, F.~H. 2008,
  \mnras, 385, 2011

\bibitem[{{Prochaska} {et~al.}(2008){Prochaska}, {Dessauges-Zavadsky},
  {Ramirez-Ruiz}, \& {Chen}}]{prochaskaetal2008}
{Prochaska}, J.~X., {Dessauges-Zavadsky}, M., {Ramirez-Ruiz}, E., \& {Chen},
  H.-W. 2008, \apj, 685, 344

\bibitem[{{Prochaska} {et~al.}(2002){Prochaska}, {Henry}, {O'Meara}, {Tytler},
  {Wolfe}, {Kirkman}, {Lubin}, \& {Suzuki}}]{prochaskaetal2002}
{Prochaska}, J.~X., {Henry}, R.~B.~C., {O'Meara}, J.~M., {et~al.} 2002, \pasp,
  114, 933

\bibitem[{{Qian} \& {Wasserburg}(2005)}]{qianwasserburg2005}
{Qian}, Y.-Z. \& {Wasserburg}, G.~J. 2005, \apj, 623, 17

\bibitem[{{Reimers} {et~al.}(2006){Reimers}, {Agafonova}, {Levshakov}, {Hagen},
  {Fechner}, {Tytler}, {Kirkman}, \& {Lopez}}]{reimersetal2006}
{Reimers}, D., {Agafonova}, I.~I., {Levshakov}, S.~A., {et~al.} 2006, \aap,
  449, 9

\bibitem[{{Reimers} {et~al.}(2001){Reimers}, {Baade}, {Hagen}, \&
  {Lopez}}]{reimersetal2001}
{Reimers}, D., {Baade}, R., {Hagen}, H.-J., \& {Lopez}, S. 2001, \aap, 374, 871

\bibitem[{{Richards} {et~al.}(1999){Richards}, {York}, {Yanny}, {Kollgaard},
  {Laurent-Muehleisen}, \& {vanden Berk}}]{richardsetal1999}
{Richards}, G.~T., {York}, D.~G., {Yanny}, B., {et~al.} 1999, \apj, 513, 576

\bibitem[{{Richter} {et~al.}(2005){Richter}, {Ledoux}, {Petitjean}, \&
  {Bergeron}}]{richteretal2005}
{Richter}, P., {Ledoux}, C., {Petitjean}, P., \& {Bergeron}, J. 2005, \aap,
  440, 819

\bibitem[{{Richter} {et~al.}(2008){Richter}, {Paerels}, \&
  {Kaastra}}]{richteretal2008}
{Richter}, P., {Paerels}, F.~B.~S., \& {Kaastra}, J.~S. 2008, Space Science
  Reviews, 134, 25

\bibitem[{{Savage} {et~al.}(1997){Savage}, {Sembach}, \& {Lu}}]{savageetal1997}
{Savage}, B.~D., {Sembach}, K.~R., \& {Lu}, L. 1997, \aj, 113, 2158

\bibitem[{{Schaye} {et~al.}(2007){Schaye}, {Carswell}, \&
  {Kim}}]{schayeetal2007}
{Schaye}, J., {Carswell}, R.~F., \& {Kim}, T.-S. 2007, \mnras, 379, 1169

\bibitem[{{Schirber} {et~al.}(2004){Schirber}, {Miralda-Escud{\' e}}, \&
  {McDonald}}]{schirberetal2004}
{Schirber}, M., {Miralda-Escud{\' e}}, J., \& {McDonald}, P. 2004, \apj, 610,
  105

\bibitem[{{Scott} {et~al.}(2000){Scott}, {Bechtold}, {Dobrzycki}, \&
  {Kulkarni}}]{scottetal2000}
{Scott}, J., {Bechtold}, J., {Dobrzycki}, A., \& {Kulkarni}, V.~P. 2000, \apjs,
  130, 67

\bibitem[{{Silva} \& {Viegas}(2002)}]{silvaviegas2002}
{Silva}, A.~I. \& {Viegas}, S.~M. 2002, \mnras, 329, 135

\bibitem[{{Simcoe} {et~al.}(2002){Simcoe}, {Sargent}, \&
  {Rauch}}]{simcoeetal2002}
{Simcoe}, R.~A., {Sargent}, W.~L.~W., \& {Rauch}, M. 2002, \apj, 578, 737

\bibitem[{{Simcoe} {et~al.}(2004){Simcoe}, {Sargent}, \&
  {Rauch}}]{simcoeetal2004}
{Simcoe}, R.~A., {Sargent}, W.~L.~W., \& {Rauch}, M. 2004, \apj, 606, 92

\bibitem[{{Simcoe} {et~al.}(2006){Simcoe}, {Sargent}, {Rauch}, \&
  {Becker}}]{simcoeetal2006}
{Simcoe}, R.~A., {Sargent}, W.~L.~W., {Rauch}, M., \& {Becker}, G. 2006, \apj,
  637, 648

\bibitem[{{Songaila}(2001)}]{songaila2001}
{Songaila}, A. 2001, \apjl, 561, L153

\bibitem[{{Songaila}(2005)}]{songaila2005}
{Songaila}, A. 2005, \aj, 130, 1996

\bibitem[{{Spite} {et~al.}(2005){Spite}, {Cayrel}, {Plez}, {Hill}, {Spite},
  {Depagne}, {Fran{\c c}ois}, {Bonifacio}, {Barbuy}, {Beers}, {Andersen},
  {Molaro}, {Nordstr{\"o}m}, \& {Primas}}]{spiteetal2005}
{Spite}, M., {Cayrel}, R., {Plez}, B., {et~al.} 2005, \aap, 430, 655

\bibitem[{{Sutherland} \& {Dopita}(1993)}]{sutherlanddopita1993}
{Sutherland}, R.~S. \& {Dopita}, M.~A. 1993, \apjs, 88, 253

\bibitem[{{Telfer} {et~al.}(2002{\natexlab{a}}){Telfer}, {Kriss}, {Zheng},
  {Davidsen}, \& {Tytler}}]{telferetal2002b}
{Telfer}, R.~C., {Kriss}, G.~A., {Zheng}, W., {Davidsen}, A.~F., \& {Tytler},
  D. 2002{\natexlab{a}}, \apj, 579, 500

\bibitem[{{Telfer} {et~al.}(2002{\natexlab{b}}){Telfer}, {Zheng}, {Kriss}, \&
  {Davidsen}}]{telferetal2002}
{Telfer}, R.~C., {Zheng}, W., {Kriss}, G.~A., \& {Davidsen}, A.~F.
  2002{\natexlab{b}}, \apj, 565, 773

\bibitem[{{Tripp} {et~al.}(2002){Tripp}, {Jenkins}, {Williger}, {Heap},
  {Bowers}, {Danks}, {Dav{\'e}}, {Green}, {Gull}, {Joseph}, {Kaiser},
  {Lindler}, {Weymann}, \& {Woodgate}}]{trippetal2002}
{Tripp}, T.~M., {Jenkins}, E.~B., {Williger}, G.~M., {et~al.} 2002, \apj, 575,
  697

\bibitem[{{Tripp} {et~al.}(1996){Tripp}, {Lu}, \& {Savage}}]{trippetal1996}
{Tripp}, T.~M., {Lu}, L., \& {Savage}, B.~D. 1996, \apjs, 102, 239

\bibitem[{{Tripp} {et~al.}(2008){Tripp}, {Sembach}, {Bowen}, {Savage},
  {Jenkins}, {Lehner}, \& {Richter}}]{trippetal2008}
{Tripp}, T.~M., {Sembach}, K.~R., {Bowen}, D.~V., {et~al.} 2008, \apjs, 177, 39

\bibitem[{{Umeda} \& {Nomoto}(2005)}]{umedanomoto2005}
{Umeda}, H. \& {Nomoto}, K. 2005, \apj, 619, 427

\bibitem[{{van Zee}(2000)}]{vanzee2000}
{van Zee}, L. 2000, \apjl, 543, L31

\bibitem[{{van Zee} \& {Haynes}(2006)}]{vanzeehaynes2006}
{van Zee}, L. \& {Haynes}, M.~P. 2006, \apj, 636, 214

\bibitem[{{van Zee} {et~al.}(1998){van Zee}, {Salzer}, {Haynes}, {O'Donoghue},
  \& {Balonek}}]{vanzeeetal1998}
{van Zee}, L., {Salzer}, J.~J., {Haynes}, M.~P., {O'Donoghue}, A.~A., \&
  {Balonek}, T.~J. 1998, \aj, 116, 2805

\bibitem[{{Verner} {et~al.}(1994){Verner}, {Barthel}, \&
  {Tytler}}]{verneretal1994}
{Verner}, D.~A., {Barthel}, P.~D., \& {Tytler}, D. 1994, \aaps, 108, 287

\bibitem[{{Wampler} {et~al.}(1996){Wampler}, {Williger}, {Baldwin}, {Carswell},
  {Hazard}, \& {McMahon}}]{wampleretal1996}
{Wampler}, E.~J., {Williger}, G.~M., {Baldwin}, J.~A., {et~al.} 1996, \aap,
  316, 33

\bibitem[{{Wolfe} {et~al.}(2005){Wolfe}, {Gawiser}, \&
  {Prochaska}}]{wolfeetal2005}
{Wolfe}, A.~M., {Gawiser}, E., \& {Prochaska}, J.~X. 2005, \araa, 43, 861

\bibitem[{{Woosley} \& {Weaver}(1995)}]{woosleyweaver1995}
{Woosley}, S.~E. \& {Weaver}, T.~A. 1995, \apjs, 101, 181

\bibitem[{{Worseck} {et~al.}(2007){Worseck}, {Fechner}, {Wisotzki}, \&
  {Dall'Aglio}}]{worsecketal2007}
{Worseck}, G., {Fechner}, C., {Wisotzki}, L., \& {Dall'Aglio}, A. 2007, \aap,
  473, 805

\bibitem[{{Zheng} \& {Malkan}(1993)}]{zhengmalkan1993}
{Zheng}, W. \& {Malkan}, M.~A. 1993, \apj, 415, 517

\end{thebibliography}

\clearpage
\Online
\appendix

\section{\ion{N}{v} systems}

\begin{figure}
  \centering
  \resizebox{\hsize}{!}{\includegraphics[bb=33 485 373 780,clip=]{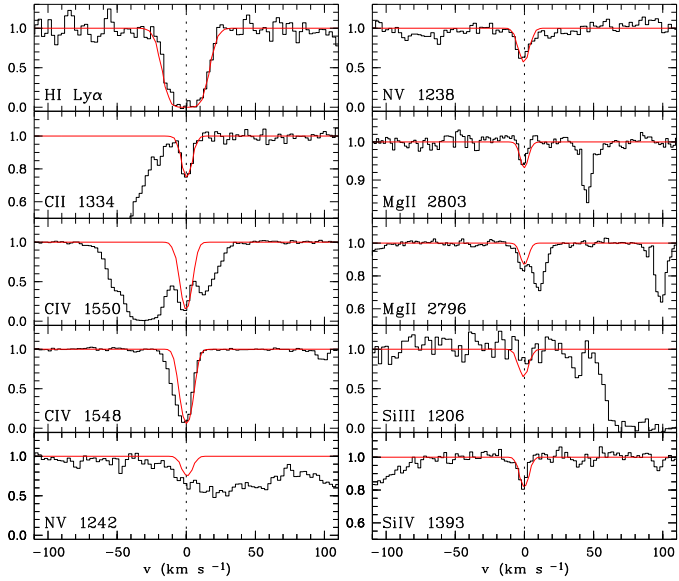}}
  \caption{Intervening system at $z=1.5771$ towards HE~0001-2340.
Smooth lines indicate the profiles for the favorite HM01 model.
  }
  \label{HE0001_1.5771}
\end{figure}

\begin{figure}
  \centering
  \resizebox{\hsize}{!}{\includegraphics[bb=33 220 373 780,clip=]{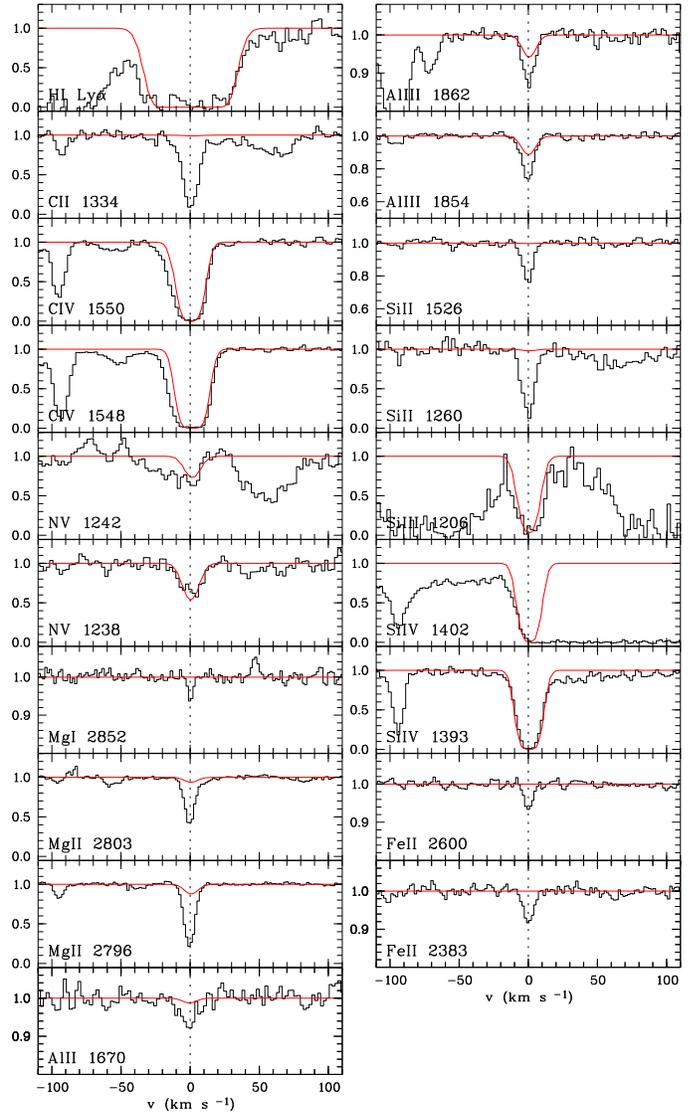}}
  \caption{Intervening system at $z=1.5855$ towards PKS~1448-232.
Smooth lines indicate the profiles for the HM01 model.
This system has a least two gas phases where the gas phase giving rise to features of low-ionization species is not included into the model.
  }
  \label{PKS1448_1.5855}
\end{figure}

\begin{figure}
  \centering
  \resizebox{\hsize}{!}{\includegraphics[bb=33 435 373 780,clip=]{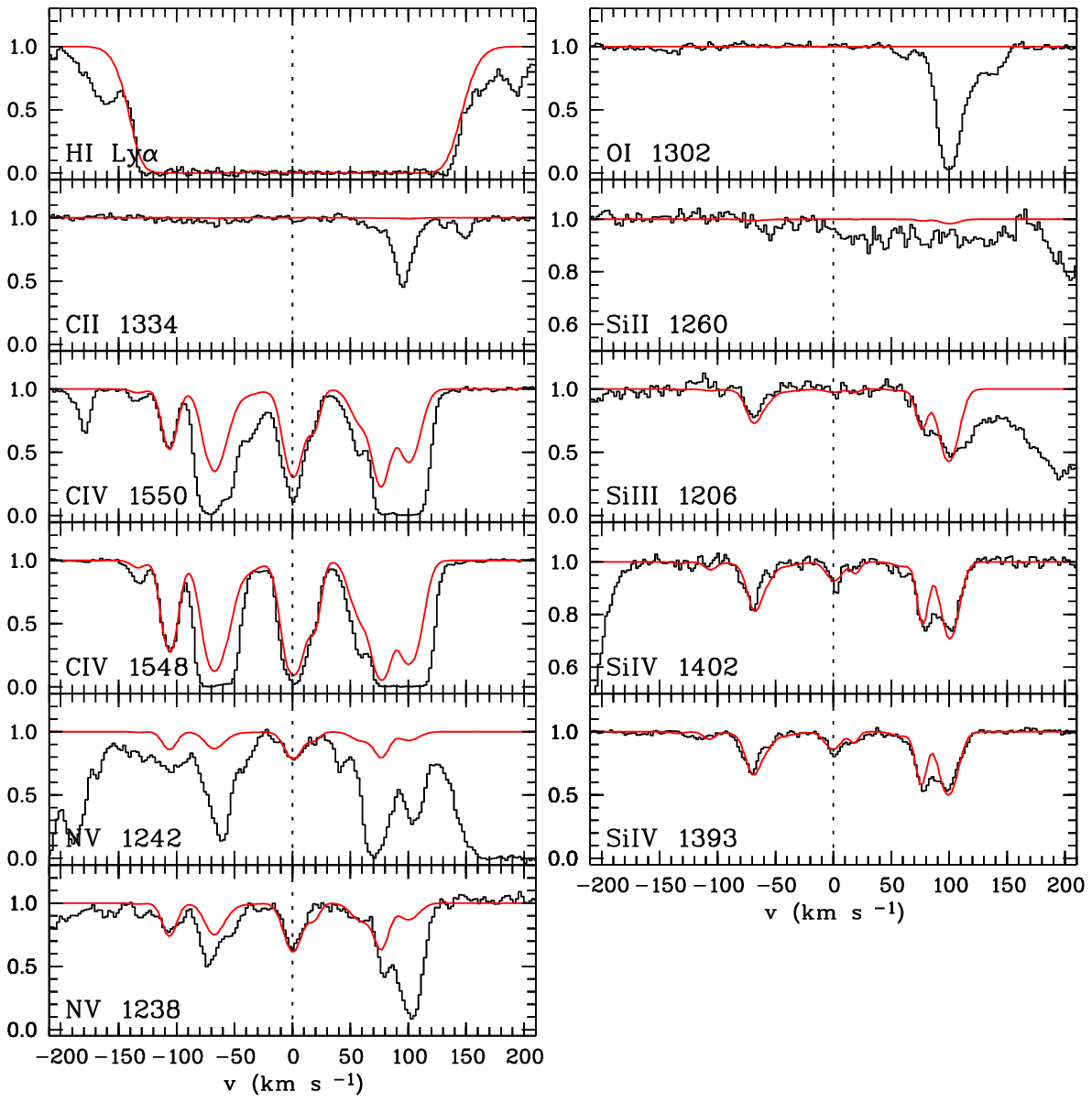}}
  \caption{Intervening system at $z=1.5966$ towards PKS~0237-23.
Smooth lines indicate the profiles for the favorite HM01 model.
  }
  \label{PKS0237_1.5966}
\end{figure}

\begin{figure}
  \centering
  \resizebox{\hsize}{!}{\includegraphics[bb=33 540 373 780,clip=]{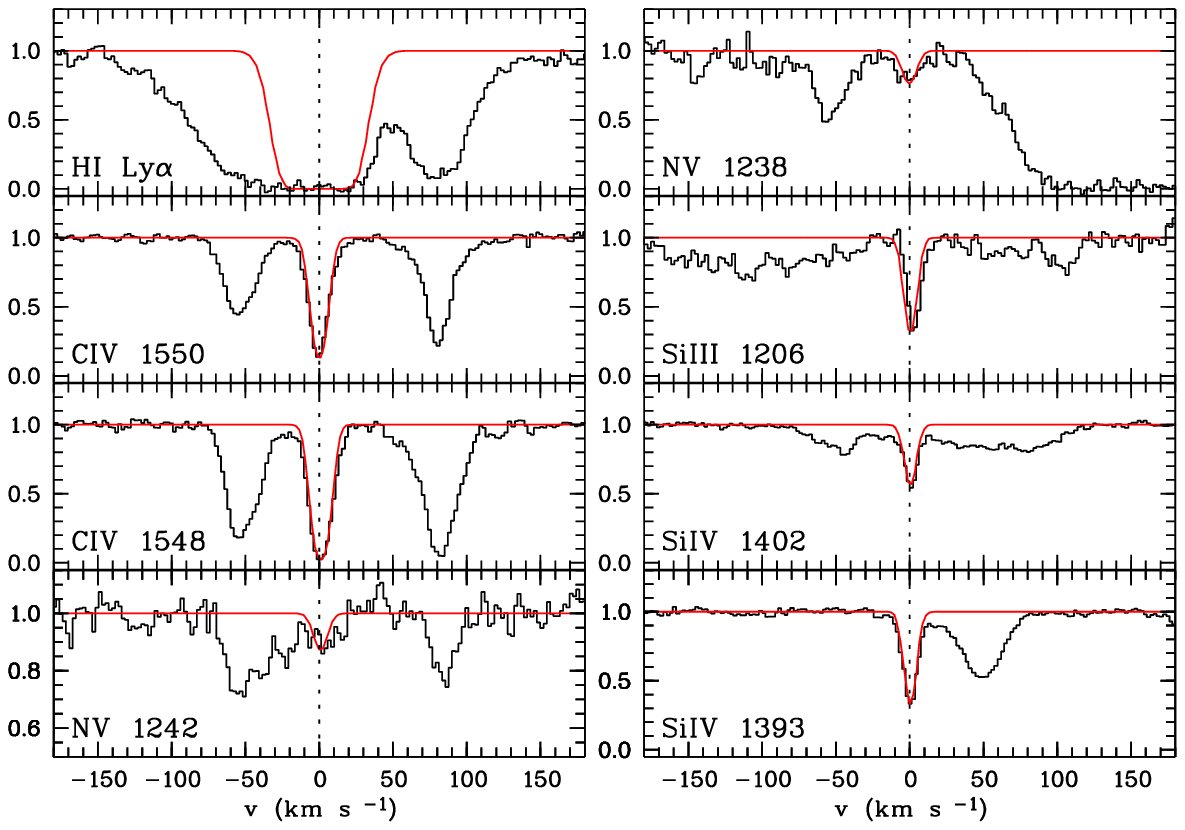}}
  \caption{Intervening system at $z=1.7236$ towards PKS~1448-232.
Smooth lines indicate the profiles for the favorite HM01 model.
  }
  \label{PKS1448_1.7236}
\end{figure}

\begin{figure}
  \centering
  \resizebox{\hsize}{!}{\includegraphics[bb=33 485 373 780,clip=]{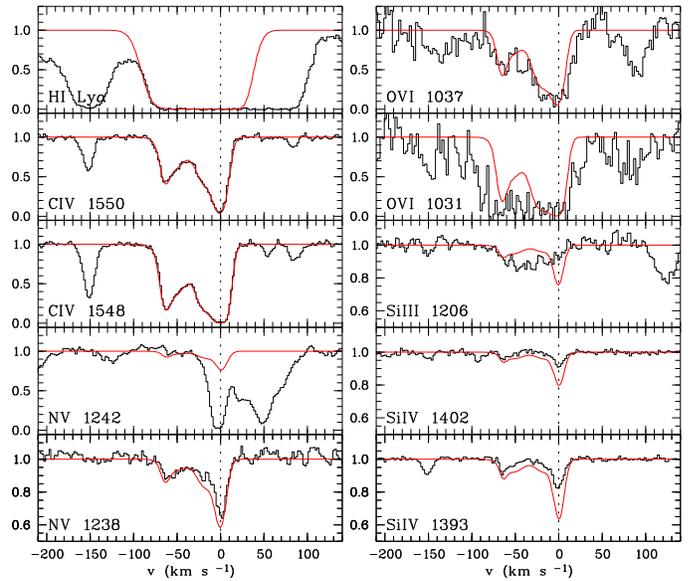}}
  \caption{Intervening system at $z=1.9656$ towards HE~2217-2818.
Smooth lines indicate the profiles for the favorite HM01 model.
  }
  \label{HE2217_1.9656}
\end{figure}

\begin{figure}
  \centering
  \resizebox{\hsize}{!}{\includegraphics[bb=33 165 373 780,clip=]{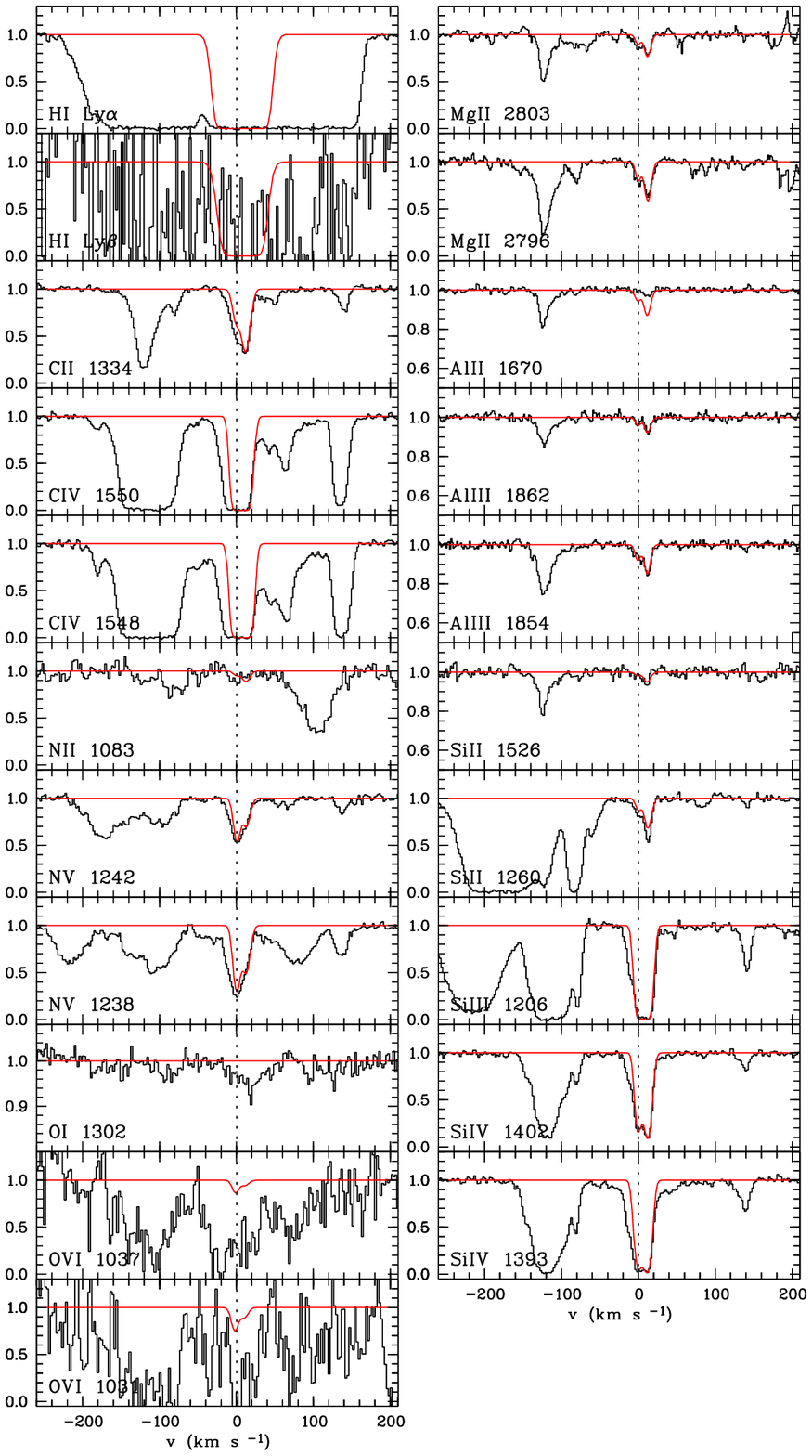}}
  \caption{Intervening system at $z=1.9744$ towards Q~0122-380.
Smooth lines indicate the profiles for the favorite HM01 model.
  }
  \label{Q0122_1.9744}
\end{figure}

\begin{figure}
  \centering
  \resizebox{\hsize}{!}{\includegraphics[bb=33 485 373 780,clip=]{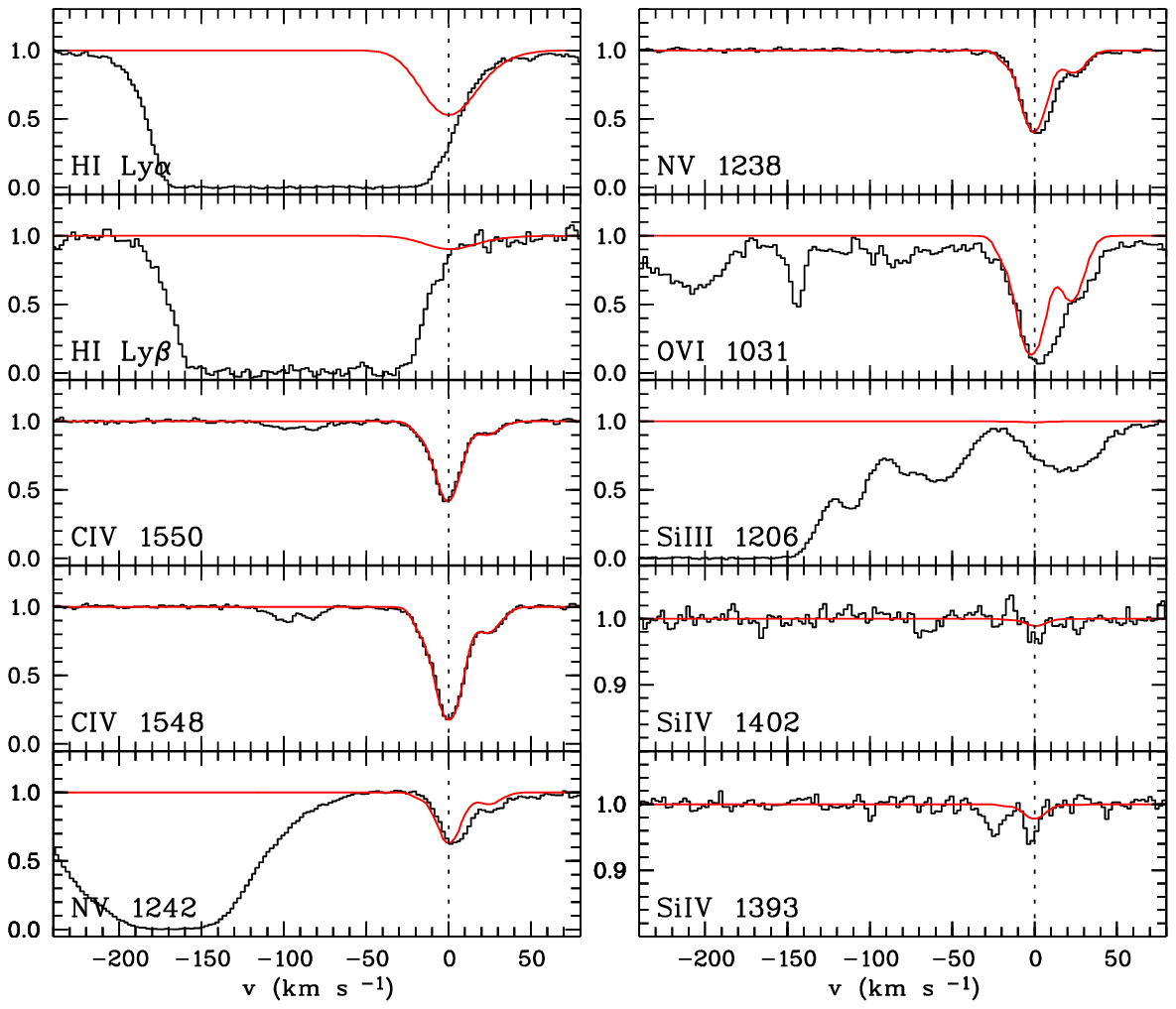}}
  \caption{Intervening system at $z=2.0422$ towards PKS~0237-23.
Smooth lines indicate the profiles for the favorite HM01 model.
  }
  \label{PKS0237_2.0422}
\end{figure}

\begin{figure}
  \centering
  \resizebox{\hsize}{!}{\includegraphics[bb=33 542 373 780,clip=]{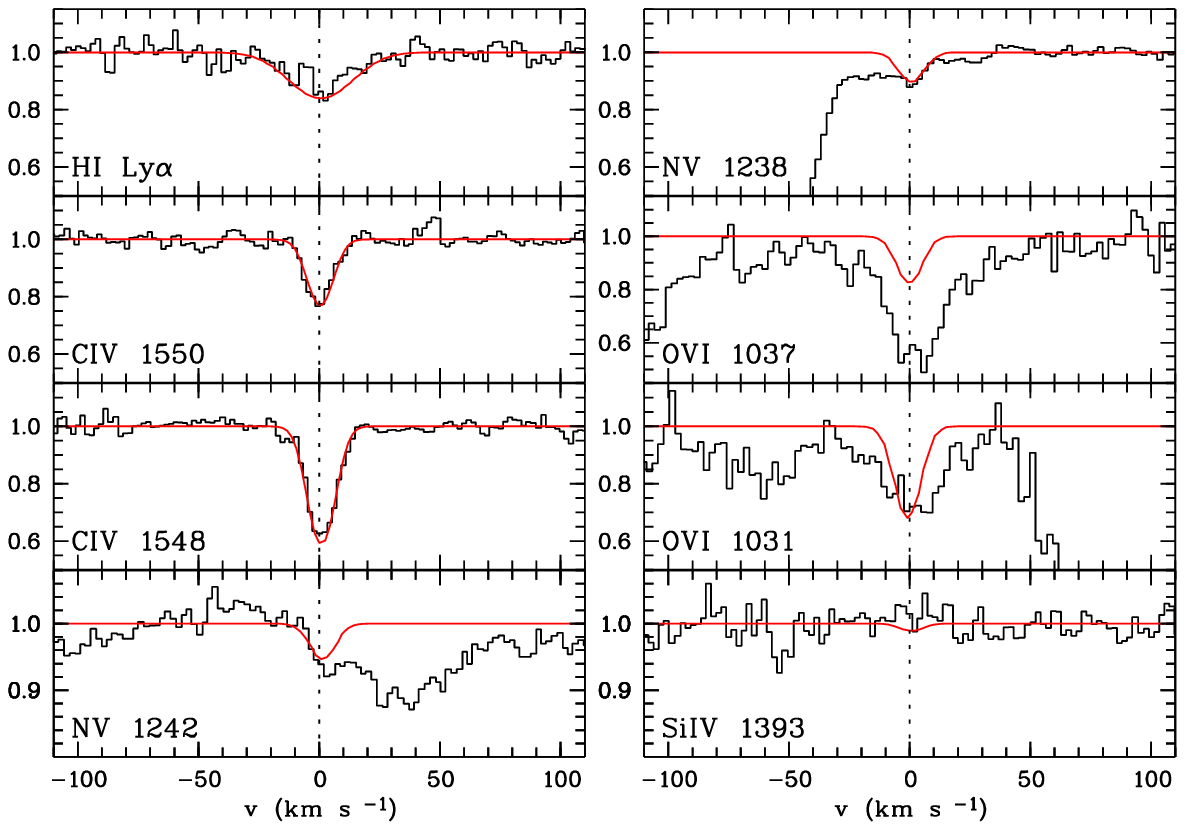}}
  \caption{Intervening system at $z=2.0626$ towards Q~0122-380.
Smooth lines indicate the profiles for the favorite HM01 model.
  }
  \label{Q0122_2.0626}
\end{figure}

\begin{figure}
  \centering
  \resizebox{\hsize}{!}{\includegraphics[bb=33 485 373 780,clip=]{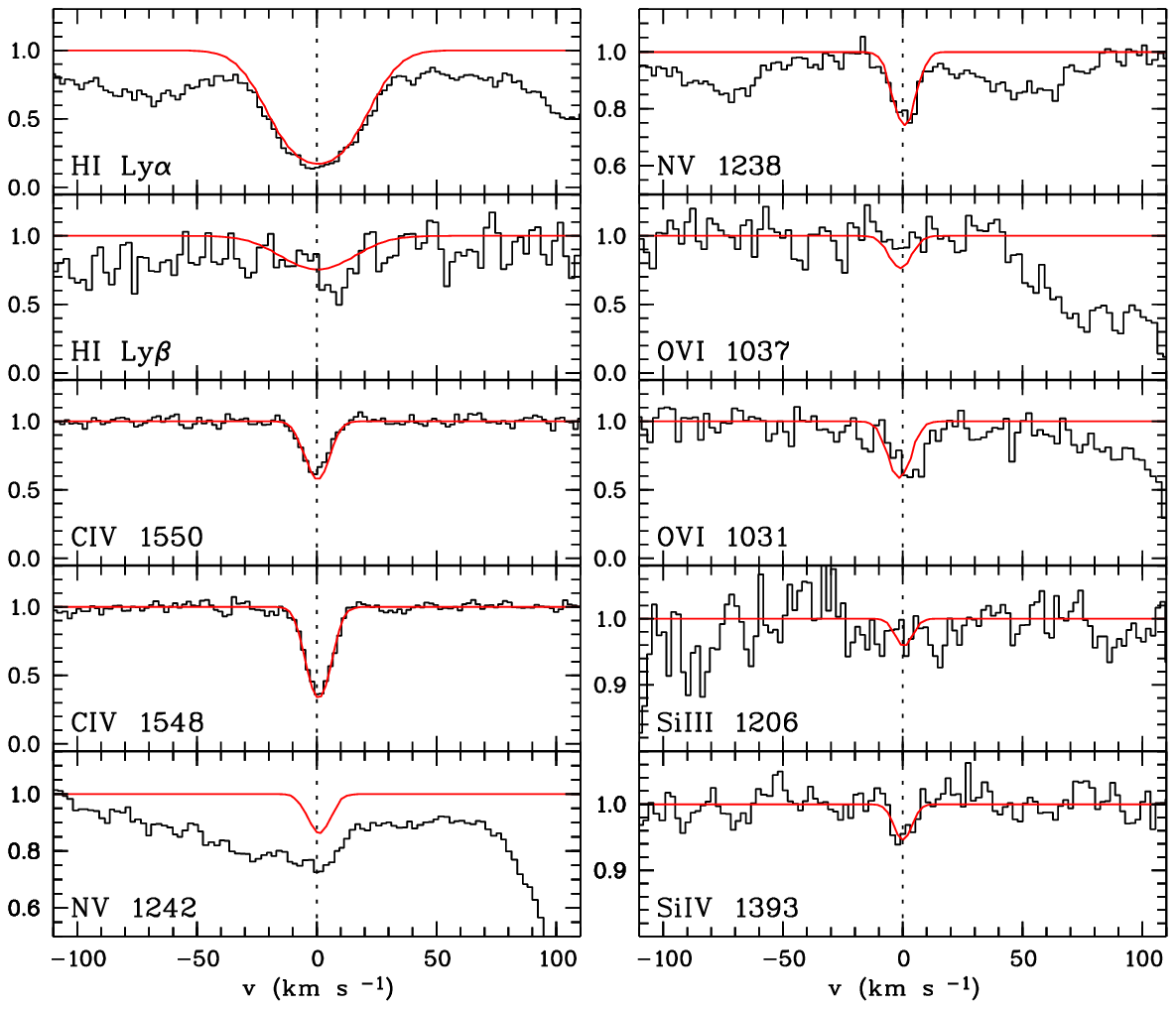}}
  \caption{Intervening system at $z=2.0764$ towards Q~0329-385.
Smooth lines indicate the profiles for the favorite HM01 model.
  }
  \label{Q0329_2.0764}
\end{figure}

\begin{figure}
  \centering
  \resizebox{\hsize}{!}{\includegraphics[bb=33 542 373 780,clip=]{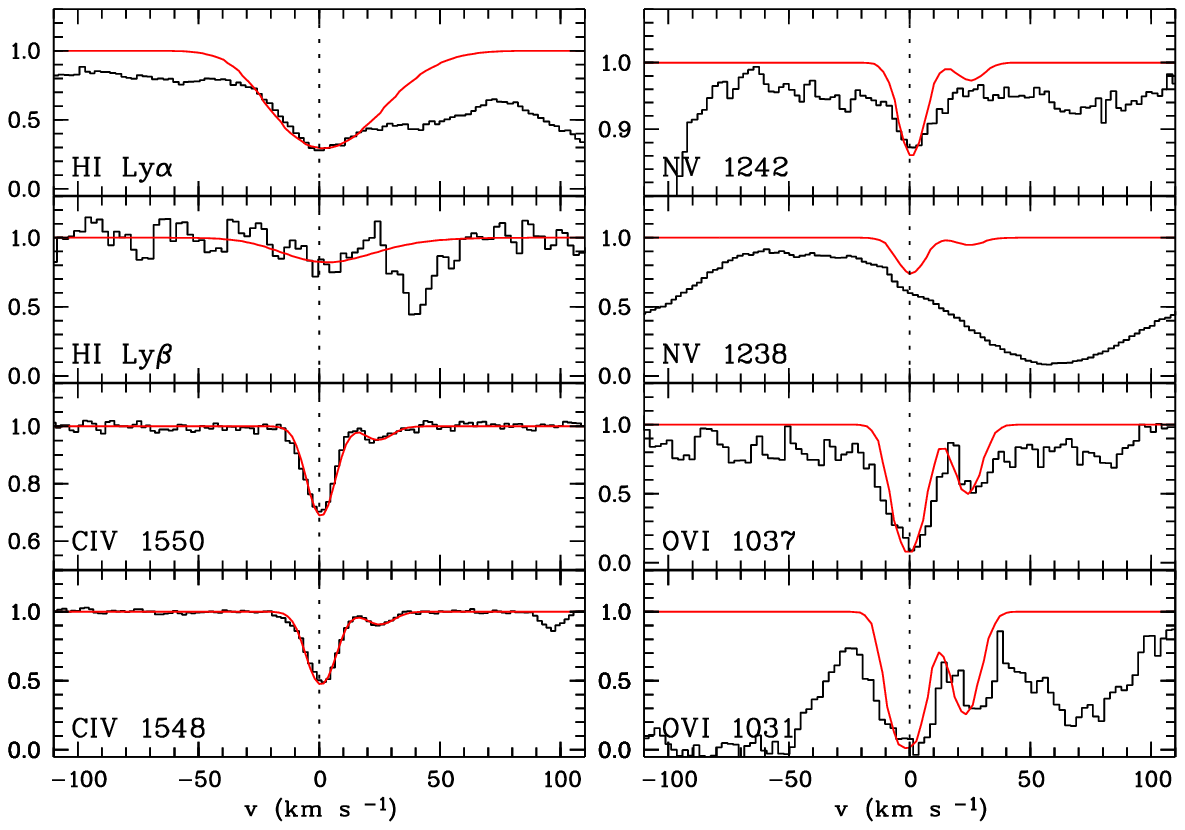}}
  \caption{Intervening system at $z=2.1098$ towards PKS~1448-232.
Smooth lines indicate the profiles for the favorite HM01 model.
  }
  \label{PKS1448_2.1098}
\end{figure}

\begin{figure}
  \centering
  \resizebox{\hsize}{!}{\includegraphics[bb=33 542 373 780,clip=]{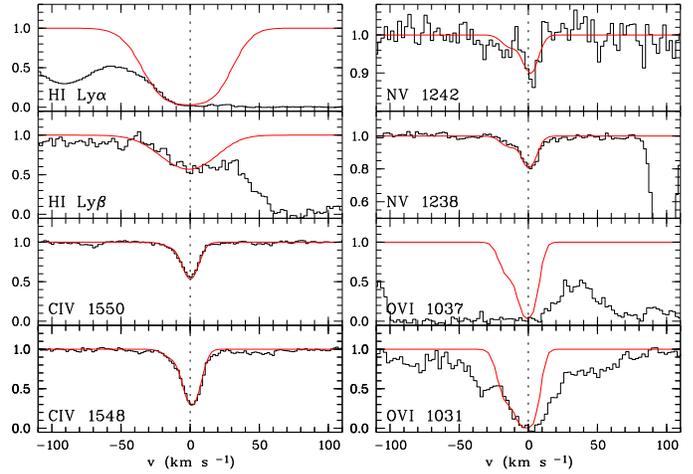}}
  \caption{Intervening system at $z=2.1162$ towards HE~1347-2457.
Smooth lines indicate the profiles for the favorite HM01 model.
  }
  \label{HE1347_2.1162}
\end{figure}

\begin{figure}
  \centering
  \resizebox{\hsize}{!}{\includegraphics[bb=33 435 373 780,clip=]{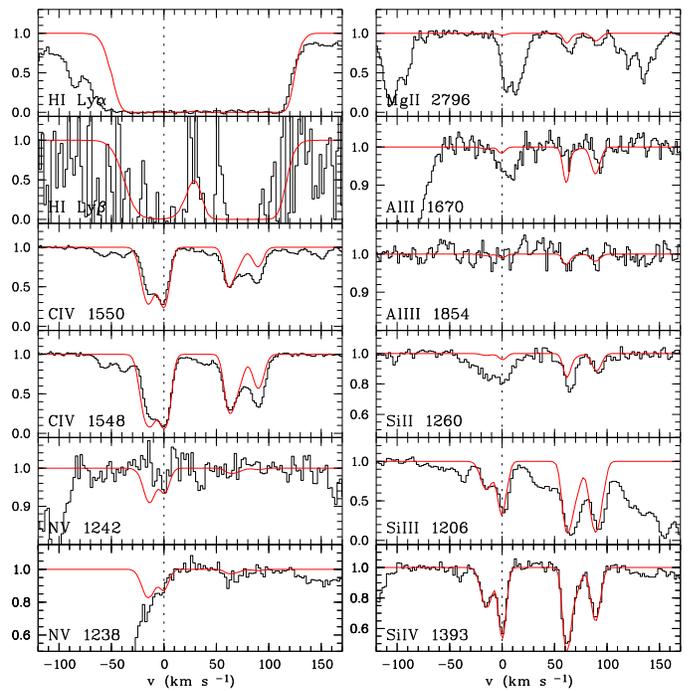}}
  \caption{Intervening system at $z=2.2212$ towards HE~0940-1050.
Smooth lines indicate the profiles for the favorite HM01 model.
  }
  \label{HE0940_2.2212}
\end{figure}

\begin{figure}
  \centering
  \resizebox{\hsize}{!}{\includegraphics[bb=33 380 373 780,clip=]{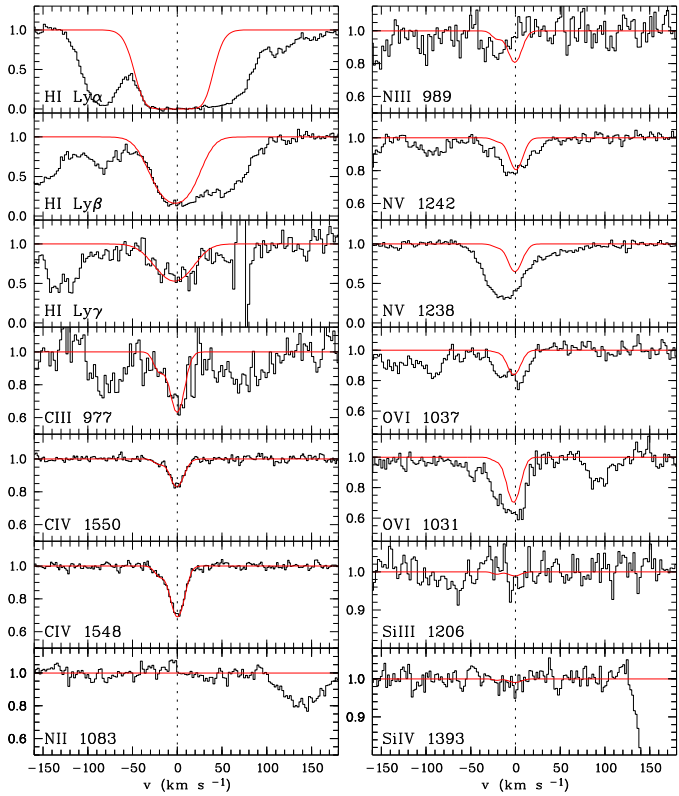}}
  \caption{Intervening system at $z=2.2354$ towards HE~1158-1843.
Smooth lines indicate the profiles for the favorite HM01 model.
  }
  \label{HE1158_2.2354}
\end{figure}

\begin{figure}
  \centering
  \resizebox{\hsize}{!}{\includegraphics[bb=33 485 373 780,clip=]{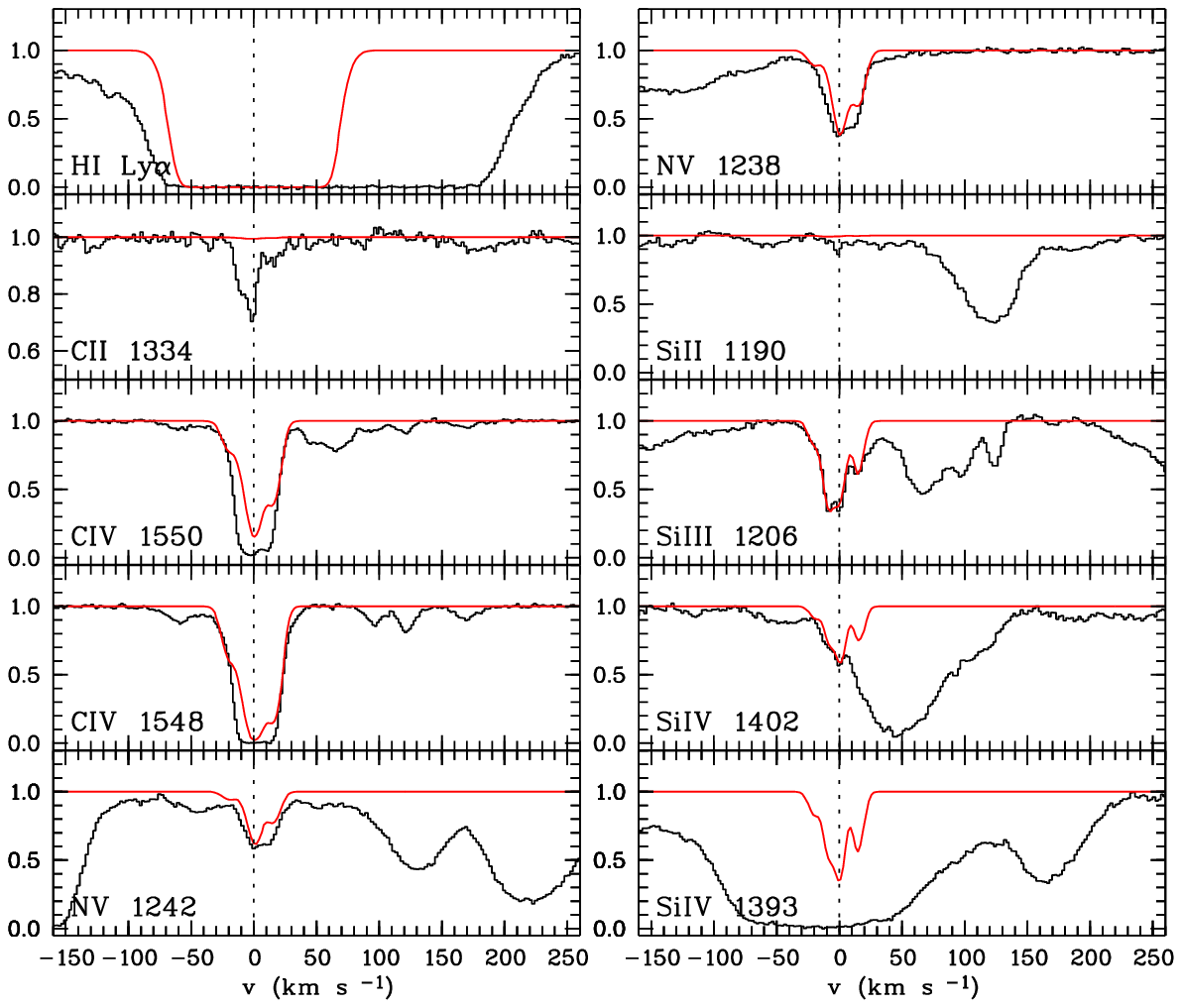}}
  \caption{Intervening system at $z=2.2464$ towards Q~0420-388.
Smooth lines indicate the profiles for the favorite HM01 model.
  }
  \label{Q0420_2.2464}
\end{figure}

\begin{figure}
  \centering
  \resizebox{\hsize}{!}{\includegraphics[bb=33 328 373 780,clip=]{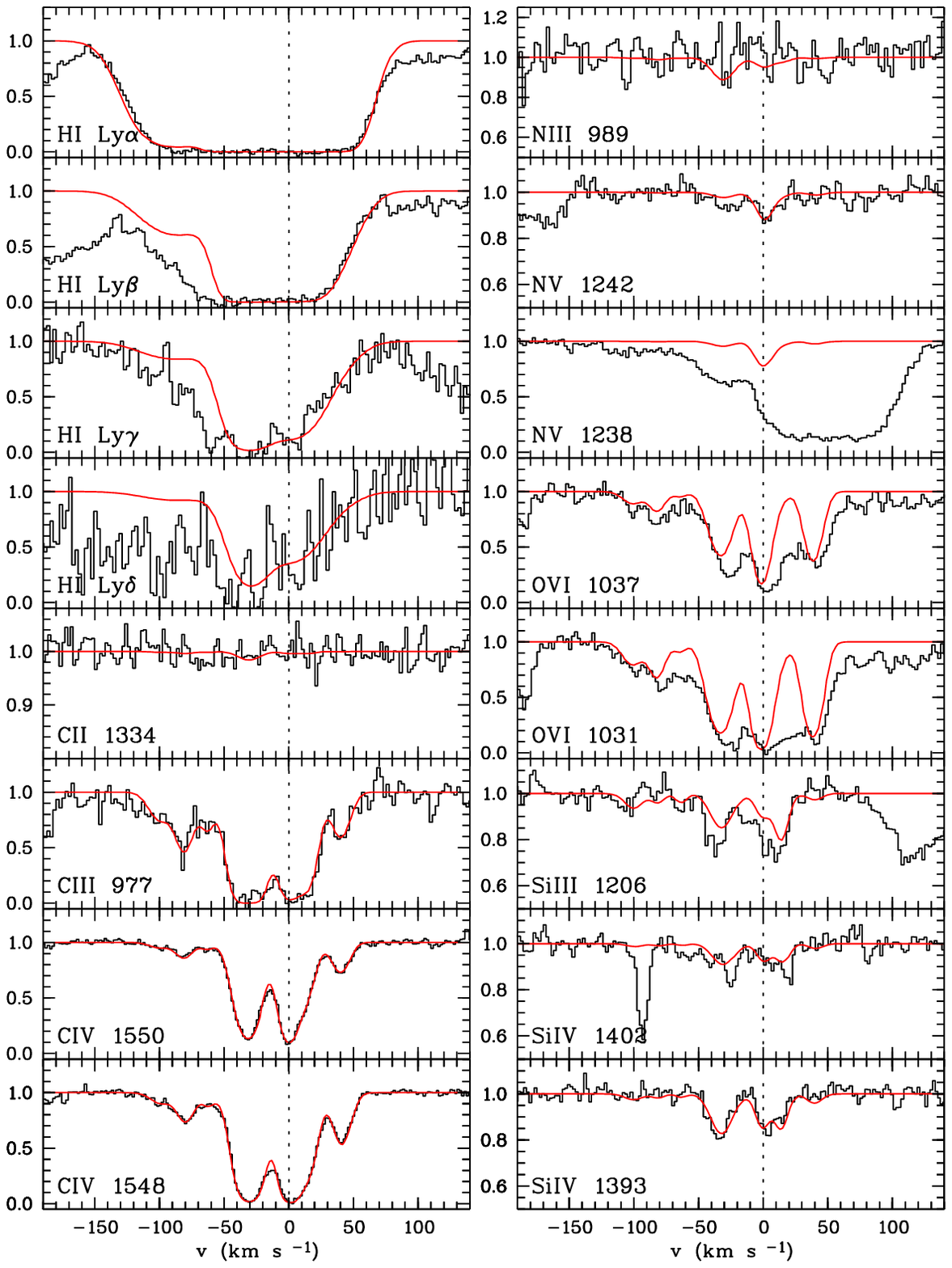}}
  \caption{Intervening system at $z=2.2510$ towards Q~0329-385.
Smooth lines indicate the profiles for the favorite HM01 model.
  }
  \label{Q0329_2.2510}
\end{figure}

\begin{figure}
  \centering
  \resizebox{\hsize}{!}{\includegraphics[bb=33 542 373 780,clip=]{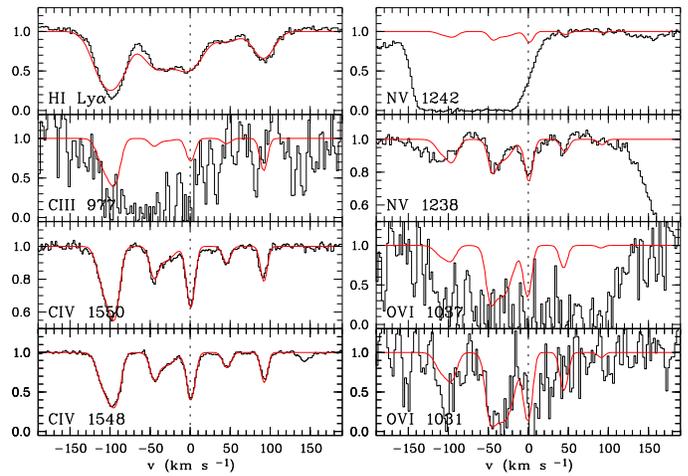}}
  \caption{Intervening system at $z=2.2753$ towards HE~2347-4342.
Smooth lines indicate the profiles for the favorite HM01 model.
  }
  \label{HE2347_2.2753}
\end{figure}

\begin{figure}
  \centering
  \resizebox{\hsize}{!}{\includegraphics[bb=33 542 373 780,clip=]{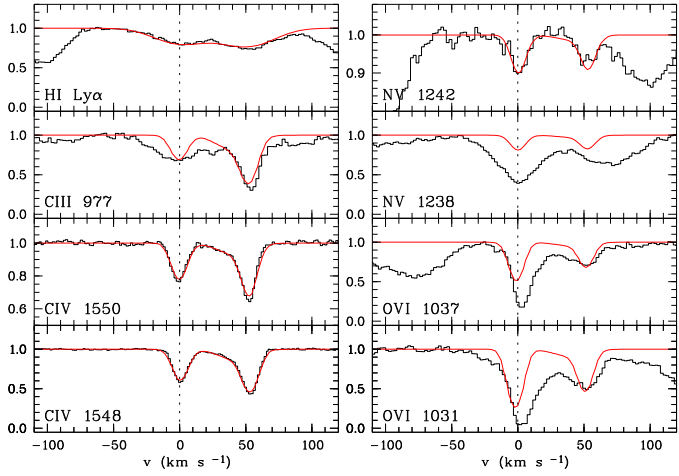}}
  \caption{Intervening system at $z=2.4686$ towards HE~0151-4326.
Smooth lines indicate the profiles for the favorite HM01 model.
  }
  \label{HE0151_2.4686}
\end{figure}

\clearpage


\begin{figure}
  \centering
  \resizebox{\hsize}{!}{\includegraphics[bb=33 381 373 780,clip=]{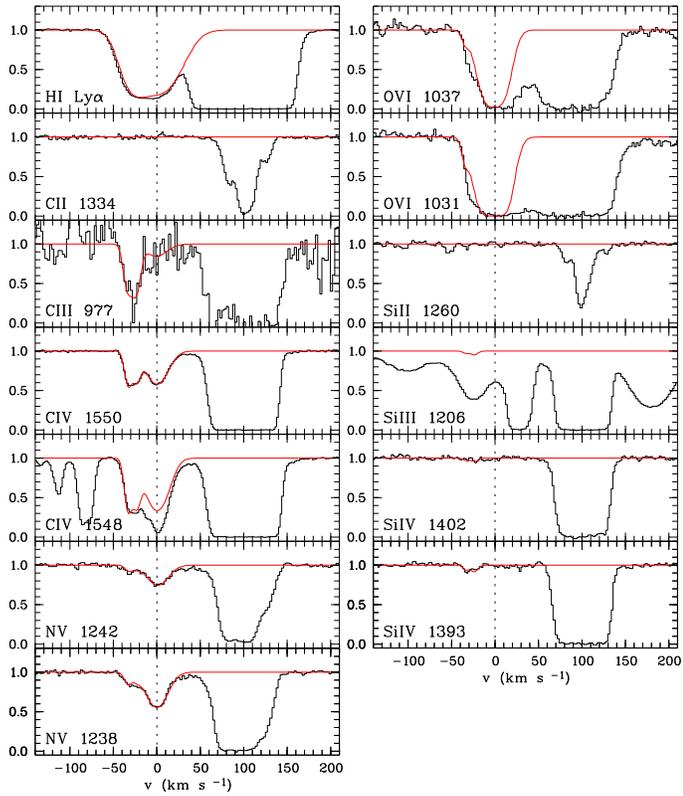}}
  \caption{Associated system at $z=2.1462$ towards HE~1341-1020.
Smooth lines indicate the profiles for the favorite HM01 model.
  }
  \label{HE1341_2.1462}
\end{figure}

\begin{figure}
  \centering
  \resizebox{\hsize}{!}{\includegraphics[bb=33 58 373 780,clip=]{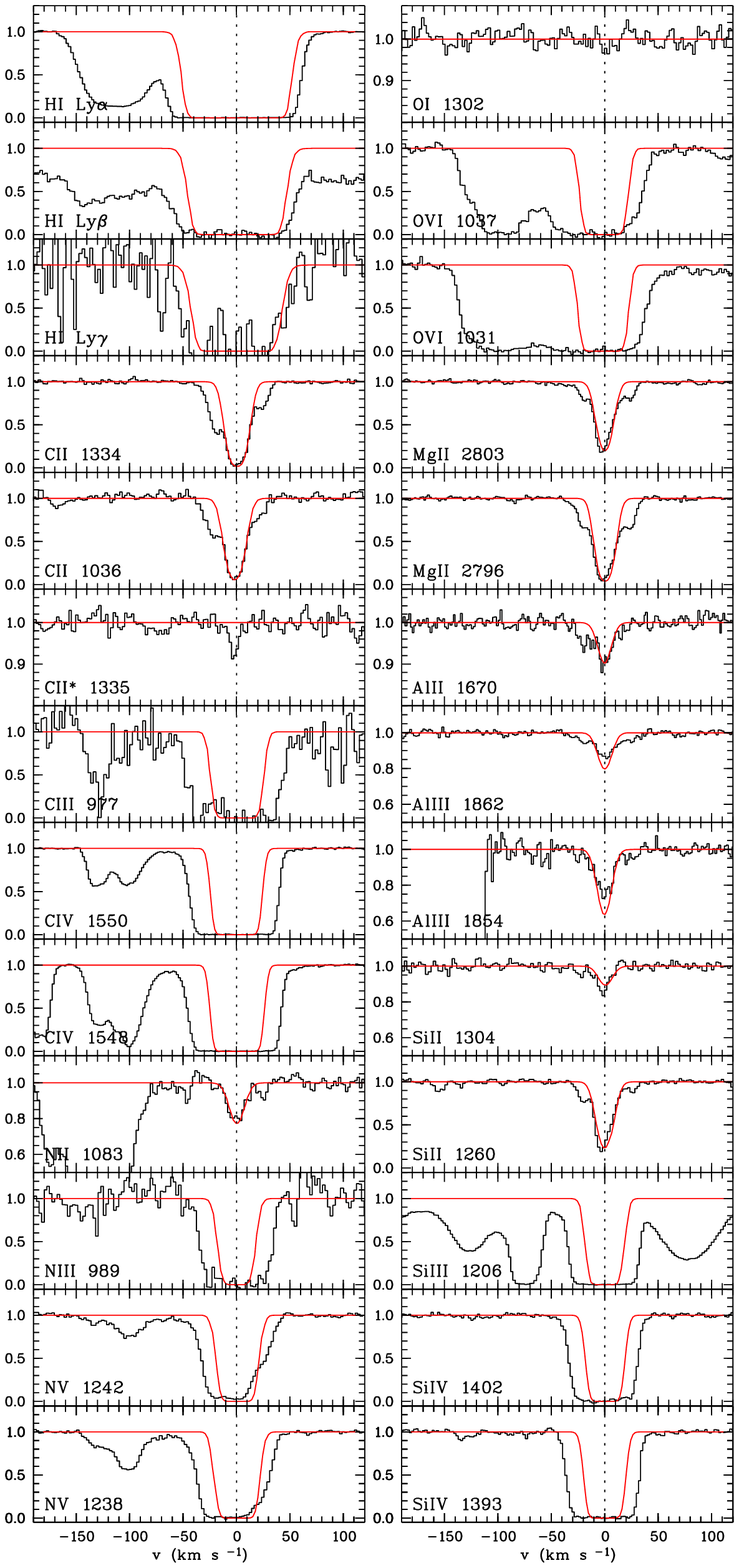}}
  \caption{Associated system at $z=2.1475$ towards HE~1341-1020.
Smooth lines indicate the profiles for the favorite model with a HM01 background and an additional power law spectrum $\propto \nu^{\,-1.0}$ strongly dominating the ionizing radiation.
  }
  \label{HE1341_2.1475}
\end{figure}

\begin{figure}
  \centering
  \resizebox{\hsize}{!}{\includegraphics[bb=33 485 373 780,clip=]{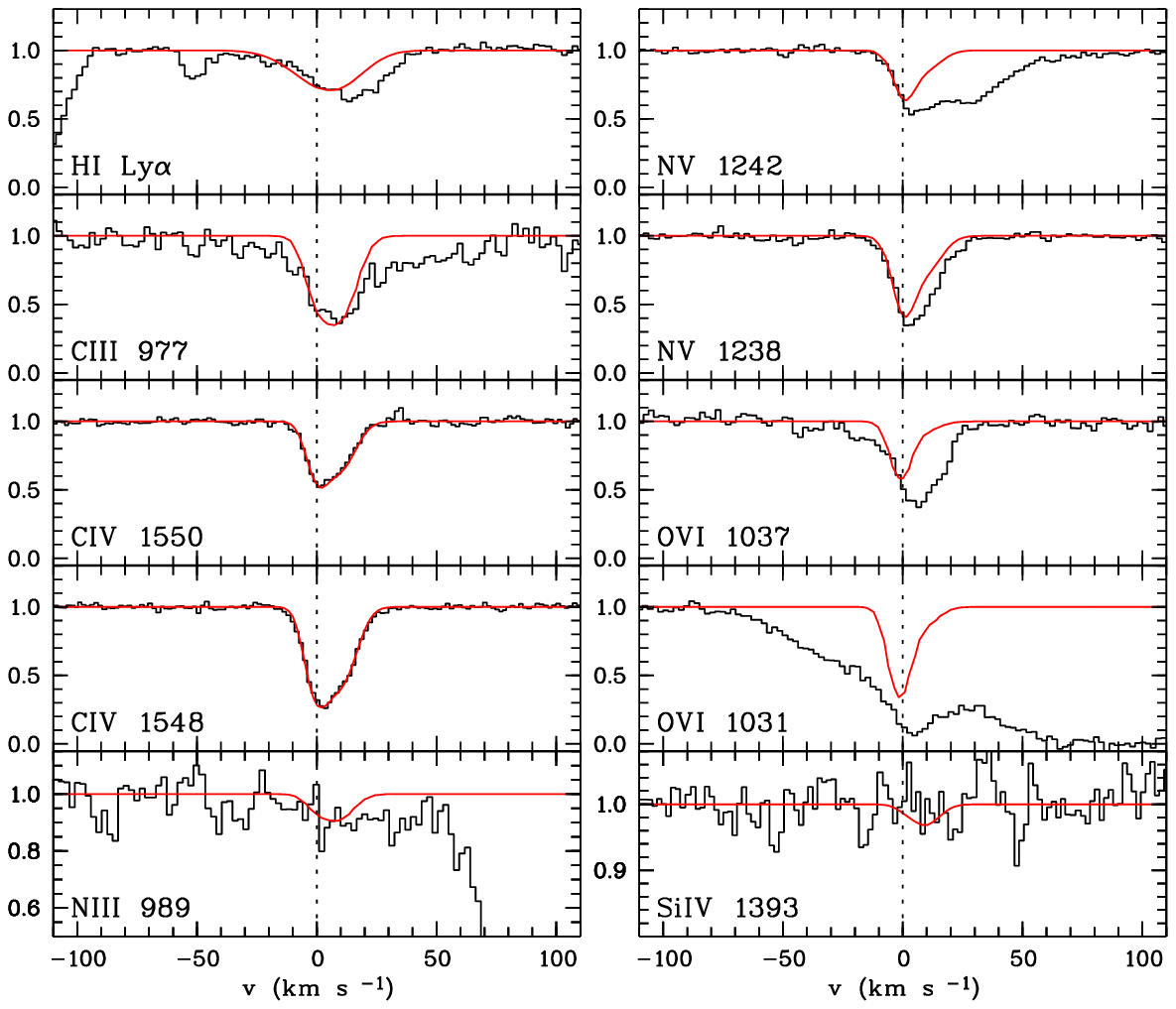}}
  \caption{Associated system at $z=2.3520$ towards Q~0329-385.
Smooth lines indicate the profiles for the favorite HM01 model plus an additional small contribution of a power law spectrum $\propto \nu^{\,-0.37}$.
  }
  \label{Q0329_2.3520}
\end{figure}

\begin{figure}
  \centering
  \resizebox{\hsize}{!}{\includegraphics[bb=33 542 373 780,clip=]{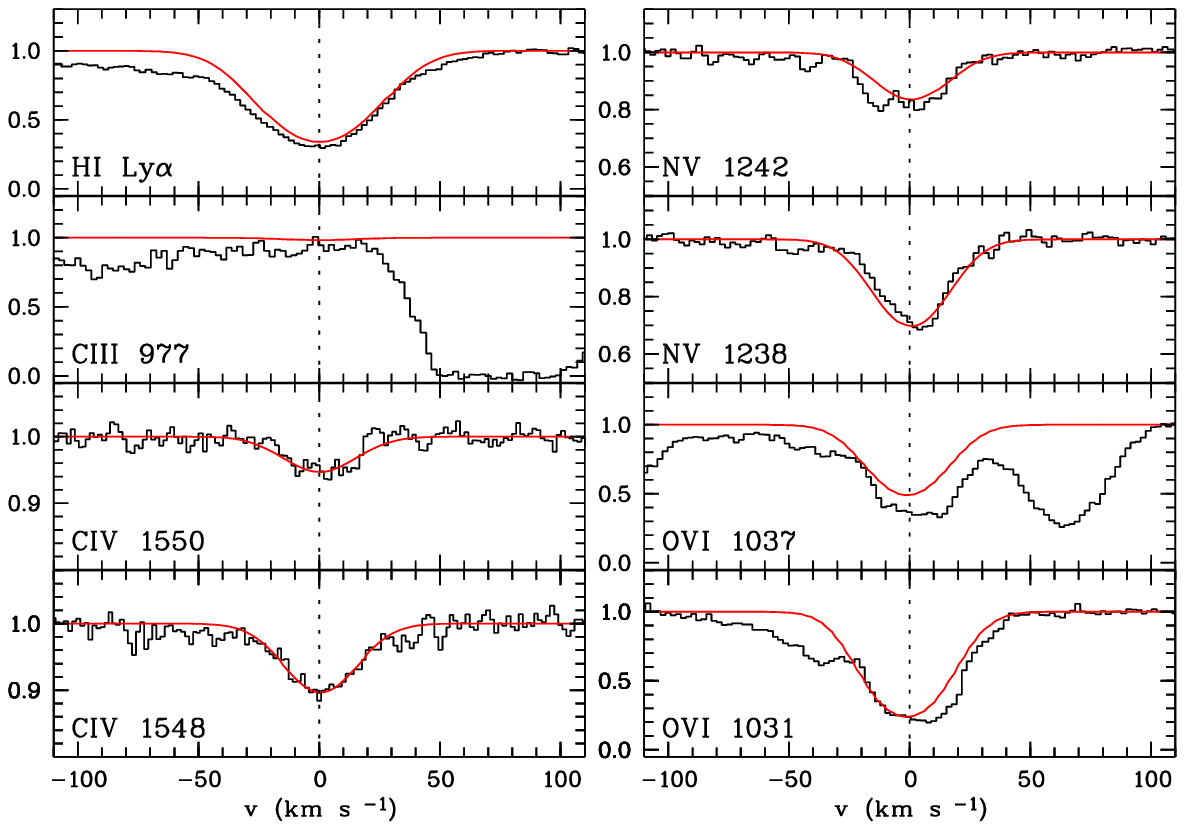}}
  \caption{Associated system at $z=2.4278$ towards HE~1158-1843.
Smooth lines indicate the profiles for the favorite model with a HM01 background and an additional power law spectrum $\propto \nu^{\,-1.0}$ strongly dominating the ionizing radiation.
  }
  \label{HE1158_2.4278}
\end{figure}

\begin{figure}
  \centering
  \resizebox{\hsize}{!}{\includegraphics[bb=33 542 373 780,clip=]{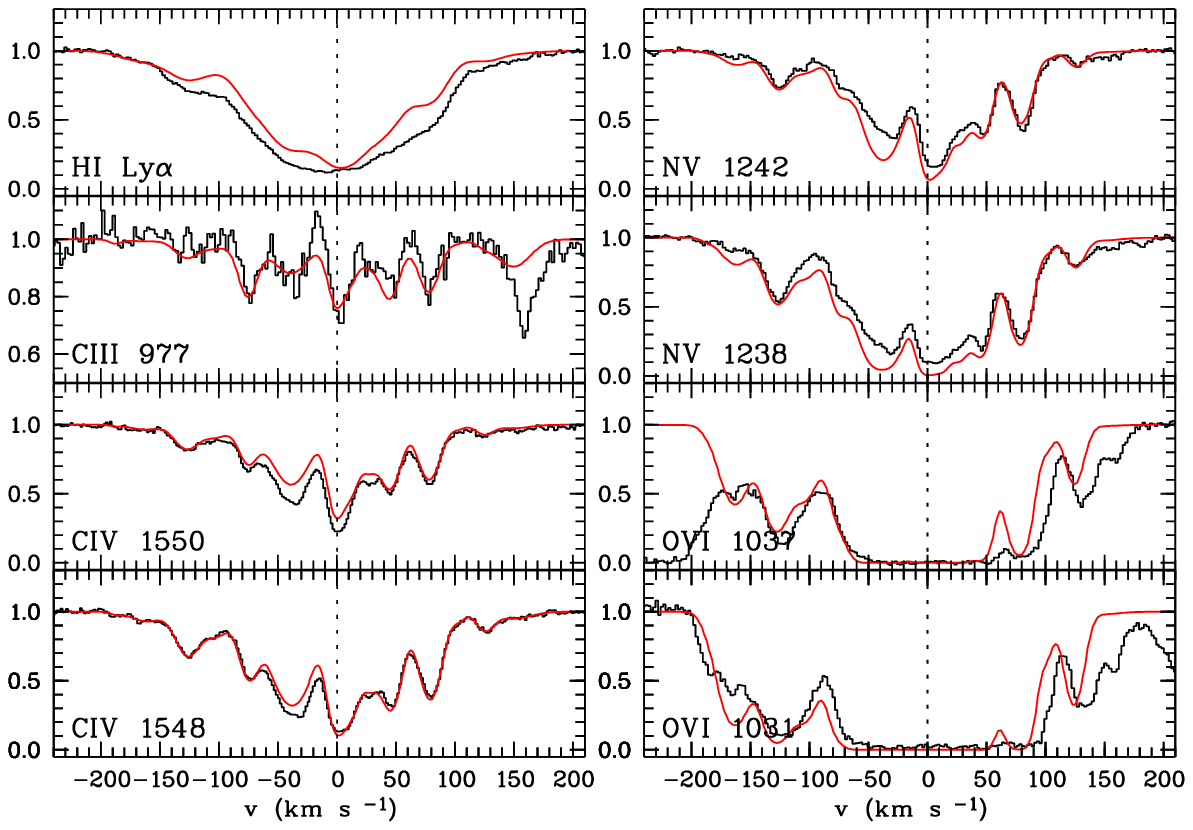}}
  \caption{Associated system at $z=2.4427$ towards HE~1158-1843.
Smooth lines indicate the profiles for the favorite model with a HM01 background and an additional power law spectrum $\propto \nu^{\,-1.0}$ strongly dominating the ionizing radiation.
  }
  \label{HE1158_2.4427}
\end{figure}

\begin{figure}
  \centering
  \resizebox{\hsize}{!}{\includegraphics[bb=33 381 373 780,clip=]{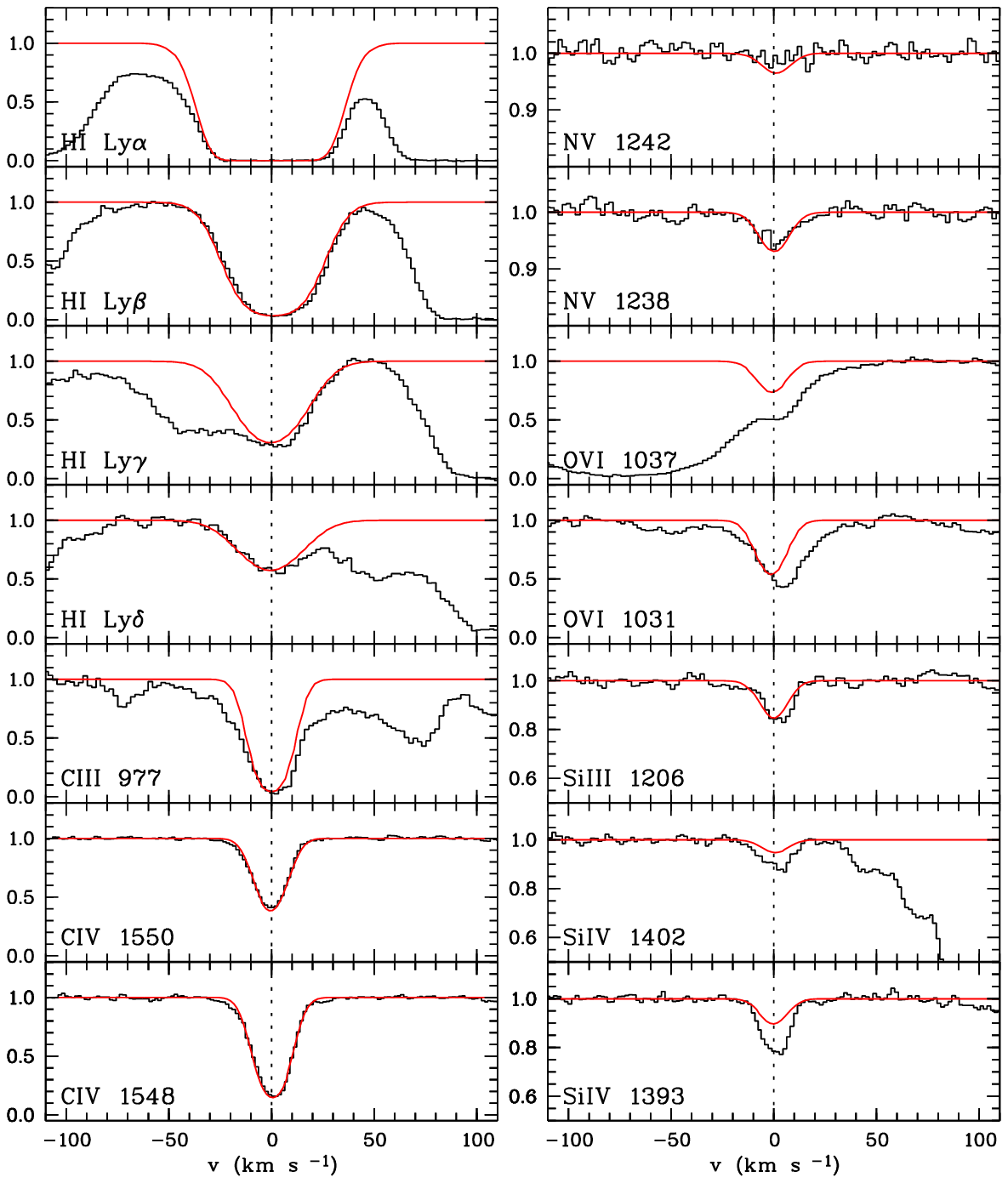}}
  \caption{Associated system at $z=2.6362$ towards Q~0453-423.
Smooth lines indicate the profiles for the favorite HM01 model plus an additional small contribution of a power law spectrum $\propto \nu^{\,-0.89}$.
  }
  \label{Q0453_2.6362}
\end{figure}

\begin{figure}
  \centering
  \resizebox{\hsize}{!}{\includegraphics[bb=33 485 373 780,clip=]{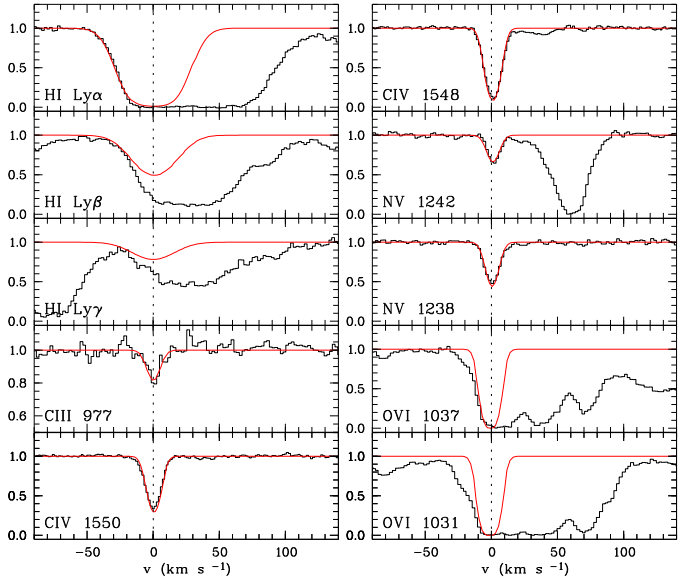}}
  \caption{Associated system at $z=2.7091$ towards PKS~0329-255.
Smooth lines indicate the profiles for the favorite HM01 model.
  }
  \label{PKS0329_2.7091}
\end{figure}

\begin{figure}
  \centering
  \resizebox{\hsize}{!}{\includegraphics[bb=33 485 373 780,clip=]{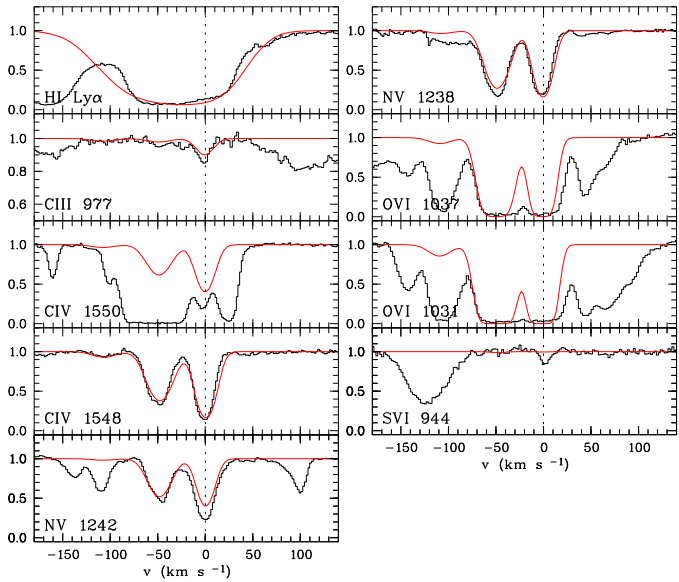}}
  \caption{Associated system at $z=2.8916$ towards HE~2347-4342.
Smooth lines indicate the profiles for the favorite HM01 model.
Due to the marginal detection of \ion{C}{iii} is the blue components, which may lead to uncertain density estimates, we slightly overestimate the blue wing of the \ion{H}{i} features in order to keep the metallicity constant in all 3 components.
  }
  \label{HE2347_2.8916}
\end{figure}

\begin{figure}
  \centering
  \resizebox{\hsize}{!}{\includegraphics[bb=33 435 373 780,clip=]{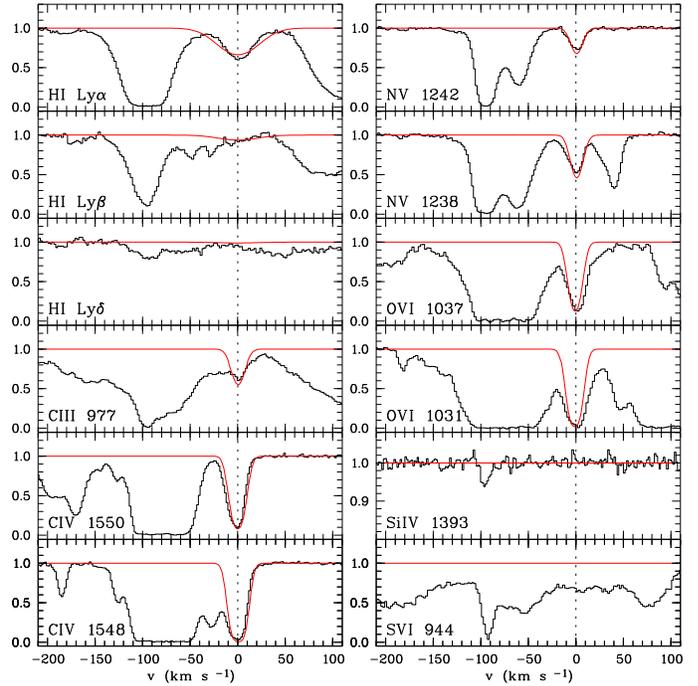}}
  \caption{Associated system at $z=2.8972$ towards HE~2347-4342.
Smooth lines indicate the profiles for the favorite HM01 model.
  }
  \label{HE2347_2.8972}
\end{figure}

\begin{figure}
  \centering
  \resizebox{\hsize}{!}{\includegraphics[bb=33 435 373 780,clip=]{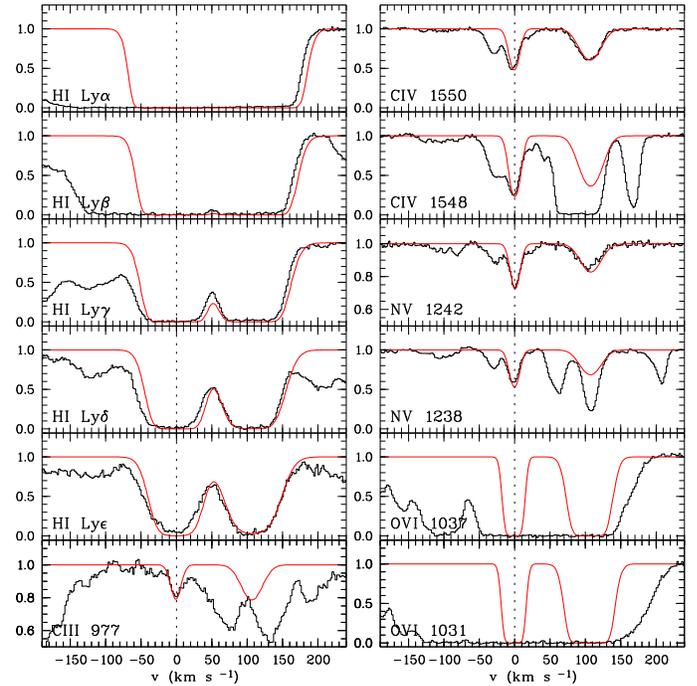}}
  \caption{Associated system at $z=2.9027$ towards HE~2347-4342.
Smooth lines indicate the profiles for the favorite HM01 model plus an additional small contribution of a power law spectrum $\propto \nu^{\,+0.56}$.
  }
  \label{HE2347_2.9027}
\end{figure}


\begin{figure}
  \centering
  \resizebox{\hsize}{!}{\includegraphics[bb=33 650 373 780,clip=]{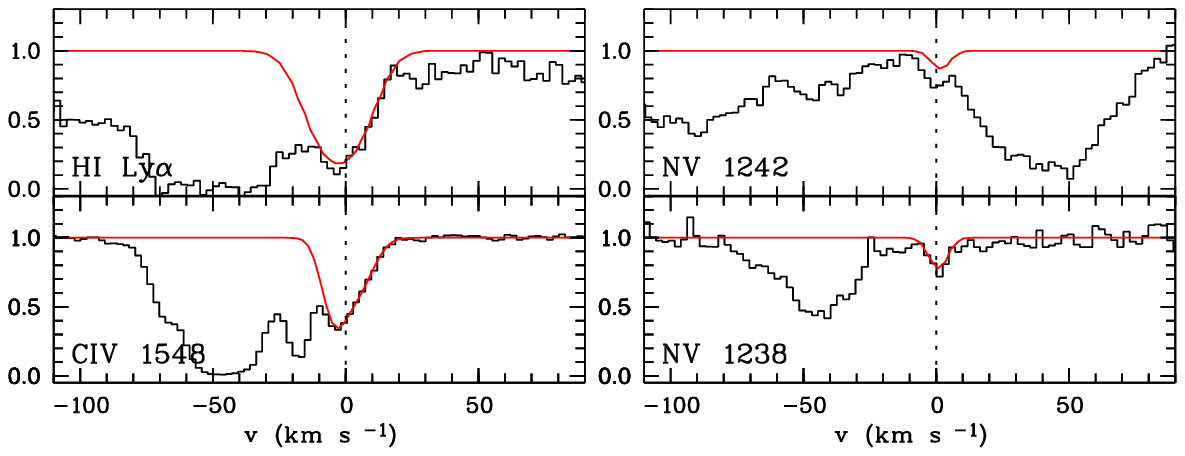}}
  \caption{Intervening system at $z=1.5814$ towards HE~0001-2340.
Due to few detected species no photoionization model can be constructed. 
Smooth lines indicate Doppler profile fits.
  }
  \label{HE0001_1.5814}
\end{figure}

\begin{figure}
  \centering
  \resizebox{\hsize}{!}{\includegraphics[bb=33 650 373 780,clip=]{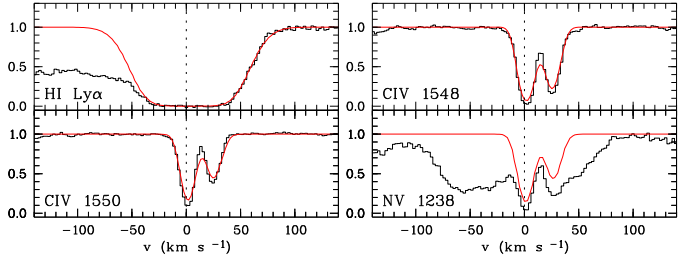}}
  \caption{Intervening system at $z=1.6109$ towards PKS~0237-23.
Due to few detected species no photoionization model can be constructed. 
Smooth lines indicate Doppler profile fits.
  }
  \label{PKS0237_1.6109}
\end{figure}

\begin{figure}
  \centering
  \resizebox{\hsize}{!}{\includegraphics[bb=33 650 373 780,clip=]{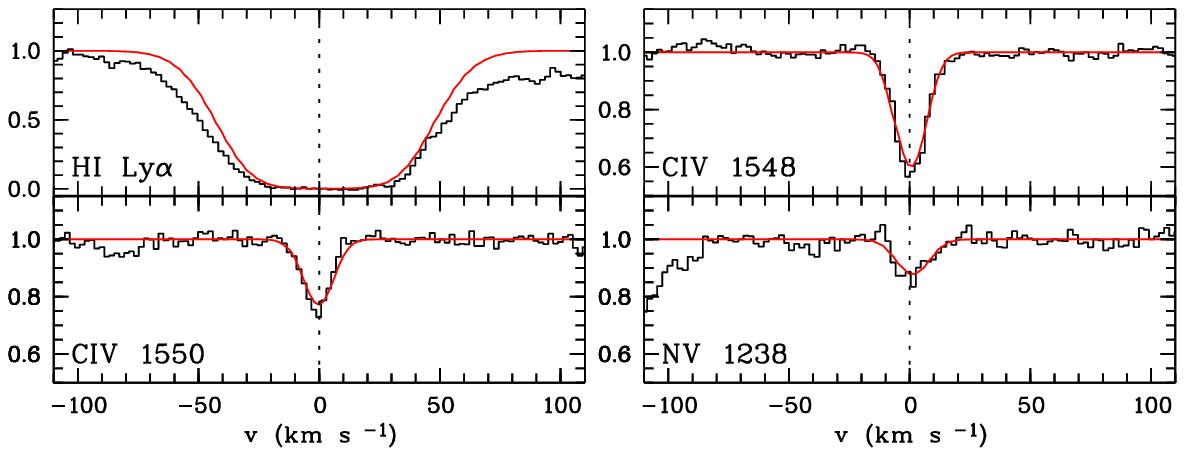}}
  \caption{Intervening system at $z=1.8748$ towards Q~0109-3518.
Due to few detected species no photoionization model can be constructed. 
Smooth lines indicate Doppler profile fits.
  }
  \label{Q0109_1.8748}
\end{figure}

\begin{figure}
  \centering
  \resizebox{\hsize}{!}{\includegraphics[bb=33 595 373 780,clip=]{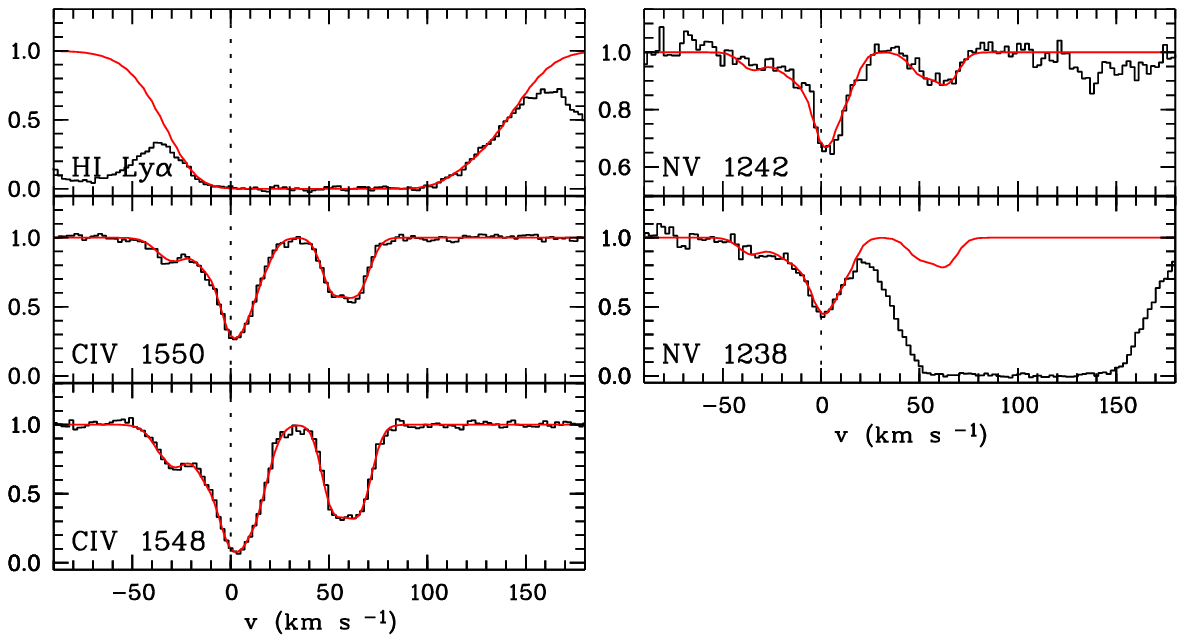}}
  \caption{Intervening system at $z=1.9144$ towards HE~1158-1843.
Due to few detected species no photoionization model can be constructed. 
Smooth lines indicate Doppler profile fits.
  }
  \label{HE1158_1.9144}
\end{figure}

\begin{figure}
  \centering
  \resizebox{\hsize}{!}{\includegraphics[bb=33 595 373 780,clip=]{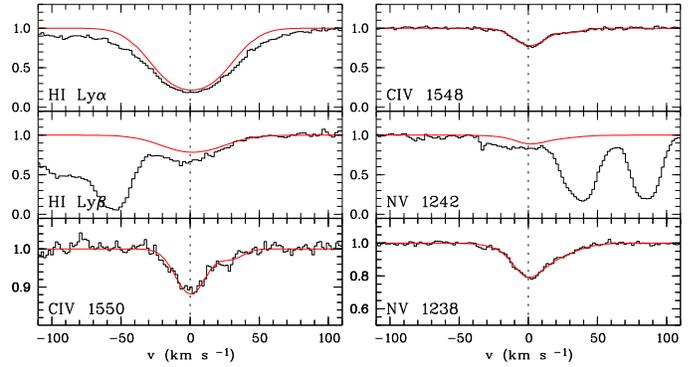}}
  \caption{Associated system at $z=2.8780$ towards HE~2347-4342.
Due to few detected species no photoionization model can be constructed. 
Smooth lines indicate Doppler profile fits.
  }
  \label{HE2347_2.8781}
\end{figure}

\clearpage
\section{Fitted line parameters}

\small
\begin{longtable}{l c c c c c r c r}
  \caption[]{\label{coldens_inter} Line parameters of intervening \ion{N}{v} absorption systems.
 The ionization parameter has been estimated adopting a \citet{haardtmadau2001} UV background at the appropriate redshift normalized to $\log J(1\,\textrm{Ryd}) = -21.15$.
}\\
  \hline\hline
 QSO & $z_{\mathrm{sys}}$ & $\log N_{\mathrm{sys}}(\ion{H}{i})$ &\# & $\log U$ & ion & $v\, (\mathrm{km\,s}^{-1})$ & $\log N$ & $b\, (\mathrm{km\,s}^{-1})$ \\
 \hline
 \endfirsthead
 \caption{continued.}\\
 \hline\hline
 QSO & $z_{\mathrm{sys}}$ & $\log N_{\mathrm{sys}}(\ion{H}{i})$ &\# & $\log U$ & ion & $v\, (\mathrm{km\,s}^{-1})$ & $\log N$ & $b\, (\mathrm{km\,s}^{-1})$ \\
 \hline
 \endhead
 \hline
 \endfoot
 \textrm{HE~0001-2340} & 1.5770 & $14.198\pm0.105$ & 1 & $-1.90$ & \ion{N}{v}    & $  -1.2 \pm  0.3$ & $13.054 \pm 0.023$ & $ 4.61 \pm  0.57$ \\
                       &        &                  &   &         & \ion{C}{iv}   & $  -3.3 \pm  0.0$ & $13.704 \pm 0.004$ & $ 6.14 \pm  0.03$ \\
                       &        &  		   &   &         & \ion{C}{ii}   & $  -0.3 \pm  0.3$ & $12.785 \pm 0.028$ & $ 2.80 \pm  0.49$ \\
                       &        &  		   &   &         & \ion{Si}{iv}  & $  -1.6 \pm  1.0$ & $11.956 \pm 0.099$ & $ 1.21 \pm  1.28$ \\
                       &        &  		   &   &         & \ion{Mg}{ii}  & $  -1.1 \pm  0.3$ & $11.366 \pm 0.041$ & $ 1.50 \pm  0.00$ \\
 \textrm{HE~0001-2340} & 1.5815 & $13.475\pm0.025$ & 1 & $\dots$ & \ion{N}{v}    & $   0.5 \pm  0.4$ & $12.718 \pm 0.071$ & $ 1.73 \pm  0.89$ \\
                       &        &  		   &   &         & \ion{C}{iv}   & $  -5.1 \pm  0.4$ & $13.100 \pm 0.099$ & $ 4.20 \pm  0.63$ \\
 \textrm{PKS~1448-232} & 1.5855 & $14.415\pm0.050$ & 1 & $-1.07$ & \ion{N}{v}    & $   0.6 \pm  0.5$ & $13.235 \pm 0.031$ & $ 8.66 \pm  0.80$ \\
\end{longtable}

\begin{longtable}{l c c c c c c r c r}
  \caption[]{\label{coldens_asso} Line parameters of associated \ion{N}{v} absorption systems.
 The ionization parameter has been estimated with an superposition of a \citet{haardtmadau2001} UV background and a QSO power law spectrum with a slope $\alpha$ where the relative intensity of the QSO radiation is varied between $0.1 \le J_{\mathrm{QSO}}/J_{\mathrm{HM01}} \le 100$. The quoted uncertainties of $\log U$ represent the resulting parameter range. If no $\alpha$ is given, a pure \citet{haardtmadau2001} background has been adopted.
}\\
  \hline\hline
 QSO & $z_{\mathrm{sys}}$ & $\log N_{\mathrm{sys}}(\ion{H}{i})$ & \# & $\alpha$ & $\log U$ & ion & $v\, (\mathrm{km\,s}^{-1})$ & $\log N$ & $b\, (\mathrm{km\,s}^{-1})$ \\
 \hline
 \endfirsthead
 \caption{continued.}\\
 \hline\hline
 QSO & $z_{\mathrm{sys}}$ & $\log N_{\mathrm{sys}}(\ion{H}{i})$ & \# & $\alpha$ & $\log U$ & ion & $v\, (\mathrm{km\,s}^{-1})$ & $\log N$ & $b\, (\mathrm{km\,s}^{-1})$ \\
 \hline
 \endhead
 \hline
 \endfoot
 \textrm{HE~1341-1020} & 2.1463 & $13.983\pm0.014$ & 1 & $-1.0$  & $-1.79\pm^{0.37}_{0.16}$ & \ion{N}{v}     & $ -30.7 \pm  0.3$ & $12.741 \pm 0.020$ & $ 8.27 \pm  0.50$ \\
                       &        &                  &   &         &                          & \ion{C}{iv}    & $ -33.3 \pm  0.5$ & $13.163 \pm 0.048$ & $ 4.63 \pm  0.49$ \\
                       &        &                  &   &         &                          & \ion{C}{iv}    & $ -23.6 \pm  0.6$ & $13.085 \pm 0.061$ & $ 4.83 \pm  0.64$ \\
                       &        &                  &   &         &                          & \ion{C}{iii}   & $ -27.7 \pm  0.9$ & $13.129 \pm 0.066$ & $ 8.36 \pm  1.63$ \\
                       &        &                  &   &         &                          & \ion{O}{vi}    & $ -27.0 \pm  0.9$ & $13.725 \pm 0.093$ & $ 7.16 \pm  1.02$ \\
                       &        &                  &   &         &                          & \ion{Si}{iv}   & $ -29.3 \pm  0.9$ & $11.904 \pm 0.036$ & $13.04 \pm  1.40$ \\
                       &        &                  &   &         &                          & \ion{Si}{iii}  & $ -23.6 \pm  0.2$ & $12.860 \pm 0.006$ & $23.01 \pm  0.23$ \\
                       &        &                  & 2 & $-1.0$  & $-0.94\pm^{0.51}_{0.22}$ & \ion{N}{v}     & $  -0.1 \pm  0.1$ & $13.530 \pm 0.004$ & $16.16 \pm  0.21$ \\
\end{longtable}

\end{document}